%% file: main.tex
\documentclass[twocolumn,floatfix]{aastex63}

\usepackage{longtable}
\usepackage{amsmath}
\usepackage{amssymb}
\usepackage{amsthm}
\usepackage{graphicx}
\usepackage[colorinlistoftodos]{todonotes}
\usepackage{natbib}

\usepackage{lineno}
\usepackage{soul}


\usepackage{subfigure}
\usepackage{makecell}
\usepackage{threeparttable}
\usepackage{changepage}
\usepackage{comment}
\usepackage{float} 
\usepackage{capt-of} 

\newcommand{\bluebright}{{\it bright \& blue}}
\newcommand{\reddishslow}{{\it reddish \& slow}}
\newcommand{\redfaint}{{\it red \& faint}}
\newcommand{\lTT}{{\it low-TT}}
\newcommand{\Top}{{\it Top}}

\definecolor{darkgreen}{rgb}{0.05,0.3,0.05}

\newcommand{\teff}{$t_\mathrm{eff}$}
\newcommand{\teffeq}{t_\mathrm{eff}}
\newcommand{\texp}{$t_\mathrm{exp}$}
\newcommand{\texpeq}{t_\mathrm{exp}}

\defcitealias{gwtc3}{GWTC}
\defcitealias{gwtc2.1}{GWTC-2.1}
\defcitealias{gwtc2}{GWTC-2}
\defcitealias{gwtc1}{GWTC-1}

\begin{document}

\title{Designing an Optimal Kilonova Search using DECam for Gravitational Wave Events}

\input{authors}


\begin{abstract}
We address the problem of optimally identifying all kilonovae detected via gravitational wave emission in the upcoming LIGO/Virgo/KAGRA observing run, O4, which is expected to be sensitive to a factor of $\sim 7$ more Binary Neutron Stars (BNS) alerts than previously. Electromagnetic follow-up of all but the brightest of these new events will require $>1$ meter telescopes, for which limited time is available. We present an optimized observing strategy for the DECam during O4. We base our study on simulations of gravitational wave events expected for O4 and wide-prior kilonova simulations. We derive the detectabilities of events for realistic observing conditions. We optimize our strategy for confirming a kilonova while minimizing telescope time. For a wide range of kilonova parameters, corresponding to a fainter kilonova compared to GW170817/AT2017gfo we find that, with this optimal strategy, the discovery probability for electromagnetic counterparts with the DECam is $\sim 80\%$ at the nominal BNS gravitational wave detection limit for O4 (190 Mpc), which corresponds to a $\sim 30\%$ improvement compared to the strategy adopted during the previous observing run. For more distant events ($\sim 330$ Mpc), we reach a $\sim 60\%$ probability of detection, a factor of $\sim 2$ increase. For a brighter kilonova model dominated by the blue component that reproduces the observations of GW170817/AT2017gfo, we find that we can reach $\sim 90\%$ probability of detection out to 330 Mpc, representing an increase of $\sim 20 \%$, while also reducing the total telescope time required to follow-up events by $\sim 20\%$.

\end{abstract}


\keywords{gravitational waves --- surveys}

\reportnum{DES-2022-0714}
\reportnum{FERMILAB-PUB-23-048-PPD}


\section{Introduction}
\input{intro}

\section{Simulation Data}\label{section:data}
\input{data}

\section{Detection Methodology}
\label{section:dectability} 

\input{methods}

\section{Exploring Parameter Space}\label{section:strategies} 
\input{results}

\section{Discussion}\label{section:discussion}

\input{discussion}

\section{Conclusion}\label{section:conclusion}

In this paper, we create families of observing strategies that optimize the probability of detecting a KN within DECam's images. We examine various filter choices, depths, area coverage, and cadence of observations in order to ensure optimal chance of detection. Given the expanded range of sensitivity in future LVK observing runs, deeper exposures will be necessary in order to be sensitive to the quickly fading counterpart.  As we do not have unlimited time for such follow ups, we examine how we can optimize our chance of detection while taking into account real world constraints.

We chose to optimize our strategies based on the probability of detecting the KN within two images that are at least 30 min apart. This constraint is put in place in order to help eliminate asteroids and other sources of noise. 
We explore two different types of observing scenarios. The first is a homogeneous covering of the sky area with a single exposure time, and the second uses deeper exposures in the higher probability sky areas and shallower exposures on the rest of the area.  
We then categorize our strategies by observational constraint, where each family of strategies is taken from the top $10\%$ or $5\%$  of \Top\ strategies. The \Top\ strategies use all available  resources, and are useful as a benchmark for the full detection capability of the DECam. 

Examining each of the realistic observing scenarios, we find we can achieve $\sim 75\%$ to $80\%$ probability of detection out to $190$ Mpc (the nominal limit of LVK BNS range) for a  wide rage of KN parameters (\reddishslow), $\sim 65\%$ for a fainter and redder KN (\redfaint) and over $90\%$ for a \bluebright\ model along the full range of distances limited to $330$Mpc. 
Additionally, we provide the mean detection probability and total telescope time required for detection and confirmation in each KN model for a given range of GW event area and distance in figure \ref{fig:distance_v_area_heatmap}. In particular, this plot might be used as a guide on how likely it is to succeed in KN detection of specific future events considering a trade-off between time budget and optimal chances.

While DECam will continue to be the optimal camera in the southern hemisphere during the next observing run, the efforts to detect the next KN optical counterpart will be greatly aided by other telescopes that are planned to be online during this time. For example, the expected addition of the Simonyi Telescope at the Vera Rubin Observatory.

\section*{Acknowledgments}
Funding for the DES Projects has been provided by the U.S. Department of Energy, the U.S. National Science Foundation, the Ministry of Science and Education of Spain, the Science and Technology Facilities Council of the United Kingdom, the Higher Education Funding Council for England, the National Center for Supercomputing 
Applications at the University of Illinois at Urbana-Champaign, the Kavli Institute of Cosmological Physics at the University of Chicago, 
the Center for Cosmology and Astro-Particle Physics at the Ohio State University,
the Mitchell Institute for Fundamental Physics and Astronomy at Texas A\&M University, Financiadora de Estudos e Projetos, 
Funda{\c c}{\~a}o Carlos Chagas Filho de Amparo {\`a} Pesquisa do Estado do Rio de Janeiro, Conselho Nacional de Desenvolvimento Cient{\'i}fico e Tecnol{\'o}gico and 
the Minist{\'e}rio da Ci{\^e}ncia, Tecnologia e Inova{\c c}{\~a}o, the Deutsche Forschungsgemeinschaft and the Collaborating Institutions in the Dark Energy Survey.

The Collaborating Institutions are Argonne National Laboratory, the University of California at Santa Cruz, the University of Cambridge, Centro de Investigaciones Energ{\'e}ticas, 
Medioambientales y Tecnol{\'o}gicas-Madrid, the University of Chicago, University College London, the DES-Brazil Consortium, the University of Edinburgh, 
the Eidgen{\"o}ssische Technische Hochschule (ETH) Z{\"u}rich, 
Fermi National Accelerator Laboratory, the University of Illinois at Urbana-Champaign, the Institut de Ci{\`e}ncies de l'Espai (IEEC/CSIC), 
the Institut de F{\'i}sica d'Altes Energies, Lawrence Berkeley National Laboratory, the Ludwig-Maximilians Universit{\"a}t M{\"u}nchen and the associated Excellence Cluster Universe, 
the University of Michigan, NFS's NOIRLab, the University of Nottingham, The Ohio State University, the University of Pennsylvania, the University of Portsmouth, 
SLAC National Accelerator Laboratory, Stanford University, the University of Sussex, Texas A\&M University, and the OzDES Membership Consortium.

Based in part on observations at Cerro Tololo Inter-American Observatory at NSF’s NOIRLab (NOIRLab Prop. ID 2012B-0001; PI: J. Frieman), which is managed by the Association of Universities for Research in Astronomy (AURA) under a cooperative agreement with the National Science Foundation.

The DES data management system is supported by the National Science Foundation under Grant Numbers AST-1138766 and AST-1536171.
The DES participants from Spanish institutions are partially supported by MICINN under grants ESP2017-89838, PGC2018-094773, PGC2018-102021, SEV-2016-0588, SEV-2016-0597, and MDM-2015-0509, some of which include ERDF funds from the European Union. IFAE is partially funded by the CERCA program of the Generalitat de Catalunya.
Research leading to these results has received funding from the European Research
Council under the European Union's Seventh Framework Program (FP7/2007-2013) including ERC grant agreements 240672, 291329, and 306478.
We  acknowledge support from the Brazilian Instituto Nacional de Ci\^encia
e Tecnologia (INCT) e-Universe (CNPq grant 465376/2014-2).

This manuscript has been authored by Fermi Research Alliance, LLC under Contract No. DE-AC02-07CH11359 with the U.S. Department of Energy, Office of Science, Office of High Energy Physics.

This material is based upon work supported by the National Science Foundation Graduate Research Fellowship Program under Grant No. 1744555. Any opinions, findings, and conclusions or recommendations expressed in this material are those of the author(s) and do not necessarily reflect the views of the National Science Foundation.
This paper has gone through internal review by the DES collaboration.
CRB acknowledges the financial support from CNPq (316072/2021-4) and from FAPERJ (grants 201.456/2022 and 210.330/2022)

{\color{black} R. Morgan thanks the LSSTC Data Science Fellowship Program, which is funded by LSSTC, NSF Cybertraining Grant \#1829740, the Brinson Foundation, and the Moore Foundation.}

A. Palmese acknowledges support for this work was provided by NASA through the NASA Hubble Fellowship grant HST-HF2-51488.001-A awarded by the Space Telescope Science Institute, which is operated by Association of Universities for Research in Astronomy, Inc., for NASA, under contract NAS5-26555.

L. Santana-Silva acknowledges the financial support from FAPESP through the grant $\#$2020/03301-5.

MESP is funded by the Deutsche Forschungsgemeinschaft (DFG, German Research Foundation) under Germany's Excellence Strategy -- EXC 2121 ``Quantum Universe'' -- 390833306.

F.~Olivares~E.\ acknowledges financial support from FONDECYT grant nr.\ 1201223.



\bibliography{AGbib}
\bibliographystyle{yahapj}
\end{document}

%% file: authors.tex

\correspondingauthor{Clecio R. Bom}
\email{debom@cbpf.br}

\author[0000-0003-4383-2969]{C.~R.~Bom}
\affil{Centro Brasileiro de Pesquisas F\'isicas, Rua Dr. Xavier Sigaud 150, CEP 22290-180, Rio de Janeiro, RJ, Brazil}
\affil{Centro Federal de Educa\c{c}\~ao Tecnol\'ogica Celso Suckow da Fonseca, Rodovia M\'ario Covas, lote J2, quadra J, CEP 23810-000, Itagua\'i, RJ, Brazil}

\author[0000-0002-0609-3987]{J.~Annis}
\affil{Fermi National Accelerator Laboratory, P. O. Box 500, Batavia, IL 60510, USA}

\author[0000-0001-9578-6322]{A.~Garcia}
\affil{Department of Physics, University of Michigan, Ann Arbor, MI 48109, USA}

\author[0000-0002-6011-0530]{A.~Palmese}\thanks{NASA Einstein Fellow}
\affil{Department of Physics, University of California Berkeley, 366 LeConte Hall MC 7300, Berkeley, CA, 94720, USA}

\author[0000-0001-5399-0114]{N.~Sherman}
\affil{Department of Physics, University of Michigan, Ann Arbor, MI 48109, USA}

\author[0000-0001-6082-8529]{M.~Soares-Santos}
\affil{Department of Physics, University of Michigan, Ann Arbor, MI 48109, USA}

\author[0000-0003-3402-6164]{L.~Santana-Silva}
\affil{NAT-Universidade Cruzeiro do Sul / Universidade Cidade de S{\~a}o Paulo, Rua Galv{\~a}o Bueno, 868, 01506-000, S{\~a}o Paulo, SP, Brazil}

\author[0000-0002-7016-5471]{R.~Morgan}
\affil{Physics Department, 2320 Chamberlin Hall, University of Wisconsin-Madison, 1150 University Avenue Madison, WI 53706-1390, USA}
\affil{Legacy Survey of Space and Time Corporation Data Science Fellowship Program}

\author{K. Bechtol}
\affil{Physics Department, 2320 Chamberlin Hall, University of Wisconsin-Madison, 1150 University Avenue Madison, WI 53706-1390, USA}

\author{T. Davis}
\affil{School of Mathematics and Physics, The University of Queensland, Brisbane, QLD 4072, Australia}
\author{H.T. Diehl}
\affil{Fermi National Accelerator Laboratory, P. O. Box 500, Batavia, IL 60510, USA} 

\author[0000-0002-7069-7857]{S.~S.~Allam}\thanks{\textit{in memoriam}}
\affil{Fermi National Accelerator Laboratory, P. O. Box 500, Batavia, IL 60510, USA}

\author[0000-0002-6119-5353]{T.~G.~Bachmann}
\affil{Department of Astronomy and Astrophysics, University of Chicago, Chicago, IL 60637, USA}

\author{B.~M.~O.~Fraga}
\affil{Centro Brasileiro de Pesquisas F\'isicas, Rua Dr. Xavier Sigaud 150, CEP 22290-180, Rio de Janeiro, RJ, Brazil}

\author[0000-0002-9370-8360]{J.~Garc\'ia-Bellido}
\affil{Instituto de Fisica Teorica UAM/CSIC, Universidad Autonoma de Madrid, 28049 Madrid, Spain}

\author[0000-0003-2524-5154]{M.~S.~S.~Gill}
\affil{SLAC National Accelerator Laboratory, Menlo Park, CA 94025, USA}

\author[0000-0001-6718-2978]{K.~Herner}
\affil{Fermi National Accelerator Laboratory, P. O. Box 500, Batavia, IL 60510, USA}

\author{C.~D.~Kilpatrick}
\affil{Center for Interdisciplinary Exploration and Research in Astrophysics (CIERA) and Department of Physics and Astronomy, Northwestern University, Evanston, IL 60208, USA}

\author[0000-0003-2206-2651]{M.~Makler}
\affil{International Center for Advanced Studies \& Instituto de Ciencias F\'isicas,  ECyT-UNSAM \& CONICET, 1650, Buenos Aires, Argentina}
\affil{Centro Brasileiro de Pesquisas F\'isicas, Rua Dr. Xavier Sigaud 150, CEP 22290-180, Rio de Janeiro, RJ, Brazil}

\author[0000-0002-5115-6377]{F.~Olivares~E.}
\affil{Instituto de Astronom\'{\i}a y Ciencias Planetarias, Universidad de Atacama, Av.\ Copayapu 485, Copiap\'o, Chile}

\author[0000-0002-7131-7684]{M.~E.~S.~Pereira}
\affil{Hamburger Sternwarte, Universit{\"a}t Hamburg, Gojenbergsweg 112, 21029 Hamburg, Germany}

\author{J.~Pineda}
\affil{Departamento de Ciencias Fisicas, Universidad Andres Bello, Avda. Republica 252, Santiago, Chile}
\affil{Millennium Institute of Astrophysics (MAS), Nuncio Monse\~nor S\'otero Sanz 100, Providencia, Santiago, Chile}

\author{A.~Santos}
\affil{Centro Brasileiro de Pesquisas F\'isicas, Rua Dr. Xavier Sigaud 150, CEP 22290-180, Rio de Janeiro, RJ, Brazil}

\author[0000-0001-7211-5729]{D.~L.~Tucker}
\affil{Fermi National Accelerator Laboratory, P. O. Box 500, Batavia, IL 60510, USA}

\author[0000-0001-8653-7738]{M.~P.~Wiesner}
\affil{Benedictine University, Department of Physics, 5700 College Road, Lisle, IL 60532, USA}


\author{M.~Aguena}
\affil{Laborat\'orio Interinstitucional de e-Astronomia - LIneA, Rua Gal. Jos\'e Cristino 77, Rio de Janeiro, RJ - 20921-400, Brazil}

\author{O.~Alves}
\affil{Department of Physics, University of Michigan, Ann Arbor, MI 48109, USA}

\author{D.~Bacon}
\affil{Institute of Cosmology and Gravitation, University of Portsmouth, Portsmouth, PO1 3FX, UK}

\author{P.~H.~Bernardinelli}
\affil{Astronomy Department, University of Washington, Box 351580, Seattle, WA 98195, USA}

\author{E.~Bertin}
\affil{CNRS, UMR 7095, Institut d'Astrophysique de Paris, F-75014, Paris, France}
\affil{Sorbonne Universit\'es, UPMC Univ Paris 06, UMR 7095, Institut d'Astrophysique de Paris, F-75014, Paris, France}

\author[0000-0002-4900-805X]{S.~Bocquet}
\affil{University Observatory, Faculty of Physics, Ludwig-Maximilians-Universit\"at, Scheinerstr. 1, 81679 Munich, Germany}

\author[0000-0002-8458-5047]{D.~Brooks}
\affil{Department of Physics \& Astronomy, University College London, Gower Street, London, WC1E 6BT, UK}

\author[0000-0002-4802-3194]{M.~Carrasco~Kind}
\affil{Center for Astrophysical Surveys, National Center for Supercomputing Applications, 1205 West Clark St., Urbana, IL 61801, USA}
\affil{Department of Astronomy, University of Illinois at Urbana-Champaign, 1002 W. Green Street, Urbana, IL 61801, USA}

\author[0000-0002-3130-0204]{J.~Carretero}
\affil{Institut de F\'{\i}sica d'Altes Energies (IFAE), The Barcelona Institute of Science and Technology, Campus UAB, 08193 Bellaterra (Barcelona) Spain}

\author[0000-0003-1949-7638]{C.~Conselice}
\affil{Jodrell Bank Center for Astrophysics, School of Physics and Astronomy, University of Manchester, Oxford Road, Manchester, M13 9PL, UK}
\affil{University of Nottingham, School of Physics and Astronomy, Nottingham NG7 2RD, UK}

\author{M.~Costanzi}
\affil{Astronomy Unit, Department of Physics, University of Trieste, via Tiepolo 11, I-34131 Trieste, Italy}
\affil{INAF-Osservatorio Astronomico di Trieste, via G. B. Tiepolo 11, I-34143 Trieste, Italy}
\affil{Institute for Fundamental Physics of the Universe, Via Beirut 2, 34014 Trieste, Italy}

\author{L.~N.~da Costa}
\affil{Laborat\'orio Interinstitucional de e-Astronomia - LIneA, Rua Gal. Jos\'e Cristino 77, Rio de Janeiro, RJ - 20921-400, Brazil
}

\author[0000-0001-8318-6813]{J.~De~Vicente}
\affil{Centro de Investigaciones Energ\'eticas, Medioambientales y Tecnol\'ogicas (CIEMAT), Madrid, Spain}

\author[0000-0002-0466-3288]{S.~Desai}
\affil{Department of Physics, IIT Hyderabad, Kandi, Telangana 502285, India}

\author{P.~Doel}
\affil{Department of Physics \& Astronomy, University College London, Gower Street, London, WC1E 6BT, UK}

\author{S.~Everett}
\affil{Jet Propulsion Laboratory, California Institute of Technology, 4800 Oak Grove Dr., Pasadena, CA 91109, USA}

\author{I.~Ferrero}
\affil{Institute of Theoretical Astrophysics, University of Oslo. P.O. Box 1029 Blindern, NO-0315 Oslo, Norway}

\author[0000-0003-4079-3263]{J.~Frieman}
\affil{Fermi National Accelerator Laboratory, P. O. Box 500, Batavia, IL 60510, USA}
\affil{Kavli Institute for Cosmological Physics, University of Chicago, Chicago, IL 60637, USA}

\author{M.~Gatti}
\affil{Department of Physics and Astronomy, University of Pennsylvania, Philadelphia, PA 19104, USA}

\author[0000-0001-6942-2736]{D.~W.~Gerdes}
\affil{Department of Astronomy, University of Michigan, Ann Arbor, MI 48109, USA}
\affil{Department of Physics, University of Michigan, Ann Arbor, MI 48109, USA}

\author[0000-0003-3270-7644]{D.~Gruen}
\affil{University Observatory, Faculty of Physics, Ludwig-Maximilians-Universit\"at, Scheinerstr. 1, 81679 Munich, Germany}

\author{R.~A.~Gruendl}
\affil{Center for Astrophysical Surveys, National Center for Supercomputing Applications, 1205 West Clark St., Urbana, IL 61801, USA}
\affil{Department of Astronomy, University of Illinois at Urbana-Champaign, 1002 W. Green Street, Urbana, IL 61801, USA}

\author[0000-0003-0825-0517]{G.~Gutierrez}
\affil{Fermi National Accelerator Laboratory, P. O. Box 500, Batavia, IL 60510, USA}

\author{S.~R.~Hinton}
\affil{School of Mathematics and Physics, University of Queensland,  Brisbane, QLD 4072, Australia}

\author{D.~L.~Hollowood}
\affil{Santa Cruz Institute for Particle Physics, Santa Cruz, CA 95064, USA}

\author[0000-0002-6550-2023]{K.~Honscheid}
\affil{Center for Cosmology and Astro-Particle Physics, The Ohio State University, Columbus, OH 43210, USA}
\affil{Department of Physics, The Ohio State University, Columbus, OH 43210, USA
}

\author[0000-0001-5160-4486]{D.~J.~James}
\affil{Center for Astrophysics $\vert$ Harvard \& Smithsonian, 60 Garden Street, Cambridge, MA 02138, USA}

\author[0000-0003-0120-0808]{K.~Kuehn}
\affil{Australian Astronomical Optics, Macquarie University, North Ryde, NSW 2113, Australia}
\affil{Lowell Observatory, 1400 Mars Hill Rd, Flagstaff, AZ 86001, USA}

\author[0000-0003-2511-0946]{N.~Kuropatkin}
\affil{Fermi National Accelerator Laboratory, P. O. Box 500, Batavia, IL 60510, USA}

\author[0000-0002-8873-5065]{P.~Melchior}
\affil{Department of Astrophysical Sciences, Princeton University, Peyton Hall, Princeton, NJ 08544, USA}

\author[0000-0001-9497-7266]{J. Mena-Fern{\'a}ndez}
\affil{Centro de Investigaciones Energ\'eticas, Medioambientales y Tecnol\'ogicas (CIEMAT), Madrid, Spain}

\author[0000-0002-1372-2534]{F.~Menanteau}
\affil{Center for Astrophysical Surveys, National Center for Supercomputing Applications, 1205 West Clark St., Urbana, IL 61801, USA}
\affil{Department of Astronomy, University of Illinois at Urbana-Champaign, 1002 W. Green Street, Urbana, IL 61801, USA}

\author[0000-0001-9186-6042]{A.~Pieres}
\affil{Laborat\'orio Interinstitucional de e-Astronomia - LIneA, Rua Gal. Jos\'e Cristino 77, Rio de Janeiro, RJ - 20921-400, Brazil}
\affil{Observat\'orio Nacional, Rua Gal. Jos\'e Cristino 77, Rio de Janeiro, RJ - 20921-400, Brazil}

\author[0000-0002-2598-0514]{A.~A.~Plazas~Malag\'on}
\affil{Department of Astrophysical Sciences, Princeton University, Peyton Hall, Princeton, NJ 08544, USA}

\author{M.~Raveri}
\affil{Department of Physics, University of Genova and INFN, Via Dodecaneso 33, 16146, Genova, Italy}

\author{M.~Rodriguez-Monroy}
\affil{Centro de Investigaciones Energ\'eticas, Medioambientales y Tecnol\'ogicas (CIEMAT), Madrid, Spain}

\author[0000-0002-9646-8198]{E.~Sanchez}
\affil{Centro de Investigaciones Energ\'eticas, Medioambientales y Tecnol\'ogicas (CIEMAT), Madrid, Spain}

\author{B.~Santiago}
\affil{Instituto de F\'\i sica, UFRGS, Caixa Postal 15051, Porto Alegre, RS - 91501-970, Brazil}
\affil{Laborat\'orio Interinstitucional de e-Astronomia - LIneA, Rua Gal. Jos\'e Cristino 77, Rio de Janeiro, RJ - 20921-400, Brazil}

\author[0000-0002-1831-1953]{I.~Sevilla-Noarbe}
\affil{Centro de Investigaciones Energ\'eticas, Medioambientales y Tecnol\'ogicas (CIEMAT), Madrid, Spain}

\author[0000-0002-3321-1432]{M.~Smith}
\affil{School of Physics and Astronomy, University of Southampton,  Southampton, SO17 1BJ, UK}

\author[0000-0002-7047-9358]{E.~Suchyta}
\affil{Computer Science and Mathematics Division, Oak Ridge National Laboratory, Oak Ridge, TN 37831}

\author{M.~E.~C.~Swanson
}
\affil{Department of Physics \& Astronomy, University College London, Gower Street, London, WC1E 6BT, UK}
\affil{Center for Astrophysics $\vert$ Harvard \& Smithsonian, 60 Garden Street, Cambridge, MA 02138, USA}

\author[0000-0003-1704-0781]{G.~Tarle}
\affil{Department of Physics, University of Michigan, Ann Arbor, MI 48109, USA}

\author[0000-0001-7836-2261]{C.~To}
\affil{Center for Cosmology and Astro-Particle Physics, The Ohio State University, Columbus, OH 43210, USA}

\author{N.~Weaverdyck}
\affil{Department of Physics, University of Michigan, Ann Arbor, MI 48109, USA}
\affil{Lawrence Berkeley National Laboratory, 1 Cyclotron Road, Berkeley, CA 94720, USA}

%% file: intro.tex
The discovery of an electromagnetic (EM) counterpart to a gravitational wave (GW) event GW170817  ushered in a new era of astrophysics and multi-messenger astronomy (\citealp{Abbott_2017,soares2017electromagnetic}; \citealp{cowperthwaite2017electromagnetic}; \citealp{Nicholl_2017}; \citealp{Coulter1556,drout2017}; \citealp{10.1093/pasj/psx118}). The discovery of the optical counterpart occurred $\sim$12hr~\citep{abbott2017multi,arcavi,coulter2017swope, marcelle2017,lipunov} after merger, allowing for extensive imaging and spectroscopic observations~\citep[e.g.][]{toros,2017PASA...34...69A,Balasubramanian21,kilpatrick22}. This enabled the community to not only confirm the existence of long-hypothesized kilonovae (KNe) and to characterize their lightcurves, but also to derive the first cosmological standard siren \citep{schutz} constraint \citep{abbott2017multi}.

The EM counterpart to GW1701817, AT2017gfo, was observed in $\gamma$-rays 2 seconds after the merger signal \cite[e.g.,][]{GRB170817A,goldstein2017,savchenko2017}, then a few hours later in optical bands as a fast-decaying blue object, with  longer-lived emission in the infrared, X-ray, and radio bands. The astrophysics community is still learning from the vast data set compiled for this event. Early analyses showed that AT2017gfo had small ejecta masses with relativistic outflow velocities (from 0.1 to 0.3 of the speed of light $c$) and was likely powered by a combination of shocked material and r-process radioactive decay, which was modulated by the presence of high opacity lanthanides synthesized during the explosion and by a highly non-spherical ejecta and jet \cite[e.g.][]{drout2017,kasen2017elements,thielemann2017,gottlieb2018}. Many questions about this event and the diversity of KNe remain open, including the dependency on the neutron star equation of state, 
the geometry of the ejecta, 
the nature of the early blue emission, 
the physics of the relativistic jet launch, 
the exact contribution of KNe to the r-process elements production, 
and the difference between KNe produced by binary neutron star (BNS) and neutron-star black-hole (NSBH) mergers.

The gravitational waveform observed from binary compact object mergers allow first-principles distance measurements. The GW distance measurement and the optical redshift of GW170817 provided constraints on the Hubble constant from a single event (\citealp{schutz}; \citealp{delpozzo}; \citealp{abbott2017multi}), and further constraints came from measurements of $H_0$ from multiple GW events  using a low precision statistical method  (i.e dark sirens; \citealt*{darksiren1}; \citealt{LVC_O2_StS,palmese20_sts,palmese21_sts,finke2021cosmology,gray2023}).

From a cosmological perspective, the detection of GW170817 provided the grounds for the application of a new approach to understand the well-known Hubble tension. Measurements of the Hubble constant $H_0$ from early-time universe observations \cite[e.g., Planck observations of the CMB in ][]{planckcolab} are in tension with measurements from observations made at late times, e.g., the Hubble Space Telescope observations of Cepheids in concert with Supernovae as standard candles \citep{riess_2021}.
These measurements have been made at increasing precision over the last few years, only adding to the open questions surrounding the tension; and thus have provoked a series of investigations of variations on the cosmological model \citep[for survey papers, see, e.g.,][]{knox2020,olympics_2021,cosmointer}. There is no obvious resolution \citep{Di_Valentino_2021}, and the Hubble tension remains unresolved.

The GW distances measured from compact binary mergers do not rely on the distance ladder, and therefore will have different systematics than the astronomical distance ladder late-time measurements of $H_0$, and possibly smaller systematics. 
With additional standard siren measurements, the precision on the Hubble constant could reach the percent-level \citep{Chen2018,bom23}. This level of precision would be an important contribution to resolving the Hubble tension (see discussion further below in this section).

The third LIGO/Virgo/KAGRA collaboration (LVK) observing run (O3) resulted in over 60 new events (see the third Gravitational-wave Transient Catalog, \cite{gwtc3}, hereafter GWTC-3).
One GW event originated from a BNS merger and two high-confidence events originated from NSBH coalescences \citep{2NSBH}, but
no electromagnetic counterparts were confirmed from any event despite extensive follow-up campaigns \citep[e.g., ][]{rob2020,alyssa2020,douglas2021, Kilpatrick_2021, SwiftO3_Oates2021, igor2019, andreoni2019growth190814, goldstein2019growth_190426}. One study \citep{Graham} proposed an AGN counterpart to the binary black hole merger GW190521, but the association to the GW event cannot be made with confidence \citep{Palmese_2021,ashton}.

LVK's O4 campaign is scheduled to start in early 2023, surveying on a factor of $2$ times more in the median Luminosity Distance and a factor of $\sim 7$ detections of BNS events than the previous O3 campaign \citep[see table 2 in ][]{petrov2022}. The expected rates of BNS are uncertain but in the range of 9-88/year \citep{Abbott2020, petrov2022}. 
It is unlikely that each subsequent KN event will obtain the same amount of follow-up resources as in O3. 
As GW detectors become more sensitive and able to detect events at larger distances, the optical follow-up of BNS events will become more challenging, including for campaigns using the Dark Energy Camera \cite[DECam,][]{decam} such as those coordinated by the DES Gravitational Wave follow-up group (DESGW). 

Here, we present improved and optimized strategies for discovery of KNe. The methodology presented here  focuses on DECam resources, which the DESGW group plans to use to follow up LVK O4 events. However, the method is generic and could be easily adapted to any telescope. 
Our primary science goal is the construction of the standard siren Hubble diagram via maximizing the number of GW-detected events with known redshifts.

Our paper is structured as follows. In Section~\ref{section:data}, we describe our GW and KN simulations.
In Section \ref{section:dectability}, we discuss how we measure the success of each observing strategy. Section \ref{sec:optimization}, we detail the various optimization options and types of KN models. In Section \ref{section:strategies}, we detail the strategies that we consider to be most successful for ensuring that DECam detects KNe with our follow-up observations. We additionally describe how this code may be used in a real-time follow up in Section \ref{section:real_observations}. Finally, we discuss and summarize our results in sections \ref{section:discussion} and \ref{section:conclusion}.

%% file: data.tex
\subsection{Simulated O4 BNS events}

We start by producing a set of simulated BNS mergers that are expected to be detectable in the upcoming LVK O4 observing run. The procedure is similar to that in \cite{petrov2022}.

\begin{figure}[htp!]
    \centering
    \includegraphics[width=0.98\linewidth]{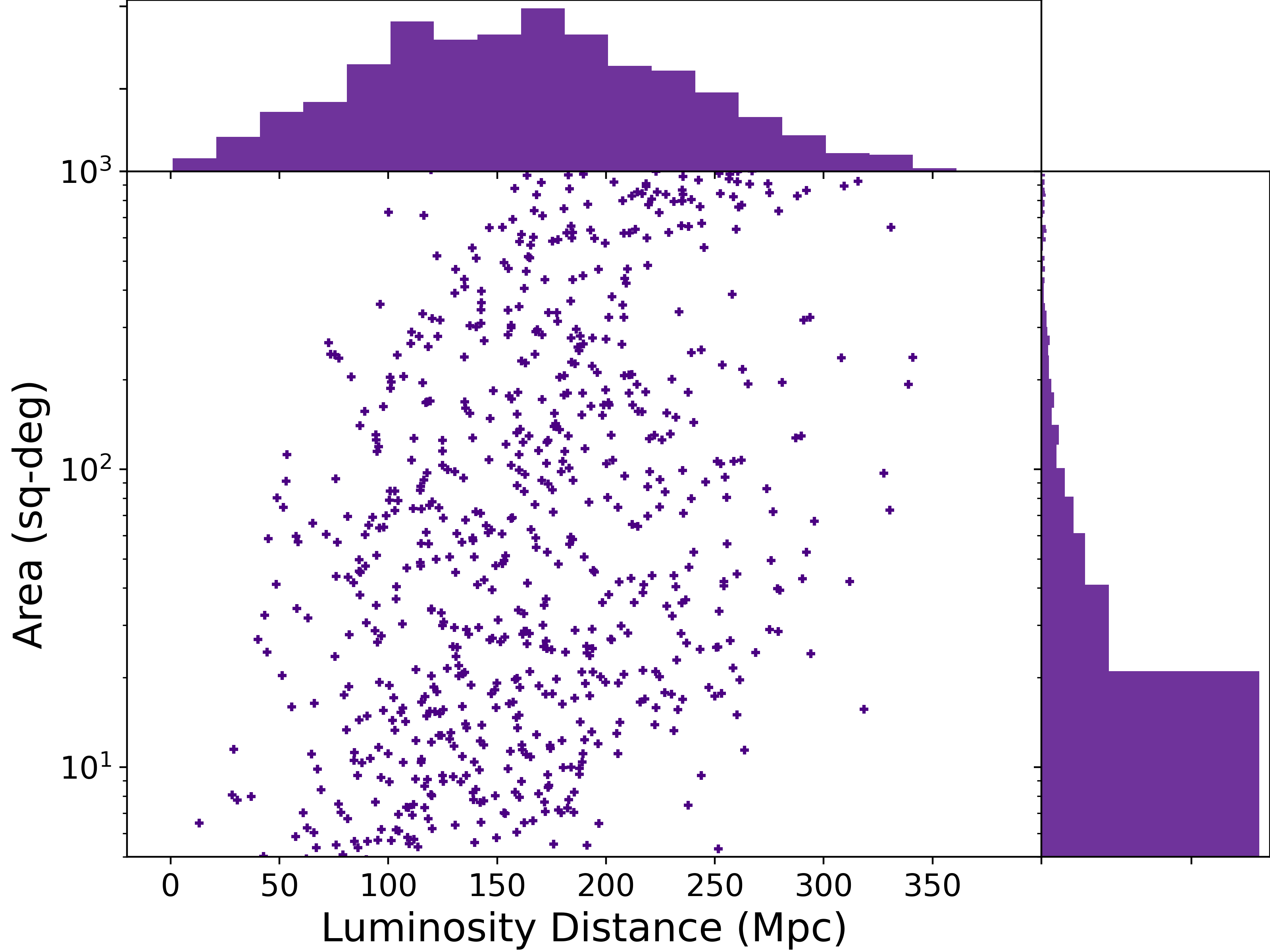}
    \caption{
	\label{fig:gw_sims}
The distance distribution and  $90\%$ credible interval sky localization area of the O4 simulations of BNS mergers used in this paper. The simulated primary masses follow a $\bar{x}=1.5 M_\odot,\, \sigma = 1.1M_\odot,\,x_\mathrm{min} = 1.1M_\odot,\,x_\mathrm{max}=3M_\odot$ Gaussian distribution, and the spins (see eq.(2) of first reference in \citet{gwtc3}) of the events are distributed uniformly between $\pm$0.05. Simulated detections are limited to events with a network SNR greater than 12. Throughout the rest of this paper, we will assume follow-up of events only with area $< 300$ degrees$^2$, which is 611 of the 860 events.
	}
\end{figure}

\begin{figure*}[htp!]
    \centering
    \includegraphics[width=0.9\linewidth]{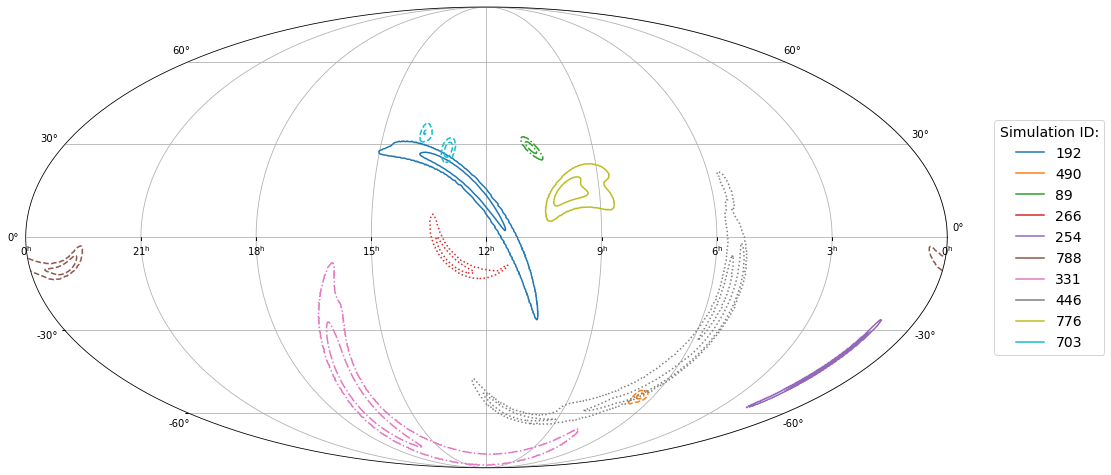}
    \caption{
	\label{fig:all_sky_gw_sims}
{A sample of $10$ randomly chosen set of simulated Gravitational wave localization maps used in this work. The inner (outer) lines represents $\%50$~($\%90$) confidence levels regions.}
    }
\end{figure*}

GW events are simulated using the \texttt{BAYESTAR} software \citep{bayestar,Singer_2016,Singer_supp}, which uses \texttt{LALSuite} \citep{lalsuite} tools. We assume sensitivity curves for Advanced LIGO, Virgo, and KAGRA as O4 sensitivities discussed in~\citet{Abbott2020}\footnote{ The curves are available in \url{https://dcc.ligo.org/LIGO-T2000012/public} and, \url{https://dcc.ligo.org/LIGO-T2000012-v1/public}}, though we assume a sensitivity for KAGRA of  a BNS range of $\sim 80 $ Mpc.\footnote{LVK has updated the assumed sensitivity for KAGRA, as seen in v2 of the above table; this makes little difference to our analysis, concerned mostly with distant events} All detectors have a duty cycle of 70\%, which is consistent with LVK predictions \cite{abbott2018prospects}. Assuming a \citet{planckcolab} cosmology, we create $10,000$ BNS events of the type O4 could theoretically observe, following a uniform in comoving volume distribution, and then inject them into the GW Search and Discovery pipeline \citep{HERNER2020100425}. 
TaylorF2 waveforms \citep{Buonanno_2009} are assumed both for injections and reconstructions. The primary mass distribution of our injections follow the neutron star mass function found in \citet{LIGOScientific:2021psn}, normally distributed with mean 1.5 $M_\odot$  and standard deviation $1.1 ~M_\odot$, truncated to be within $1.1 M_\odot \le M_\mathrm{NS}\leq 3M_\odot$ in order to stay consistent with the Kasen models' parameter space. The NS spin distribution was uniformly distributed  between -0.05 and 0.05. After injecting the BNS mergers, a matched-filter search retrieves the detected events. We consider as detections those events for which a single--detector signal--to--noise ratio SNR$>4$ is reached by at least 2 detectors and the overall network SNR is $>12$, resulting in 860 detected events. The measured SNR is added with Gaussian noise. Finally, we produce \texttt{BAYESTAR} skymaps for the detected events.  
In Figure~\ref{fig:gw_sims} we present the area (90\% credible interval) and luminosity distance (integrated over the whole sky) for all of the simulations used in this analysis.

\cite{petrov2022} argues that the alerts produced by the LVC during O3 are better modeled by dropping the two detector coincident detection requirement and using a minimum SNR for BNS events of $> 8$. These criteria would have the effect of increasing the number of low SN, and therefore large sky area, events in our simulations. Figure~\ref{fig:gw_sims} shows median luminosity distance of $\sim150$ Mpc and sky area of $\sim20$ sq-degrees, whereas \cite{petrov2022} find $352\pm10$ Mpc and $1820^{+190}_{-170}$ sq-degrees, a difference from previous work that they attribute to the changing of the SNR requirements. There are 249 from the total of 860 events that have 90\% sky area $> 300$ sq-degrees. To change our sample to have a median sky area of $\sim350$ sq-degrees we would need an additional $\approx360$ events, all of which would need $90\%$ sky area $>350 $ sq-degrees. In our strategy definitions and proposed decision-making process to trigger limited Target-of-Opportunity resources, we choose a sky area limit of $<300$ sq-degrees, so we chose not to consider these additional high-sky-area events here. We do not make any other selections in the events.

\subsection{Kilonova physical models}


We model KNe using the time-evolving theoretical Spectra Energy Distributions (SEDs) of KN atmospheres from \cite{kasen2017elements}, hereafter referred to as the Kasen models. 
The Kasen models are parameterized by the mass ejected in the explosion, $M_{\mathrm{ej}}$, the abundance of lanthanide elements in the ejecta, $X_{\mathrm{lan}}$, and the velocity of the ejecta, $v_{\mathrm{ej}}$.  The  models are a set of $329$ time-dependent SEDs on a grid of discretized parameters: 0.001 M\textsubscript{\(\odot\)} $\leqslant$ $M_{\mathrm{ej}}$ $\leqslant$ 0.1 M\textsubscript{\(\odot\)}, $0.03c \leqslant v_{\mathrm{ej}} \leqslant 0.3c$, and $1\times {10}^{-9}\leqslant {X}_{\mathrm{lan}}\leqslant 1\times {10}^{-1}$. 

There are other KN models available in the literature \cite[e.g.][]{bulla2019, darbha_2020, hotokezaka_2020, wollaeger_2021, gillanders_2022}.
In comparison to these, the Kasen models do not build in the dependence on the geometry of the mergers and the viewing angle (see, e.g., \citealt{stewart2022} for a visualization of an asymmetric KN supported by accretion disk simulations), but instead provide atmosphere models to be used in building a geometric model.

\subsection{Kilonova lightcurve simulations}

The lightcurve simulation pipeline is similar to that used \citet{rob2020}, which constrains the physical KNe properties of GW190814.
The SuperNova ANAlysis software (SNANA: \citealt{kessler_2009,kessler_2019}) enables the simulation of light curves of KNe as they would be measured by DECam. SNANA produces light curves by simulating fluxes and uncertainties in observations by incorporating information about cadence, image zero points, and noise levels in search and template images.
The  light curves are in absolute magnitudes and are converted to observed magnitudes using a given cosmology. 
In particular, SNANA chooses a grid of 15 redshifts at $ 0.003 \leqslant z \leqslant 0.2$, which is used to transform the Kasen model SEDs without evolution. 
The Kasen model SED is redshifted to a $z$ on our grid. SNANA also take into consideration the reddening by the dust in the Milky way \citep{kessler_2019,1994ApJ...422..158O}. Later, the SEDs are convolved with DECam transmission curves accounting for atmosphere, telescope, filters, and CCDs. Given a cadence, the light curves are calculated by sampling the magnitude grid.

The $329$ Kasen models, $\alpha_j$, each with a range of  time after burst, $\tau$ ($ 0.00 \leqslant \tau \leqslant 16.7$ days), at the SNANA grid of redshifts, $z_i$  ($0.003 \leqslant z \leqslant 0.2$), describe  $829,080$ simulated magnitudes, $m_\lambda(\lambda, \tau, \alpha_j, z_i)$, for each of the 4 filters that we use ($g,r,i,z)$. 

\begin{table}[htp!]
\centering
\caption{ 
    \label{tab:teff} 
    Observational conditions (photometric filter, limiting magnitude, and effective exposure time) averaged from DECam follow-ups of previous observations. Note that the higher the \teff\ value (which ranges from 0 to 1), the better the observational conditions.}
\begin{tabular}{cccc}

\hline \hline 
Filter &  $m_0$ & \teff\ (bright time) & \teff\ (dark time) \\ \hline
$g$ & 23.4 & 0.05 & 0.7 \\
$r$ & 23.1 & 0.15 & 0.8\\
$i$ & 22.5 & 0.45 & 0.7\\
$z$ & 21.8 & 0.6  & 0.6\\
\hline \hline
\end{tabular}
\end{table}

\subsection{DECam limiting magnitudes}

We define the limiting magnitude as the magnitude at which we can measure a point source, e.g. a star, with $0.1$ mag error, which corresponds to  signal-to-noise ratio, SNR $\sim 10$. We then define the limiting magnitude $m_0$ for a total effective exposure time of $90$ sec, i.e. $\texpeq \times \teffeq = 90 \mathrm{s}$, where $\teffeq$ is a unitless quantity that scales the exposure time to the `effective' exposure time when taking into account sky conditions (higher $\teffeq$ being better sky conditions). Then, to scale up for different exposures, we construct:
\begin{equation}\label{eq:mlim}
m_\mathrm{lim}=m_0+1.25\log {\left(\teffeq \times \frac{\texpeq}{90 \mathrm{s}}\right)},
\end{equation}
which reflects S/N going as the square root of time in a sky noise limited observation.
We use the $m_0$ measured in the DES data by \citet{osti_1250877}, thus the $90 \mathrm{s}$ normalization factor, and present those values in Table~\ref{tab:teff}, and the derived limiting magnitudes in Table~\ref{tab:mlim}. The \teff\ is closely related to observational conditions during the night. Therefore, we break observing into nights of bright time and dark time and use \teff\ from previous target of opportunity (ToO) programs in DECam, in particular the observations from past GW follow-up events (\citealp{rob2020}, \citealp{alyssa2020}). Although we split our tests into only dark and bright time, rather than on gray nights, which correspond to $\sim 50\%$ of the nights available in telescopes, we later show in Section \ref{section:bright-and-dark-nights}, that the difference in performance between dark and bright and therefore gray is negligible due to the adaptability of the presented method. The differences are in the particular configurations chosen, e.g. the filters and exposures times selected in different observational conditions.
Throughout this paper, we focus on detectability on dark nights unless otherwise stated.

\begin{table}[htp!]
\centering
\caption{
\label{tab:mlim} 
DECam $m_{\mathrm{lim}} (10\sigma)$, dark time }
\begin{tabular}{ccccc}

\hline \hline 
&\multicolumn{3}{c}{Exposure time (sec)}\\
Filter &  $10$ &  $100$ & $1200$ & $3600$ \\
g & 22.0 & 23.3 & 24.6 & 25.2 \\
r & 21.8 & 23.0 & 24.4 & 25.0\\
i & 21.1 & 22.4 & 23.7 & 24.3 \\
z & 20.3 & 21.6 & 22.9  & 23.5 \\
\hline \hline
\end{tabular}
\tablecomments{
$\Delta m_\mathrm{lim} (10\sigma-5\sigma) = 0.75$ mags.}
\end{table}

\begin{table*}[htp!]
\small
\caption{ 
    \label{tab:obs_params} 
    Observational parameter space $\Theta$ explored.}
\begin{adjustwidth}{-2cm}{}
\centering
\begin{tabular}{@{}lllllll@{}}

\hline \hline 
Time after burst (Days) &

$0.5,1.0,1.5,2.0,2.5,3.0,3.5,4.0$\\
 \hline 
Scenario 1\\ 
\quad Filter & $g,r,i,z$ \\
\quad 2d credible sky area covered & $0.9,0.85,0.8,0.75,0.7$  \\
\quad Exposure time (sec) & $60,90,120,200,300,600,1200,3600$ \\
  \hline 
Scenario 2 (with inner \& outer region) \\ 
\quad Filter & $g,r,i,z$ \\
\quad 2d credible sky area covered &\\
\qquad Outer: 0.9  & Inner: $0.8, 0.7, 0.5, 0.3$ \\
\qquad Outer: 0.8 & Inner: $0.7, 0.5, 0.3$\\
\qquad Outer: 0.7 & Inner: $0.5, 0.4, 0.3$ \\
\quad Outer/inner exposure time pairs (sec)& 60/90;\quad 90/120;\quad 120/200; 200/300;\quad 
\\ & 300/600; \quad 300/1200;\quad  600/1200; \quad 600/2400;
\\ & 1200/2400;\quad 2400/3600;\quad 3600/5400 \\
                                
\hline \hline
\end{tabular}
\end{adjustwidth}
\end{table*}

\subsection{Cadences \& observational parameters}

We use of SNANA to find effective search strategies that maximize candidate detection, considering realistic conditions, including the maximum duration of the night, intervals between observations, and sky brightness, while also minimizing telescope time and enabling earlier discovery. This is at base a trade-off of exposure times versus sky area coverage, but with additional complications of filter choice and the time since the event occurred. We limit our telescope time expenditures to $8$ hours per night and assume Blanco/DECam has telescope/readout slew time of $30\mathrm{s}$ between the exposures, which is true for small slews, such as the ones less than $\sim$ 10 degrees.  In fact, long slews $\sim 100$ deg, which might be necessary to cover disjoint gravitational wave maps, take on the order of $3$ minutes. Thus, if there are two disjoint regions, which would add one or possibly two long slews depending on the observational conditions, it would add a negligible amount of total telescope time, relatively. 
We test four filters, $g,\,r,\,i,\,\mathrm{and}\,z$, starting $12$ hours after the trigger, and going to 4 days post-merger in half-day increments with several exposure times. These are summarized in Table~\ref{tab:obs_params}.  

 Note that the time required to respond a GW alert depends upon several factors including human decision-making to trigger, when the event is visible in the sky, observation planning, and time to ask for and obtain the ToO interrupt. The time required to get the telescope on-sky also depends on when the alert is given. It is possible that it may be significantly less than 12 hours until we are on-sky, however, for this study we opted for a conservative option of 12 hours, by which time for any significant and observable event we should be able to observe.
  
 We define two exposure time scenarios. In Scenario 1 we cover the area of a given GW event with a single set of exposure times and in Scenario 2, we explore the use of two different exposure times for a single search. The latter is  motivated by the need to cover the high probability area sky with deep exposures, while covering the larger low probability localization area outskirts with shorter exposure images. We  designate central high probability areas as the ``inner region'' 
 and the rest of the area inside the the localization region as the ``outer region''.   The last section of Table \ref{tab:obs_params} presents the combinations considered for the inner region, ranging from $30\%$ to $80\%$ sky probability coverage, 
 for 3 different values of the total (deep+shallow) sky map probability coverage, from 70\% to 90\%.
 For instance, a combination of $40\%$ probability for the inner region and $70\%$ total coverage means that the $40\%$ highest probability region is covered with higher exposure time and the $70-40=30\%$ left over is covered with the shorter exposures. Each of these combinations is considered for all possible deep and shallow exposures presented in Table \ref{tab:obs_params}. 
For scenarios where we cover the sky area twice in a single night, we additionally take into consideration the KN variability a few hours after the first search. 
 
 \subsection{Simulation data summary}
 
 We simulate a set of GW detections $S={S_1,\ldots,S_n}$, where $n=611$ and $S_i$ is the $i^{th}$ simulation with distance $d_i$. For each $S_i$ we evaluate each of the parameter sets $\Theta$ in the two scenarios. In Scenario 1 there are 8 passes since burst, 4 filters, 8 exposure times, and 5 sky area probability coverages: $8\cdot 4\cdot 8\cdot 5 = 1280$ possible parameter sets. In Scenario 2 there are $8\cdot 4\cdot 11\cdot 10 = 3520$ possible parameter sets to be evaluated. Over both, 4800 observation models are evaluated.  Each model is evaluated with the machinery described in Section 2, resulting in SNANA kilonova measured magnitudes. We have done this for each of the 329 Kasen models. 

%% file: methods.tex


\subsection{Discovery probability}

We define the probability of detection, $p_{\alpha_j}$, of the KN model $\alpha_j$ for the $j^{th}$ combination of $(M_{\mathrm{ej}},\log(X_{\mathrm{lan}}),v_{\mathrm{ej}})$, with observed magnitude $m_\lambda=m(\lambda, \tau, \alpha_j, z_i)$ in a given filter $\lambda$, for a given exposure time \texp\ weighted by observing condition \teff\  for an event at mean redshift, $\bar{z}$, over the SNANA grid of redshifts, $z_i$:

\begin{align}
p_{\alpha_j} &\propto   \sum_i 
\begin{cases}
pr(\alpha_j) \cdot  pr(z_i), & \text{if}\, \mathrm{m}_\lambda < \mathrm{m}_\mathrm{lim} \\
0, & \text{if}\, \mathrm{m}_\lambda \geq \mathrm{m}_\mathrm{lim}
\end{cases} \,\label{eq:prob_detect}\\
\mathrm{where}\nonumber \\
\alpha_j & \equiv   j^{th} \mathrm{\, combination \, of \, } (M_{\mathrm{ej}},\log(X_{\mathrm{lan}}),v_{\mathrm{ej}}), \nonumber \\
p_{\alpha_j} &\equiv p_{\alpha_j}(\mathrm{detection}|\tau, \teffeq,\Theta,\bar{z}),\nonumber  \\
\mathrm{m}_\lambda &\equiv \mathrm{mag}_\lambda(\lambda, \tau,\alpha_j,z_i), \nonumber \\
\mathrm{m}_\mathrm{lim} &\equiv \mathrm{mag}_\mathrm{lim}(\lambda, \teffeq,\Theta),\quad \mathrm{and} \nonumber \\
pr(z_i) &= \exp\left(-\frac{(z_i-\bar{z})^2}{2 \sigma_{\bar{z}}^2}\right), \nonumber 
\end{align}

\noindent where
$\tau$ is time after the GW detection, $\Theta$ contains the specific observation strategy characteristics including  $\texpeq$, 
  filter and area coverage as described in Table~\ref{tab:obs_params},  $\mathrm{m}_\mathrm{lim}$ is the limiting magnitude of the observation  for a $10\sigma$ detection, and  $pr(\alpha_j)$ are the model priors defined by each kilonova model described in the Section \ref{section:models} and Table \ref{tab:priors}. 
The summation is over all redshifts for a given model, $j$ at a given $\tau$.  Eq.~\ref{eq:prob_detect} represents a Gaussian prior for choosing light curve models from the grid of SNANA defined redshifts given the GW distance. 
Explaining it differently, Eq.~\ref{eq:prob_detect} is, for a given Kasen model, examining whether the resulting apparent magnitude is less than the limiting model with a Gaussian prior on the redshift, using the GW event mean redshift, $\bar{z}$, and the grid of SNANA redshifts the KN could be at, weighted by the GW event variance in redshift, $\sigma_{\bar{z}}^2$.


The GW localization maps present the probability that the event is located at a given sky position, the luminosity distance at that position and its uncertainty. Therefore we define the total probability of detecting an event  as



\begin{align}
P_d &= \frac{\int_{\hat{\Omega}} d\Omega\,\,  d_{L}(\Omega)\, p(\Omega) }{\int_{\hat{\Omega}} d\Omega\,\, d_{L}(\Omega)}\, \times \sum_{j=1}^{329} p_{\alpha_j}\,,\label{eq:prob_area}\\
\mathrm{where}\nonumber \\
P_d &\equiv P(\mathrm{discovery}|\hat{\Omega},\tau,t_\mathrm{eff},\Theta,\bar{z})\nonumber
\end{align}

\noindent and $\hat{\Omega}$ is the entire sky area observed in the follow-up,  the $d\Omega$ is the voxel, $p(\Omega)$ is the probability in the voxel, and $d_L({d\Omega})$ is the luminosity distance to the voxel.  The sum over $p_{\alpha_j}$ includes the priors and thus indicates the model used. 
The sky coverage and exposure times determine the total telescope time for a given KN detection. 
We note that we use weighted spatial probability, rather than 2-d on-sky probabilities, therefore we have the  $d_L$ weighting in Eq~\ref{eq:prob_area} and the attendant implication that we can have a higher detection probability than on-sky credible area covered.
We can now evaluate $P_d$ for a given set of $\Theta$. 

\subsection{Confirmation probability}

In order for an object to be confirmed as a kilonova candidate, we require it be detected twice (in two observing `epochs'). This requirement can be lifted if, for example, there are sufficient spectroscopic resources available to follow-up all of the candidates found after a single epoch. Generally this is not the case, however. The second detection  eliminates spurious detections, including image artifacts, asteroids and other possible contaminants \citep{rob2020, Shandonay_2022}. 
Another reason to consider the detections independent is that we typically observe while working on post-processing and making target selection for spectroscopy in the data from the previous epoch.


Given that we want to make two detections to positively identify kilonova candidates, we define the probability of confirming the transient with two independent detections as:




\begin{align}
P_c  \equiv P (\mathrm{confirmed}) = & P(\mathrm{discovery}|\hat{\Omega},\tau_1,t_\mathrm{eff},\Theta,\bar{z})\\ \nonumber  & \times
  P(\mathrm{discovery}|\tau_2,t_\mathrm{eff},\Theta,\bar{z}) \label{eq:prob_area}
\end{align}

\noindent where,
\begin{equation}
P(\mathrm{discovery}|\tau,t_\mathrm{eff},\Theta,\bar{z}) = \sum_{\alpha_j} p_{\alpha_j}.\nonumber    
\end{equation}

\noindent We design the strategies along the paper optimizing for $P_c$ given a set of constraints. Due to implementation choices and to speed up the numerical optimization we use $P_c^{\star} 
 \equiv P_{d,1} \cdot P_{d,2}$ which gives us equivalent results in terms of strategy~\footnote{In this form $P^{\star}(\mathrm{confirmed})$ has an extra spatial/volume probability term multiplied $V_p$ in which $P_c^{\star}=V_p \cdot P_c$ is the same for both $P_{d1}$ and $P_{d2}$. Therefore, this is irrelevant from an optimization viewpoint, since $\mathrm{arg max} (P_c^{\star}) =\mathrm{arg max} (P_c)$.} such as telescope time or the duration of the night. We further present our main findings as a function of discovery probability, which is the most relevant outcome for the proposed strategies.

\section{Optimizations and Kilonova Types}
\label{sec:optimization}

%

After calculating $P_c$ for each O4 event in two observing scenarios each with a grid of observational parameters we can evaluate what works best to optically find the kilonova.
The answer to this depends on the science goal. Our primary science objective is standard siren cosmology, so we aim to identify optical counterparts to every kilonova-progenitor detected by the LVK. To this end, we chose to focus on optimizations that require two detections at least 30 minutes apart, in order to remove spurious detections due to asteroids. In this section we discuss the optimizing the strategy given the science goal, and then discuss the detailed metric, which involves exploring the meaning of covering the space of Kasen models. 

\subsection{Optimizations}

We derive $P_c$ for each of the 1280 parameter combinations of Scenario 1 and for each of the 3520 combinations of Scenario 2, resulting in 4800 total for each simulated merger detection event, $S_i$. Not all of these combinations have an appreciable $P_c$ much above zero, as most of the predicted magnitudes are below sky noise.

 We use $P_c$ as the variable to optimize on.  We choose the highest $P_c$ for each sim $S_i$, look up the set of observational parameters  $\Theta_i$ for it, and define this the  \textit{Top} strategy.
Choosing the highest $P_c$ is the simplest optimization, but for our evaluation we need at least two more, \textit{Reference} and \textit{Low Telescope Time}.

\begin{enumerate}
 
\item 
\textit{Top Strategy} is the $\Theta_i$ producing the highest $P_c$ for each $S_i$, the observational parameters producing the highest probability of confirmation for every O4 simulation. 
\item 

\textit{Low Telescope Time (\lTT)} is the $\Theta_i$ combination that uses the lowest telescope time given while retaining a $P_c$  within 10\% of the highest confirmation probability strategy, by definition \textit{Top}.  
For example,  if the \Top\ strategy finds $P_c=0.85$, then a \lTT\ strategy will have $P_c \ge 0.75$, usually with a much reduced telescope time. We will find it interesting to vary the threshold away from $10\%$.

\item 

\textit{Reference Strategy} has the 90$\%$ probability sky area observed in $i$ and $z$ bands  with 90 second exposures on the first two nights after the merger. This strategy models previous DECam searches, in particular the extensive search of \cite{rob2020}, and has been used as the DECam strategy for the predictions of \cite{chase2022}. 

\end{enumerate}

The \textit{Top} strategy  uses as much telescope time as possible to explore the volume, given the parameter exploration presented in Table \ref{tab:obs_params}. All strategies work within the constraint of requiring two passes over two 8 hour nights.

\subsection{Bayesian Average Models}
\label{section:models}

\begin{table}[htp!]
\small
\centering
\caption{
    \label{tab:priors}  
    Kilonova Gaussian model priors. 
}
\begin{tabular}{@{}lcccccc@{}}
\hline \hline
  &  \multicolumn{2}{c}{Blue \& Bright} & \multicolumn{2}{c}{Reddish \& Slow} &  \multicolumn{2}{c}{Red \& Faint}   \\
  &  $\mu$ & $\sigma$ &  $\mu$ & $\sigma$ &  $\mu$ & $\sigma$\\ \hline
$M_{\mathrm{ej}} (M\textsubscript{\(\odot\)})$ & 0.025 & 0.001 & 0.025 & 0.01 & 0.035 & 0.15  \\
$\log(X_{lan})$ & -5.0 & 1.0 & -5.0 & 10.0 & -2.0 & 5.0 \\
$v_{\mathrm{ej}}$ ($c$) & 0.25 & 0.01 & 0.25 & 0.10 & 0.15 & 0.30 \\
\hline 
Peak $ \left< M_i \right> $ &  \multicolumn{2}{c}{$-16.3$} & \multicolumn{2}{c}{$-15.7$}  & \multicolumn{2}{c}{$-14.6$}    \\
\hline \hline
\end{tabular}
\tablecomments{Table of means, $\mu$, and standard deviations, $\sigma$, of the Gaussian priors on the Kasen models parameter ranges. The Kasen models have parameter ranges of:  \\
\centerline{0.001 M\textsubscript{\(\odot\)} $\leqslant$ $M_{\mathrm{ej}}$ $\leqslant$ 0.1 M\textsubscript{\(\odot\)}}
 \centerline{$-9\leqslant \log({X}_{\mathrm{lan}})\leqslant -1$} 
\centerline{$0.03c \leqslant v_{\mathrm{ej}} \leqslant 0.3c$}
}
\end{table}

Not every Kasen model atmosphere is equally likely to be a good model for a real KN light curve. Most models for GW170817  are 2+ component models, as in \cite{Kilpatrick_2017, villar2017, coughlin2018, gillanders_2022}. If the Kasen models define a linear space of KN models, then the Kilpatrick models are in that space; if the Kasen models are eigenmodels of KN, then the models in Kilpatrick et al are defined by the eigenvalues multiplying the eigenmodels. Both the values of the non-zero eigenvalues and the number of non-zero eigenvalues are highly model dependent.
Dropping the eigenvector language, it is clear that in the current situation of very few well studied KNe,  the number of components in models describing KNe candidates
is uncertain.

One of the most common ways to define detection efficiency in the literature is to set up a grid of KN models, for example over viewing angle, as done for a $\chi^2$ analysis and then calculating the fraction of models detected given an observation. This makes the detection probability explicitly dependent on non-physical choices of the grid breadth and grid spacing. In our case the grid would be the 329 Kasen models, even though these models were meant to extend past the range of models likely to describe real KN. In a Bayesian framework, each of these models would come with a prior describing our belief in their applicability.

\begin{figure}[htp!]
	\centering
  	\includegraphics[width=0.98\linewidth]{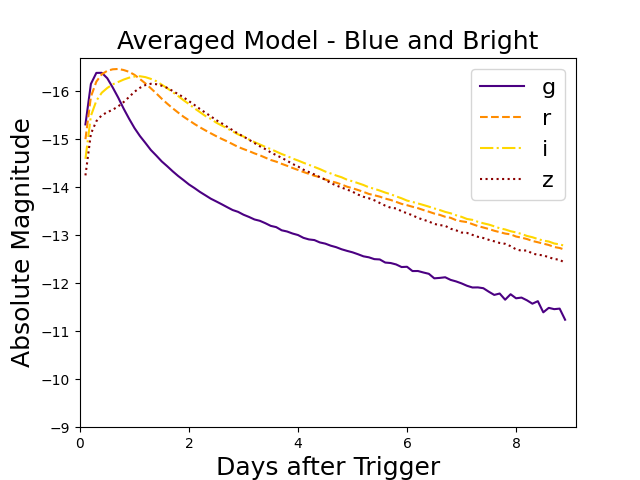}
	\includegraphics[width=0.98\linewidth]{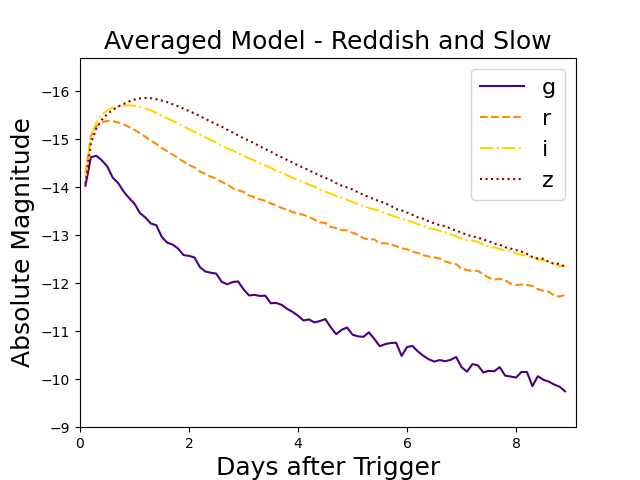}
   	\includegraphics[width=0.98\linewidth]{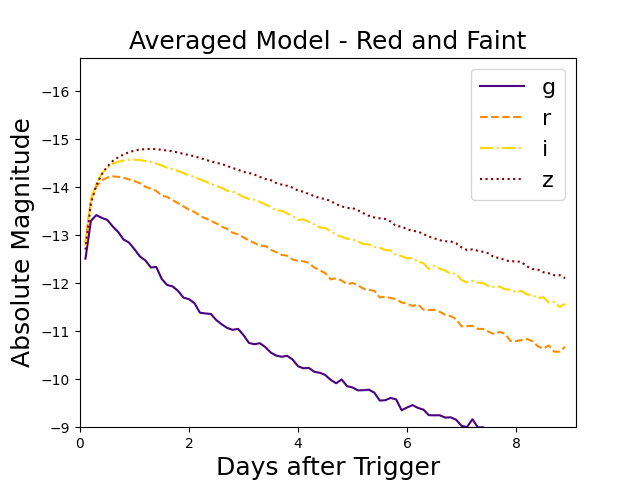}
    \caption{
        \label{fig:abs_mag}
    The absolute magnitudes $x$ days after the trigger for the weighted average of three sets of priors in the KNe parameter space models considered. 	}
\end{figure}

We will employ the useful idea of a Bayesian model average. We evaluating the entire grid of Kasen models, but instead of a uniform weighting we place a Bayesian prior, $pr(\alpha)$, on each model.
The Bayesian average model detection probability is implicit in eq~\ref{eq:prob_detect}, but can be thought of as:
\begin{equation}\label{eq:BAM}
\left< P_d \right> =  \sum_{j=1}^{329}\, P_d \cdot\, pr(\alpha_j)    
\end{equation}
where $\sum pr(\alpha_j) = 1$. 
Here we will use Gaussian priors to produce three Bayesian average models, \bluebright, \reddishslow, \& \redfaint, as given in Table~\ref{tab:priors}.
It is useful to guide the intuition to form the Bayesian average model absolute magnitude,
\begin{equation}
M_\lambda(\lambda, \tau) = \sum_{j=1}^{329}\, M_{\alpha_j}(\lambda, \tau, \alpha_j) \cdot\, pr(\alpha_j)  \, ,
\end{equation}
of which peak $M_i$ is also given in Table~\ref{tab:priors} and light curves shown in Fig~\ref{fig:abs_mag}.
This mean quantity, while illuminating, is incomplete as the Bayesian formalism is designed to make the uncertainties explicit- the curves are to thought of as the median value of a band of light curves weighted by the prior.

The \bluebright\ model is defined as the means and uncertainties of the blue component model in \citet{Kilpatrick_2017} interpreted as Gaussian priors. Since \citet{Kilpatrick_2017} gives no uncertainties for the blue model, 
we assume for $M_{\mathrm{ej}}$ a relatively narrow 0.001, and for $v_{\mathrm{ej}}$ of 0.01. Where \citet{Kilpatrick_2017} has the blue model lanthanide fraction evolving from $\log{X_\mathrm{lan}} = -4$ to -6 as the opacity falls due to KN atmosphere expansion, we take $\log{X_\mathrm{lan}} = -5$ with an uncertainty of 1.0.
This model is, in average, blue and reaches a peak luminosity a half day after trigger in $g$-band. It is also, in average, the brightest of the three models and $0.8$ magnitudes brighter than GW170817's peak $M_{g,r,i,z} \approx -15.5$.

The \reddishslow\ model is defined as in \bluebright, except that we take the $\sigma$ to be ten times the uncertainties there, to reflect our ignorance of the KNe population. This results in a prior allowing the entire $\log{X_\mathrm{lan}}$ range of the Kasen models to contribute. Such wide priors make our model, in average, to be redder and slower to peak than GW170817, though with the same $\left< M_i \right>$. Most of our results use this model.

The \redfaint\ model is defined as the means and ten times the uncertainties of the red component model in \citet{Kilpatrick_2017} interpreted as Gaussian priors. The predominantly lanthanide-rich Kasen models contribute. This results in a model that is, in average, redder and fainter than GW170817, with a peak $\left< M_i \right>$ fainter by 1 magnitude, and a $\left< M_r \right>$ fainter by  1.4 magnitudes. This model aligns with what is expected by most models for viewing KN in the orbital plane, where the lanthanide-poor material is hidden from view.

\begin{figure*}[htp!]
    \centering
    \includegraphics[width=0.45\linewidth]{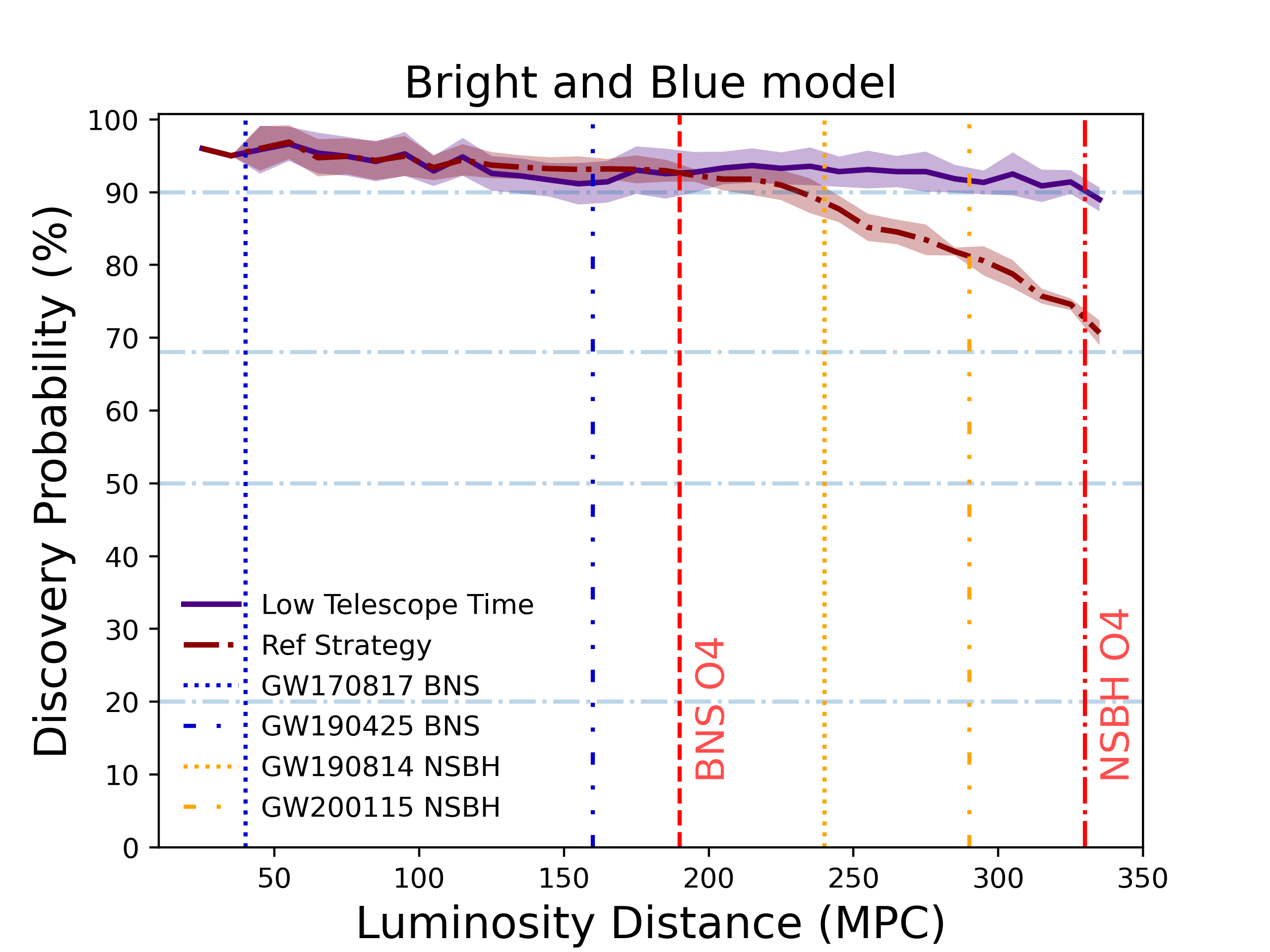}
    \includegraphics[width=0.45\linewidth]{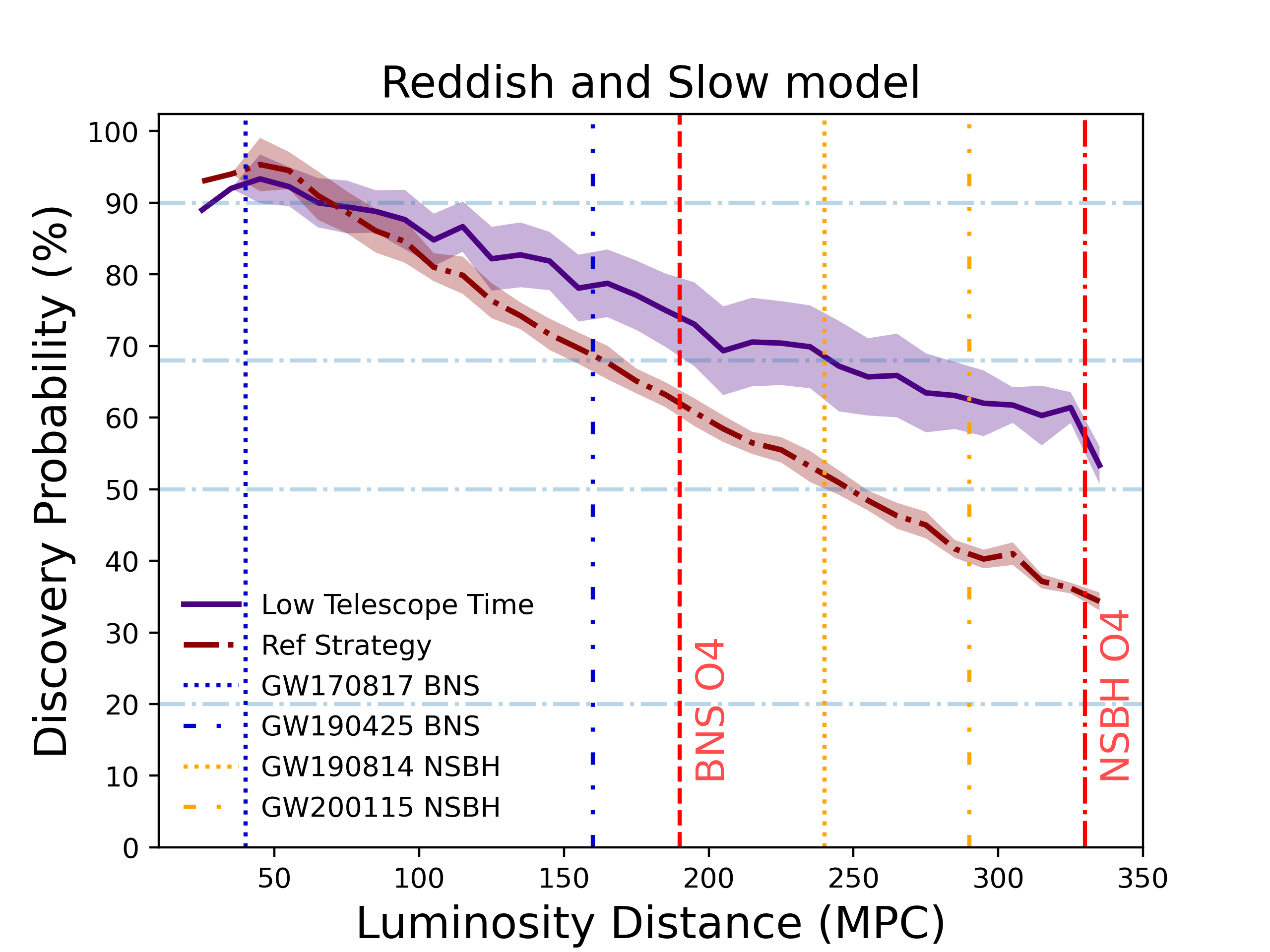}
    \includegraphics[width=0.45\linewidth]{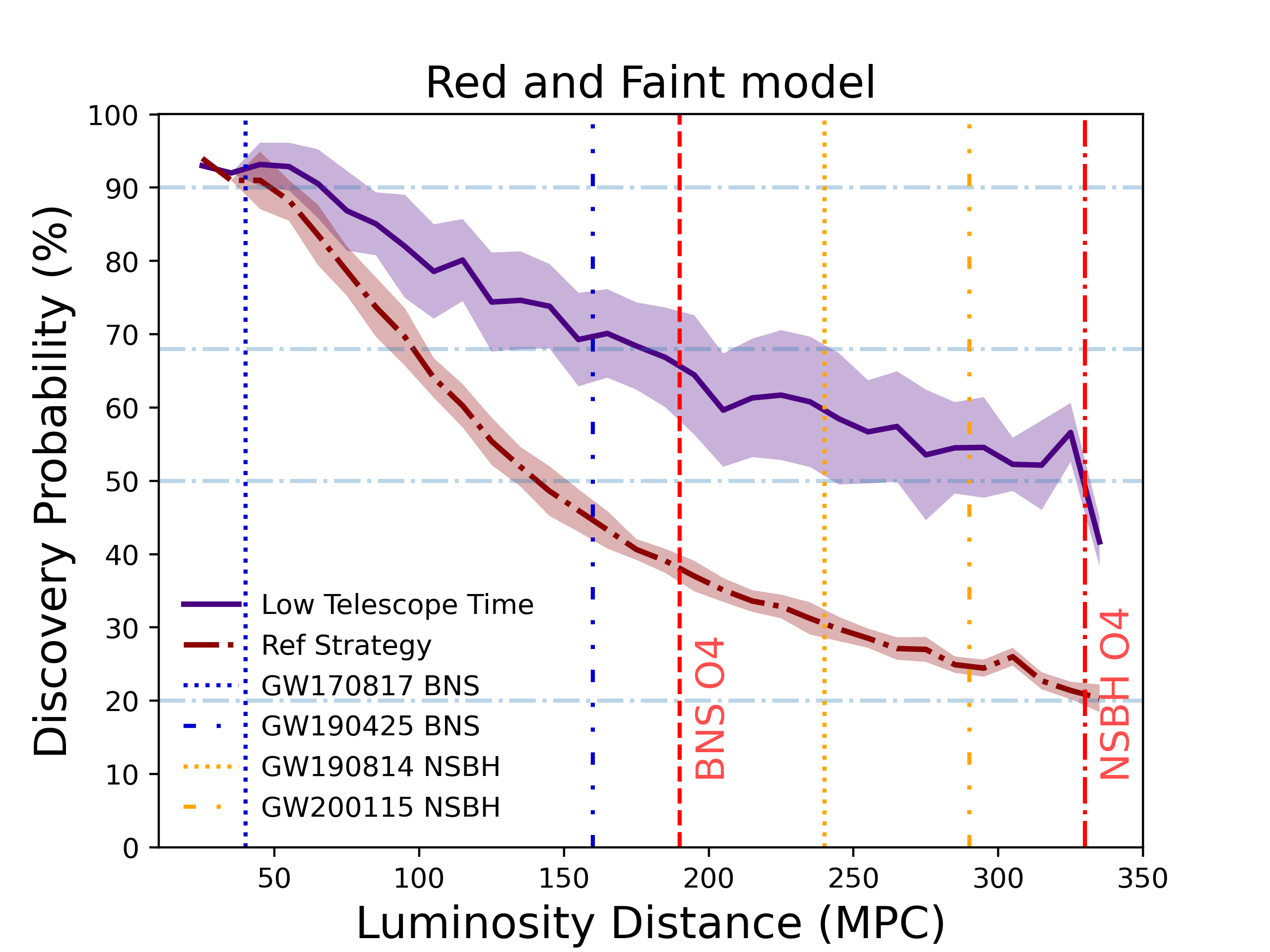}
    \includegraphics[width=0.45\linewidth]{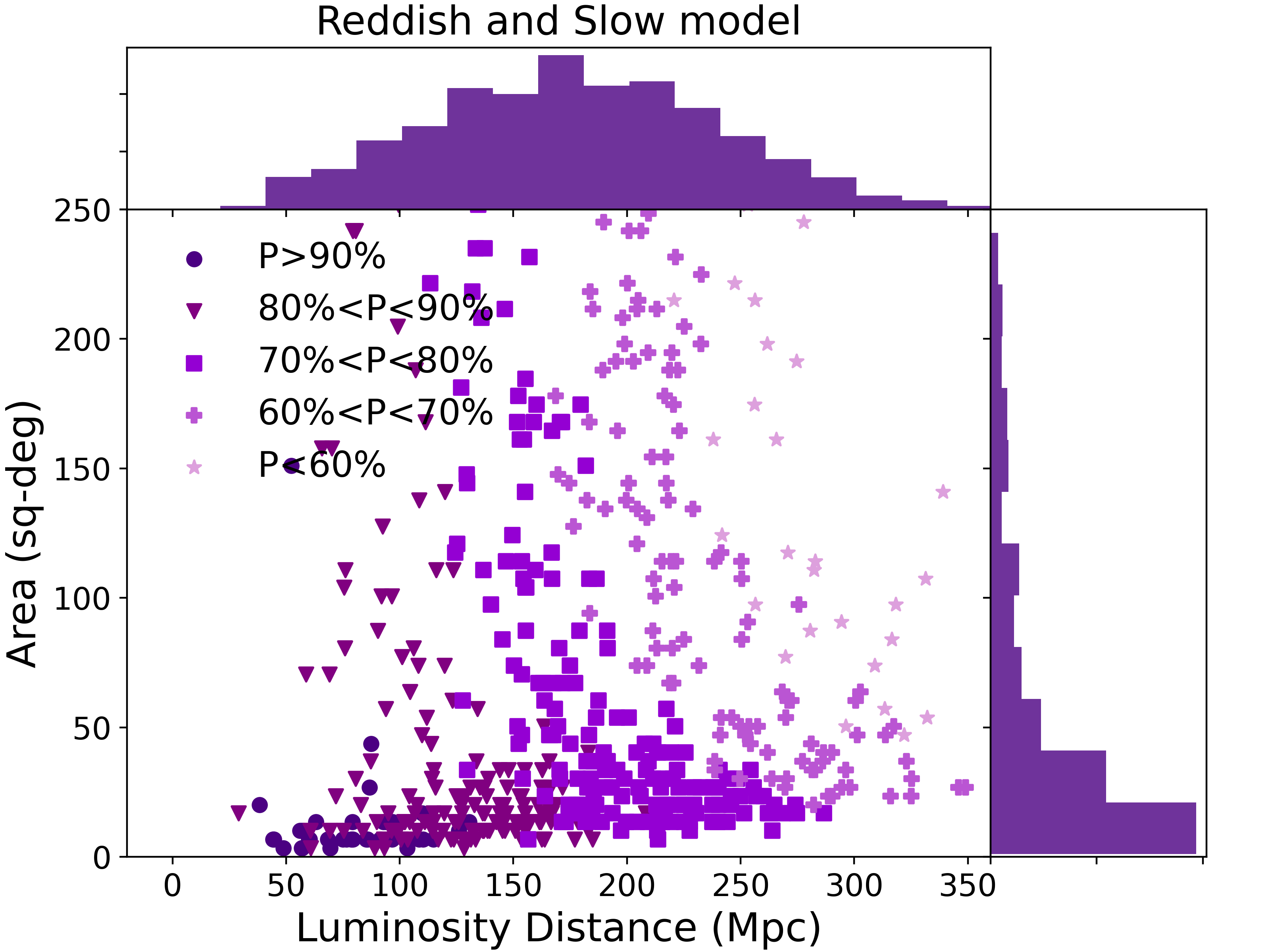}
    \caption{ 
        \label{fig:prob-v-dist}
	Discovery probability vs luminosity distance using the \lTT\ observing strategy compared
        to the \textit{Ref} strategy for the 611 events with $90\%$ probability area $< 300$. Each plot describes a different KNe model where (Upper Left:)
        \bluebright, 
        (Upper Right:) 
        \reddishslow,
        (Bottom Left:) 
         \redfaint.
        The shaded region about the median line represents the $68\%$ confidence interval of the scatter among
        the simulations. We also mark the distance to the upper limit of the LVK O4 projected range as well as noted GW events for reference.
	Bottom Right: distribution of simulations with for the \lTT\ strategy, binned by discovery probability in our wide prior model, \reddishslow. The marginal histograms are discovery weighted.}
\end{figure*}

 We will use our three models in various ways to evaluate the optimized strategies.   The \bluebright\ model will be the easiest to detect, as  the light curves peak brighter than $-16$ in $g,r,i,z$. The \redfaint\ model will be the hardest, as the absolute magnitude peaks only brighter than $-13$ in $g$, barely $-14$ for $r$, approaching $-15$ for the the observationally more difficult $i,z$. The \reddishslow\ model is intermediate. In this paper, we often use \bluebright\ to inform the reader's intuition to low inclination angle KN. We use \redfaint\ to show the effect of a faint KN on the strategies. 
 It can be argued that the \redfaint\ represents the most likely KN from a NSBH merger, as those events are expected to be redder and with low absolute luminosity compared to NSNS mergers \citep{anand2021optical}.
We often use \reddishslow\ as an
intermediate case that shows the behavior of choices in strategy optimization well.


There remains considerable uncertainty in the KN population statistics. While our three models are comparable to those in, e.g.,
\cite{zhu-wu2021, sagues_2021, coughlin_2020, petrov2022}, the population study of \cite{setzer2022} has a two peaked absolute magnitude distribution, at $i$-band absolute magnitudes of $-15$ and $-12$. In that study, GW170817 is a $95^{th}$ percentile event in luminosity, and the $50^{th}$ percentile luminosity corresponds to our $\left< M_i \right>$
for \redfaint. The \citet{setzer2022} study is for a random distribution of inclination angles and thus corresponds to the intrinsic KN population, whereas it is known that GW observatories predominately select merging compact objects with inclination angles near $\approx 30^{circ}$ \citep[see, e.g., ][]{finn1993, nissanke2010, schutz2011}. The observed KN population from GW event followup will mostly be from the first peak of Setzer et al. While a KN sample with $M \approx -12$ would be challenging even for a 4m telescope, the first peak $M \approx -15$ is inside our model ranges.

We  take our priors from \citet{Kilpatrick_2017} because we wish to emphasize the uncertainties in the models in this study. The uncertainties in the models reported by \citet{villar2017}, for example, are much smaller.  For observations in O4 the models and uncertainties in \citet{coughlin2018} are likely more accurate and constraining for GW170717. However,  we prefer to keep a wide variety of possible KNe. We suggest using the 2-component lightcurve red and blue models  as the counterparts to our \bluebright\ and \redfaint\ models, interpreted as a low and large inclination angle models and to use the 1-component model with $10\sigma$ as the counterpart to our \reddishslow, interpreted as a maximally uncertain KN population model.



\subsection{Strategies and KN models}

We begin our study of the strategies. As the community routinely uses the probability of detecting a KN once, calling this discovery, we will be showing $P_d \equiv P_{d,1}$ in most of the succeeding plots.  Note that the strategies are all optimized on $P_c$ and in our language, “chance of discovery” is to be interpreted as “fraction of models detected given our priors on the space of models”.

Reducing our GW population population statistics to only distance, we can describe the NS-NS merger population in  Figure~\ref{fig:gw_sims} as having a median $d_L$ of 150 Mpc, a $75\%$-tile of 225 Mpc, and having only $1\%$ at $d_L >300$ Mpc. We take 200 Mpc as a characteristic distance. 

We show in  Figure~\ref{fig:prob-v-dist} the \lTT\ strategy  discovery probability as a function of distance and compare it to the reference strategy.
Starting with the easy case model, \bluebright, we see that the \lTT\ detection probability has a ceiling at $\approx 90\%$.
The reference strategy $P_d$  falls after 200 Mpc, whereas the \lTT\ $P_d$ remains  high out to 330 Mpc.
The \redfaint\ model is difficult for the reference strategy after about 100 Mpc, whereas the \lTT\ has 
$P_d > 50\%$ out to 330 Mpc. The \reddishslow\
model is intermediate and the \lTT\ performs well. The luminosity distance vs area distribution shows the probability weighted histograms on the margins. Not surprisingly, small spatial localizations are both are easiest to make identifications for, but less obvious is that the most likely distances for detection is flat between 125 Mpc and 175 Mpc.

The exposure time must be balanced
by the sky area to be covered: in our maximum 8 hours per night DECam using 1 hour exposures can survey 24 square degrees per night. Figure~\ref{fig:gw_sims} shows that the number of events with sky area less than
24 square degrees drops rapidly at distances greater than 200 Mpc.

Turning to the the distribution of telescope time required per event in Figure~\ref{fig:prob-v-dist}, we will explore the motivation and design of the \lTT\ strategy. For the best (by construction) $P_c$ strategy \Top\, the mode of the time required is $\approx 13-15$ hours for \reddishslow\ \& \redfaint, but saturates at $3-5$ hours for \bluebright. 
The \Top\ strategy likes to use all the time available over two nights to maximize detection.
The \lTT\ strategy is the lowest telescope time within 10\% of \Top's $P_c$, but clearly one can tune how much loss in $P_c$ one is willing to accept. We have compared choosing $5\%$,$10\%$ and $15\%$ (see Table~\ref{tab:cumulative}) and not surprisingly the best choice depends on the KN model chosen. For the  \bluebright~model, a threshold of $5\%$ gives a strategy that outperforms the Reference strategy in $P_d$, in particular after 200 MPc, while using less total telescope time considering following up $90\%-100\%$ of the best-localized events after our initial cut of $300$ sq degrees. This is, in fact, one of the key advantages of our proposed method: we scale up the exposure times with the distance while the Reference strategy keeps the same exposure times of $90s$,  thus we keep high performance in farther away events and use only the necessary time for close events. Another critical point to make optimized use of time is the fact that our proposed method allows the use of shallower exposures in the outer regions that contain less probability, while focusing more deep exposures in the core probability region. This configuration, named scenario 2 in Table \ref{tab:obs_params}, was preferred as discussed later in the Section \label{sec:exp_time} and is the one presented in the plots unless otherwise stated. This is particularly important in \lTT~strategy, since it selects, most of the time, $60$s exposures in \bluebright~model in the outer regions and therefore saves time compared to Reference.  
 For the \reddishslow\ and \redfaint\ models, we prefer to set the threshold at 10\%. This produces in the \reddishslow\ case a strategy that uses a factor of 3 less telescope time than \Top, and the \redfaint\ case a factor of 2 less, for a loss in $P_c <10\%$. The gain in $P_d$ over the reference strategy is particularly dramatic for \redfaint\ at higher distances.

%
%

\begin{table}[htp!]
\centering
\caption{
    \label{tab:cumulative} 
    Average telescope time per event in hours required for two detections, discovery and confirmation. The events used were all that had $90\%$ probability area $< 300$ sq-deg. From 860 events, 611 are retained after this cut. The $50\%$, $90\%$ or $100\%$  columns give average times for events ordered by the statistics of the size of the $90\%$ localization area, low to high. 
    }
\begin{tabular}{lrrr}

\hline \hline 
    Strategy & \multicolumn{3}{c}{Telescope time (hours)}  \\
    \hline \hline
        \hline
    \bluebright &  $50\%$ &  $90\%$ & $100\%$ \\
Reference & 0.4 & 1.1 & 1.6 \\
\Top\ & 2.3 & 3.0 & 3.2\\
\lTT\ ($5\%$) & 0.4 & 0.9 & 1.3 \\
\lTT\ ($10\%$) & 0.4 & 0.9 & 1.2 \\
\lTT\ ($15\%$) & 0.2 & 0.7 & 0.9 \\
\hline
\textit{wide prior,} \reddishslow\\
Reference & 0.4 & 1.1 & 1.5 \\
\Top\ & 10.5 & 12.1 & 12.2\\
\lTT\  ($5\%$) & 5.2 & 6.8 & 7.0 \\
\lTT\  ($10\%$) & 3.2 & 4.3 & 4.5 \\
\lTT\  ($15\%$) & 2.2 & 3.0 & 3.2 \\

\hline 
\redfaint \\
Reference & 0.4 & 1.1 & 1.6 \\
\Top\ & 10.4 & 12.3 & 12.6\\
\lTT\  ($5\%$) & 7.4 & 9.1 & 9.2 \\
\lTT\  ($10\%$) & 5.4 & 6.3 & 6.4 \\
\lTT\  ($15\%$) & 4.1 & 4.8 & 4.8 \\

\hline \hline
\end{tabular}
\end{table}

\begin{figure}[htp!]
    \includegraphics[width=0.95\linewidth]{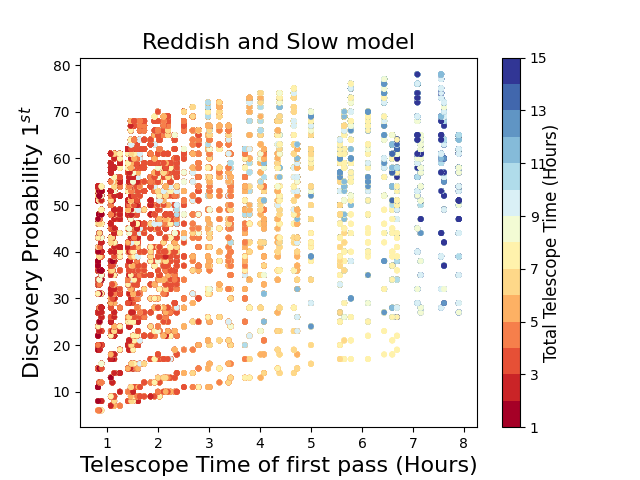}
    \includegraphics[width=0.95\linewidth]{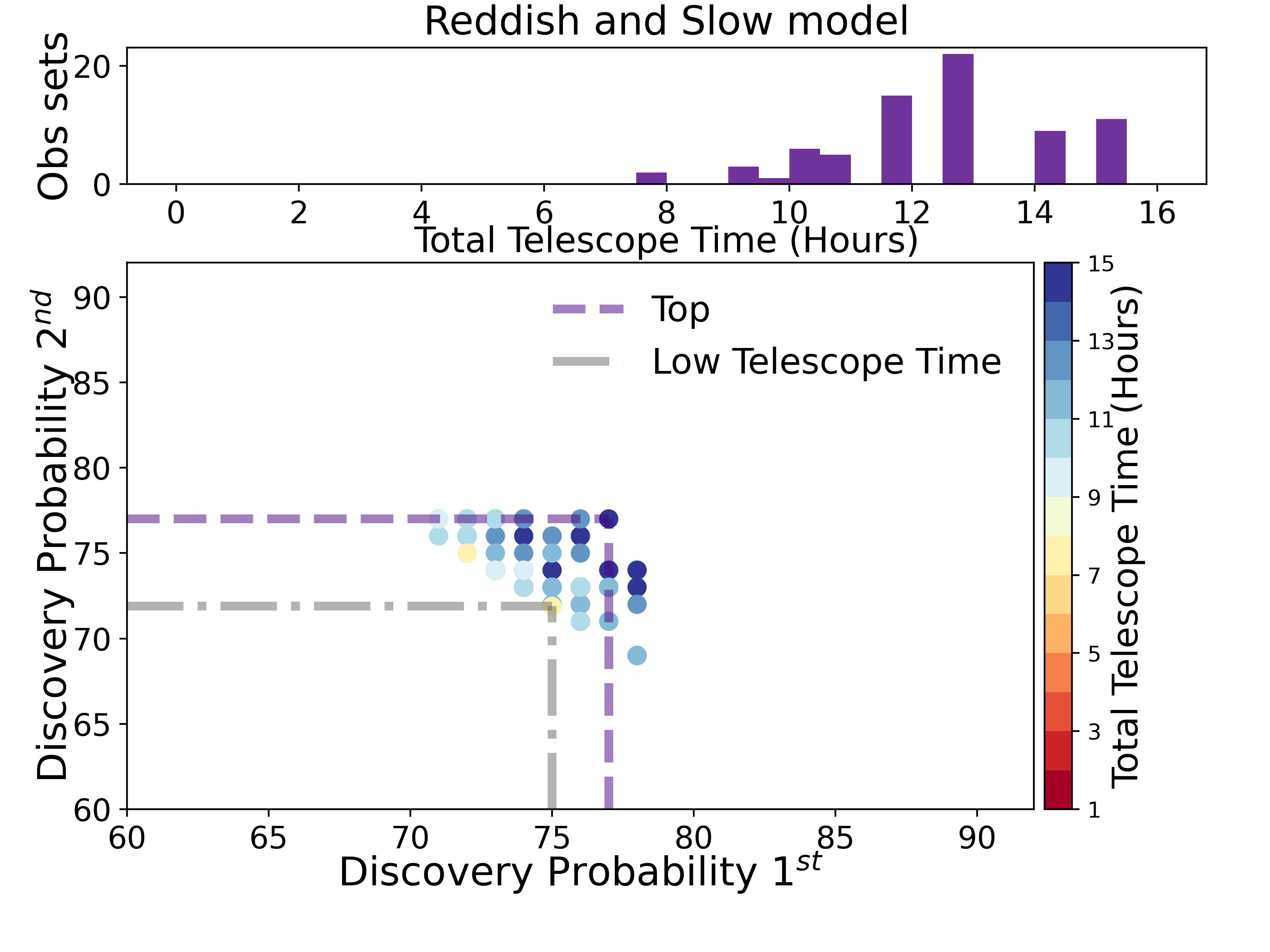}
    \caption{ 
        \label{fig:pd_vs_pd_spray} Discovery propabilities of diferent $\Theta_i$ sets of configurations on a single event in scenario 2. The event is located at 160 Mpc with $90\%$ sky area of 168 sq-degrees in a \reddishslow\ kilonova model peaking at $0.9$ days after the burst in $i$ and $1.2$ in $z$
        Upper: Discovery probability vs time to completion of first pass denoted for each configuration $\Theta_i$. Total telescope time in hours is color coded. 
        Bottom: The best 10\% of Confirmation Probability from the upper plot, displayed in a $P_{d,1}$ vs $P_{d,2}$ plot with total telescope time color coded. The $\Theta_i$ corresponding to \Top\ and \lTT\ strategies are marked. The histogram gives the distribution of telescope times for the $\Theta_i$ in the lower plot.
         }
\end{figure}
\begin{figure*}[htp!]
    \centering
    \includegraphics[width=0.89\linewidth]{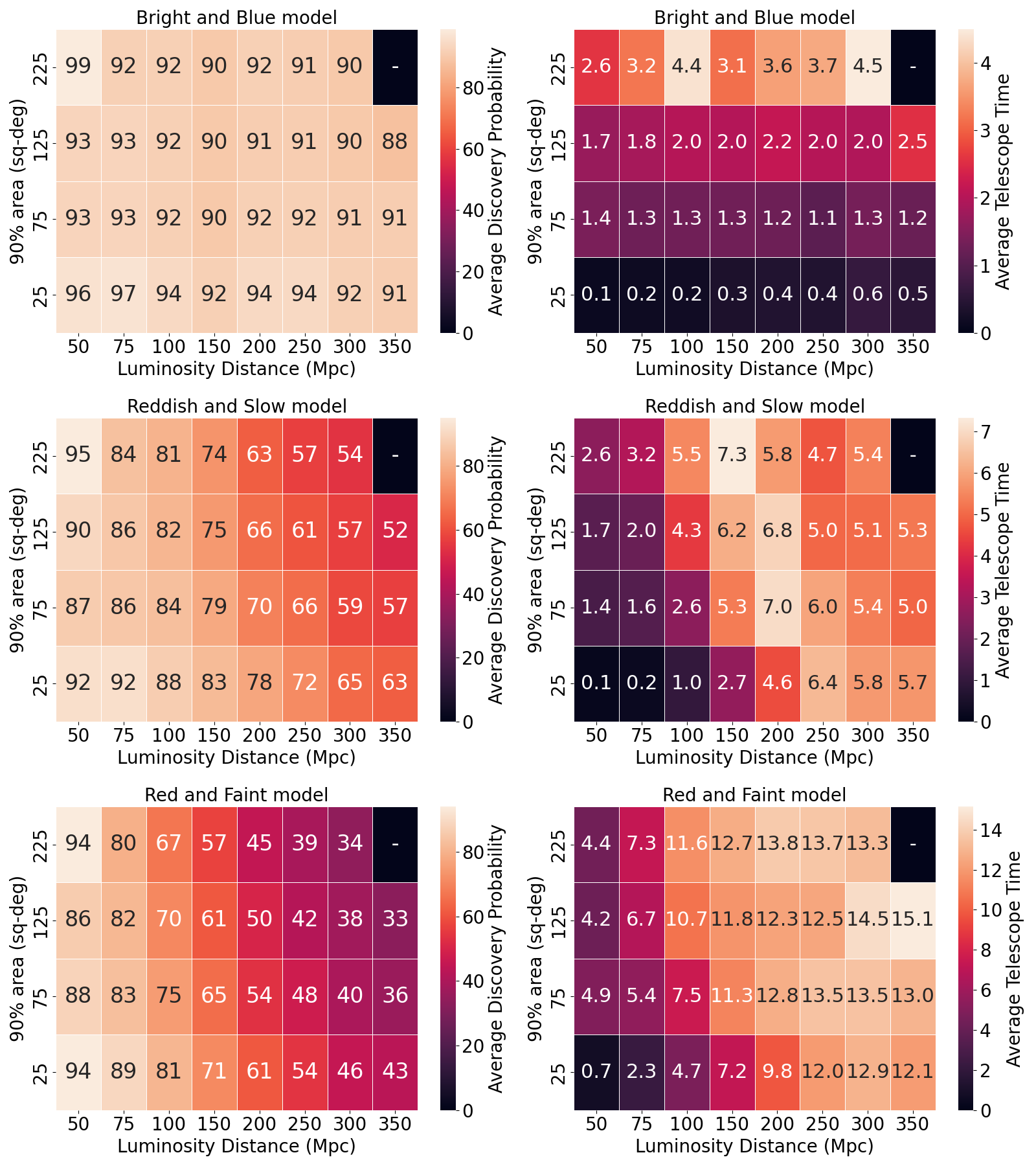}
    \caption{ 
       \label{fig:distance_v_area_heatmap}
        \textit Using the \lTT case {Left column:} Detection probability for a given luminosity distance and sky area for \reddishslow, \redfaint, \bluebright\ in   (top), (middle), and (bottom), respectively. 
        \textit{Right column:} Average required telescope time per event area as a function of luminosity distance, ordered the same as in the left column. 
	}
\end{figure*}


The plots in Figure~\ref{fig:pd_vs_pd_spray} show how the \lTT\ strategy selects its optimized choice as compared to \Top. All 4800 $P_{d,1}$ for the parameter combinations in Table~\ref{tab:obs_params}, assuming scenario 2, for a single event simulation are shown, color coded by the required telescope time. The highest $P_{d,1}$ tend to use the highest telescope times, but there are many high  $P_{d,1}$ parameter combinations $\Theta_i$ that use much less telescope time. The lower plot has the \Top\ and \lTT\ strategies marked. Generically, our strategies prefer balanced $P_{d,1}$ and $P_{d,2}$ to maximize $P_c$. For the event presented in Figure~\ref{fig:pd_vs_pd_spray} the \lTT~configuration chosen for the first detection was core exposure $200$s($120$s) using $i$ band, the second  detection was done with $300$s($200$s) in the core (outer) region, using $z$. The core (outer) region is $80\%$($90\%$) during day 1 after the burst. The \Top~ configuration prefers to explore the core (outer) region with $600$s ($300$s) exposures in the first detection and $300$s ($200$s) in the second detection both cases using filter z. The core (outer) area represents $70\%$($90\%$) regions. It is worth noticing that this \reddishslow models peaks in $i$ band around one day after the burst and in $1.2$ days in $z$ bands, however this model present a slow decaying in $z$ allowing a significant probability of discovery $2$ days after the burst.

There is the question of ``when is it good enough?'', of diminishing
returns. We can adopt the \reddishslow\ model compared to the reference strategy, for example, and work with the best $90\%$ of events. Then the  \lTT\ at a cost of $\times4$ more telescope time detects $20\%$ more KN
and Top t the cost of $\sim 12\times$ more
telescope time detects only $ \sim 30\%$ more KN.
Whether one is willing to accept the cost depends on the science case. 
For the standard siren cosmology case which wants to maximize the number
of KN detected, one would prefer the maximum return of \Top, but might be willing to accept the rate of \lTT. For the science case of studying the astrophysics of KN, where one wants to select good objects for detailed astrophysical study, gathering the next 10 expected in
O4 might well be worth expending what \Top\  requires for the right events.

Returning to the performance of the strategies, we delve deeper in Figure~\ref{fig:distance_v_area_heatmap} for the \lTT\ strategy. 
In this plot we present detection probability as a function of sky area. Here the distance weight in equation~\ref{eq:prob_area} becomes important. In each event, the detection probability calculated for each voxel in the skymap is weighted by the distance to that voxel to form a distance weighted detection probability for the whole event. This accurately weights the probability for the often smaller distances in the lower probability areas of the sky map (see, e.g., \cite{singer2016}). Thus the detection probability for \bluebright\ at $d_L=50$ Mpc can be 99\% when we only cover the 90\% sky probability region.

The easiest way to understand Figure~\ref{fig:distance_v_area_heatmap} is to start with the right column and especially the \bluebright\
model. We have seen from Figure~\ref{fig:prob-v-dist} that \lTT\ is very effective at discovery for this model
and the two dimensional projection shown in Figure~\ref{fig:distance_v_area_heatmap} shows little variation
about that high efficiency. What variation is present is the expected loss of efficiency with increasing
distance and area. In the right column for the \bluebright\ case, it is clear that the  telescope time required  increases with
increasing distance and area. The increase with area is very nearly the ratio of the sky areas.

The \redfaint\ model efficiency plot shows that the  dominant variation is a decrease in efficiency with
increasing distance as the faint objects fall below the limiting magnitude for the maximum exposure time.
Table~\ref{tab:obs_params} shows a maximum exposure of 1 hour so in our 8 hour maximum night,
DECam can cover 24 sq-degrees. The variation in sky area is not as dramatic as with distance and shows the
success of the inner/outer split in Scenario 2, when deep exposures over the high probability area are combined
with shallower exposures over the lower probability area.
(Note that eq~\ref{eq:prob_detect} was evaluated separately in each area.) The corresponding telescope time plot behaves as expected with increasing telescope
time with increasing distance and area. 

The intermediate \reddishslow\ model behaves as expected in
efficiency. It is more $m_{\mathrm{lim}}$ dominated than \bluebright\ and doesn't require  $m_{\mathrm{lim}}$ as deep as \redfaint. The
smallest area bin is likely using 1 hour to maximize area coverage.   The corresponding telescope time plot cell shows 5.7 hours for
the average event, likely less than 8 hours because of the events with less than 24 sq-degree sky area. The
telescope time plot as a whole shows a surprise. There is a peak in the time at intermediate distances and
area, and then the time falls with further distance. 
One can see that some of the same behavior
in the top row of the \redfaint\ model telescope time plot.  The explanation is that there are strategies
(here meaning $\Theta_i$) that use high telescope time to maximize $P_{d,1}$ but these take so long to
cover the area that $P_{d,2}$ is compromised by the fading of the KN, and thus $P_c$ is lowered.  A higher
$P_c$ is obtained by using a shorter time to cover the area 
to cover the area before the object fades, thus maximizing the product of $P_{d,1}\cdot P_{d,2}$.

%% file: results.tex
We have constructed 4800 parameter-set evaluations of DECam kilonova magnitudes produced by SNANA for $611$
Bayestar simulation-detected O4 binary neutron star merger events. We have developed a methodology that uses
discovery and confirmation probabilities and a set of optimization rules to produce strategies optimized for
the discovery of KNe under certain constraints. We have discussed 
maximized discovery (\textit{Top}) and minimal telescope time (\lTT) already, and will discuss several more strategies in \ref{section:bright-and-dark-nights}.
Here we will show detailed behavior of the strategies.


 \begin{figure}[htp!]
    \includegraphics[width=0.90\linewidth]{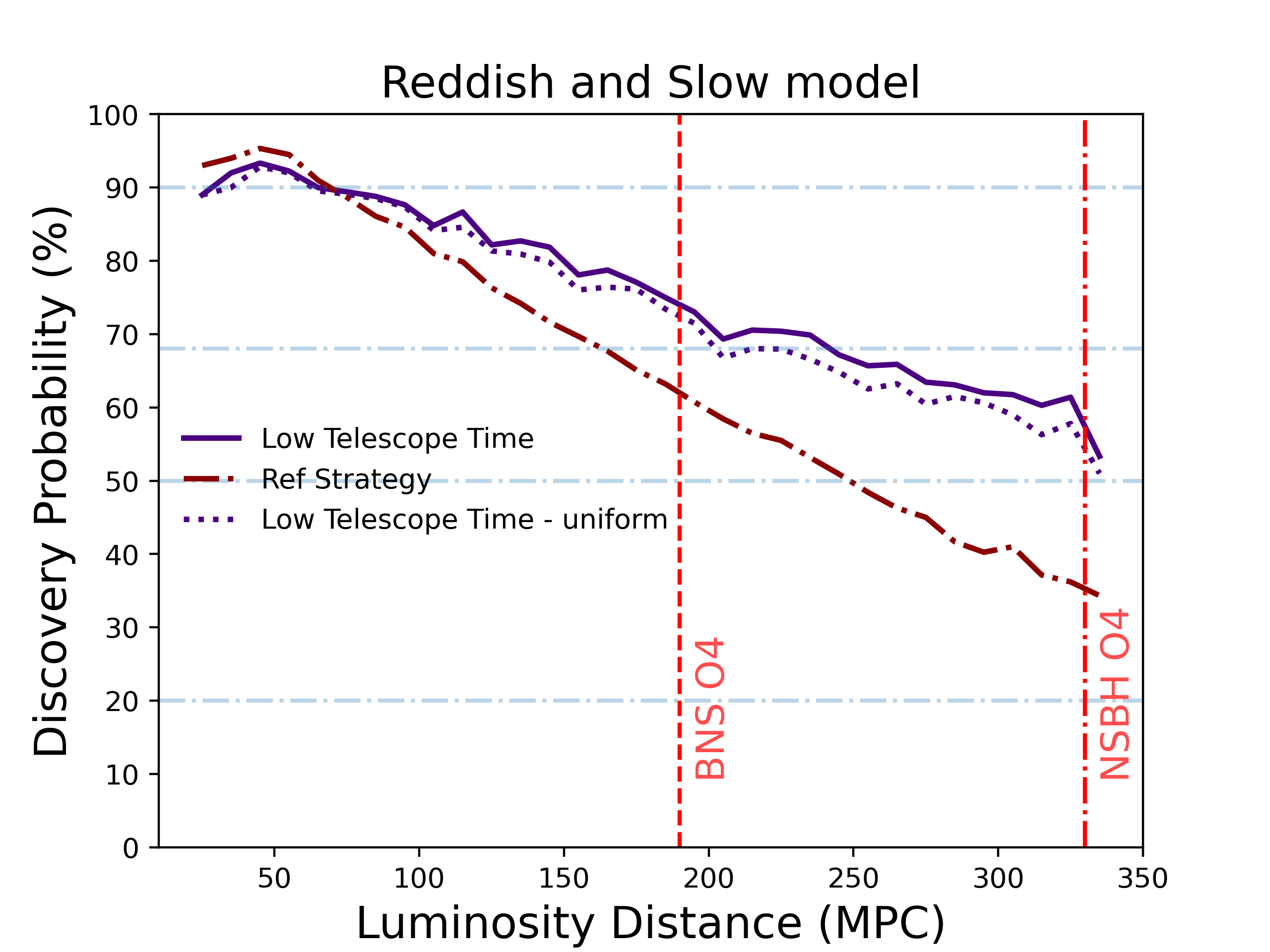}
    \includegraphics[width=0.99\linewidth]{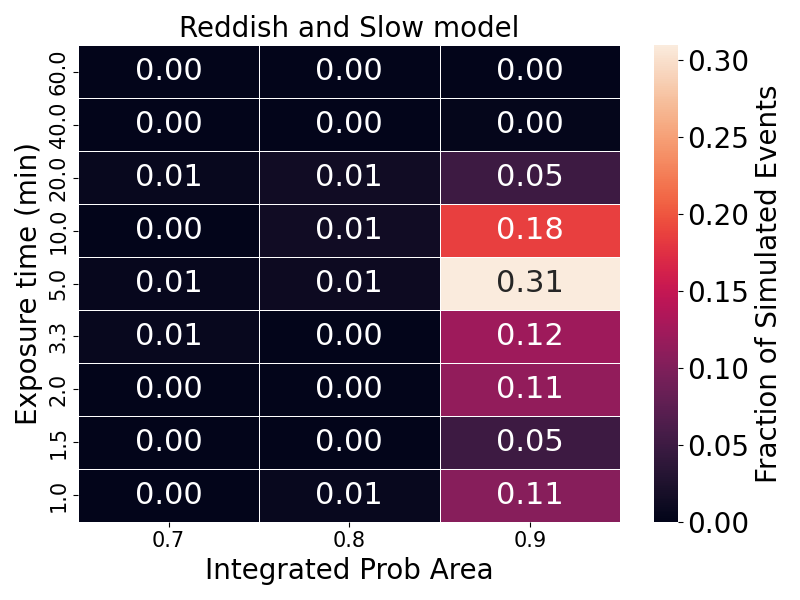}
    \includegraphics[width=0.99\linewidth]{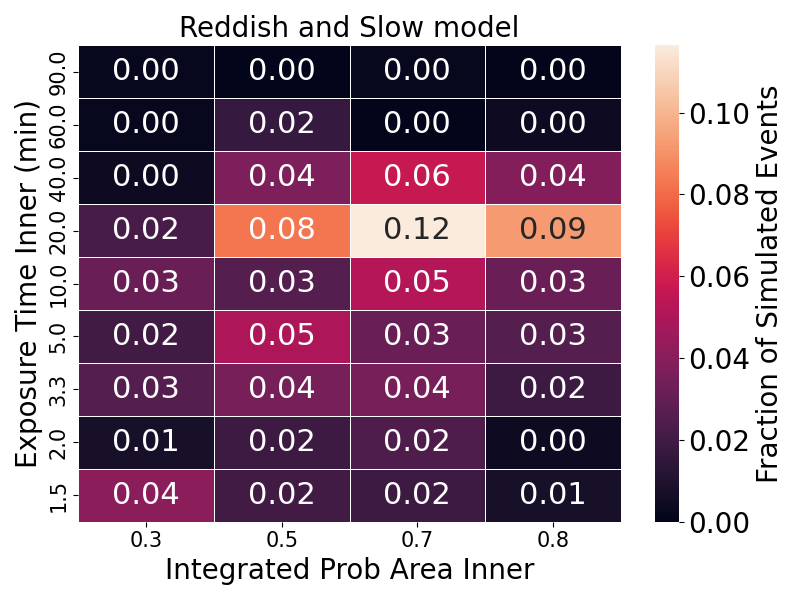}
    \caption{
        \label{fig:exposures}
	Upper: discovery probability vs distance for \lTT\ comparing Scenario 1 and Scenario 2, as well as \textit{Reference}, which only uses Scenario 1. The middle and bottom plots show the exposure times that were most common when using Scenario 2, which breaks up the area into a shallower outer region and a deeper inner region of the spatial sky area. 
}
\end{figure}

We use the \reddishslow\ (GW1701817-blue, $10\times\sigma$) KN model for the results in this section unless otherwise stated.
Likewise, we will 
use the \lTT\ strategy for the results in this section unless otherwise noted. To recap this strategy,
 we go through all combinations of inner \& outer region sizes and exposure times as listed in Table \ref{tab:obs_params} and present the combinations that give the lowest telescope time within $10\%$ of the highest confirmation probability strategy.



\subsection{Exposure times}
\label{sec:exp_time}

One of the dominant features of the observing parameters described in Table~\ref{tab:obs_params} is the splitting of Scenario~1 and Scenario~2. In Scenario~1, the sky is covered with uniform exposure times. In Scenario~2, we allow the splitting of the sky localization area into an outer region and a longer exposure inner region. Scenario~2 is best thought of as an homogeneous pass with a deeper exposure in the high probability region. Putting aside the distance weighting, covering a sky localization area sets a ceiling on the discovery probability to the probability contained in that area, and thereafter it is maximizing the limiting magnitude in the sky area.
Our optimal selections almost always prefer Scenario~2 over Scenario~1 as implicitly or explicitly the total telescope time is constrained.

Figure \ref{fig:exposures} shows the percentage of simulations that preferred each exposure time broken into inner and outer areas. The mode of the 2-d distribution for the inner region is 20 minute exposures, with $31\%$ of events using that exposure time and $84\%$ of events using 20 minutes or less. For the outer region, the mode is 5 minutes, with $33\%$ of events using that exposure time and $74\%$ of events using 5 minutes or less. This is model dependent. The \bluebright\ model uses 90 seconds 99\% of the time in the inner region and 60 seconds 99\% of the time in the outer region. Most often, 87\% of the time, this is a very small inner region of 0.3, and in 93\% of the simulations chose the 0.9 outer probability area. This is the lowest exposure time combination for Scenario 2 in Table~\ref{tab:obs_params}, and it is likely that the strategies would have used a shorter exposure time if available, although it is notable that the strategy did not prefer the available 60 sec homogeneous pass in Scenario 1. 
The \redfaint\ model has a more complicated exposure time pattern for the inner area. In order of use, 2400, 600, 3600, 1200 second exposures are used in 26\%, 18\%, 16\%, \& 15\% of the simulations, respectively. The outer area exposure time has a mode at 1200 seconds of 25\%, with 13\% and 14\% for 2400 and 600 seconds respectively, and 17\% of simulations use 300 seconds.

For distances out to 125 Mpc, the difference between Scenario 1 and Scenario 2 is minimal, but at around 300 Mpc using scenario 2 gains $\sim$ $5\%$-$10\%$  in discovery probability. In other words, the slope in probability vs distance is shallower for the two-zone scenario than it is for the one-zone scenario. Therefore,  we believe that using the deeper in the center approach will in general be more successful the more distant the event. 

\begin{figure}[htp!]
    \centering
    \includegraphics[width=0.95\linewidth]{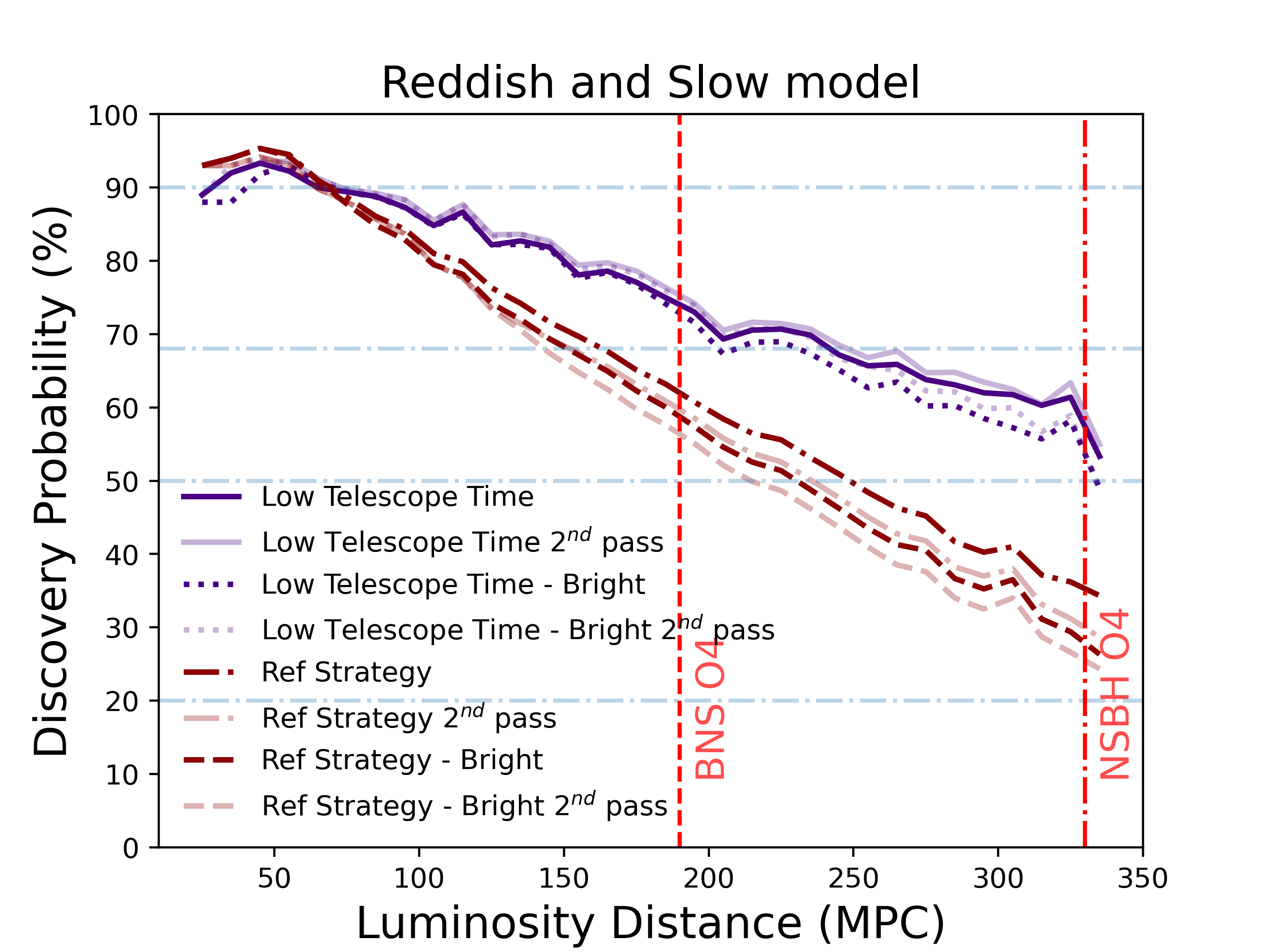}
    \includegraphics[width=0.99\linewidth]{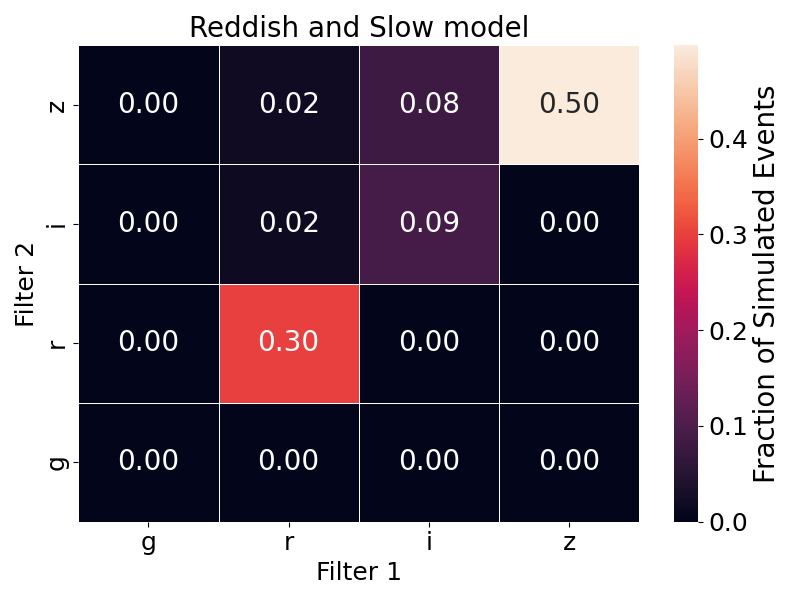}
    \includegraphics[width=0.99\linewidth]{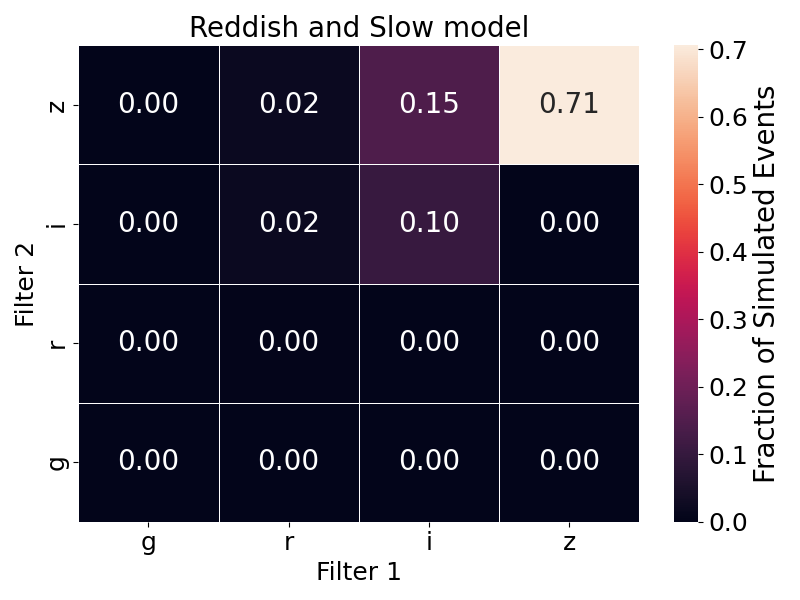}
    \caption{ 
        \label{fig:deep_vs_wide_blue_dark}
        Upper: discovery probability for the different strategies adopted in bright and dark time.
        Middle: heatmap showing the percentage of simulations using each filter in the first and second passes in dark time for the \lTT\ 10\% strategy and the \reddishslow\ model.
        Lower: heatmap showing the percentage of simulations using each filter in the first and second passes in bright time for the \lTT\ 10\% strategy and the \reddishslow\ model. 
	}
\end{figure}

\subsection{Filter Choice and bright/dark nights}
\label{section:bright-and-dark-nights}


In following-up an LVK event, there is a similar chance of observing in dark conditions as in bright conditions. In Figure~\ref{fig:deep_vs_wide_blue_dark} we show the effect of dark versus bright time. The bright/dark distinction is handled in our methodology via a change in $t_\mathrm{eff}$ as seen in Table~\ref{tab:teff}. Bright time lowers discovery probability by $5-10\%$, with the loss being mitigated by filter choice and exposure times.
The filters used in the two passes are,in dark time, $rr (33\%), zz (50\%), ii+iz (17\%)$.
In bright time the filters are $zz (71\%), ii+iz (25\%)$. The filter choices are strategy dependent, and, for comparison, the \Top\ strategy used in dark time  $rr (12\%), ri (7\%), rz (27\%), iz (29\%)$, and $zz (17\%)$ and in bright time  $iz (45\%)$ and $zz (41\%)$. The filter choices are also model dependent, driven by the color evolution of the \reddishslow\ model as seen in Figure~\ref{fig:abs_mag}. The $g$ filter is never going to be favored in this model, $i$ will be picked the first night, $z$ on the second, except that it is easier to go deeper in $r$ than in $i$. We can predict that the filter selection for \redfaint\ model will be nearly the same but that the \bluebright\ model would predominately use $g,r$. Notably, it is not straightforward for us to say which filters we use in our best strategy.

\begin{figure}[htp!]
	\centering
	\includegraphics[width=0.95\linewidth]{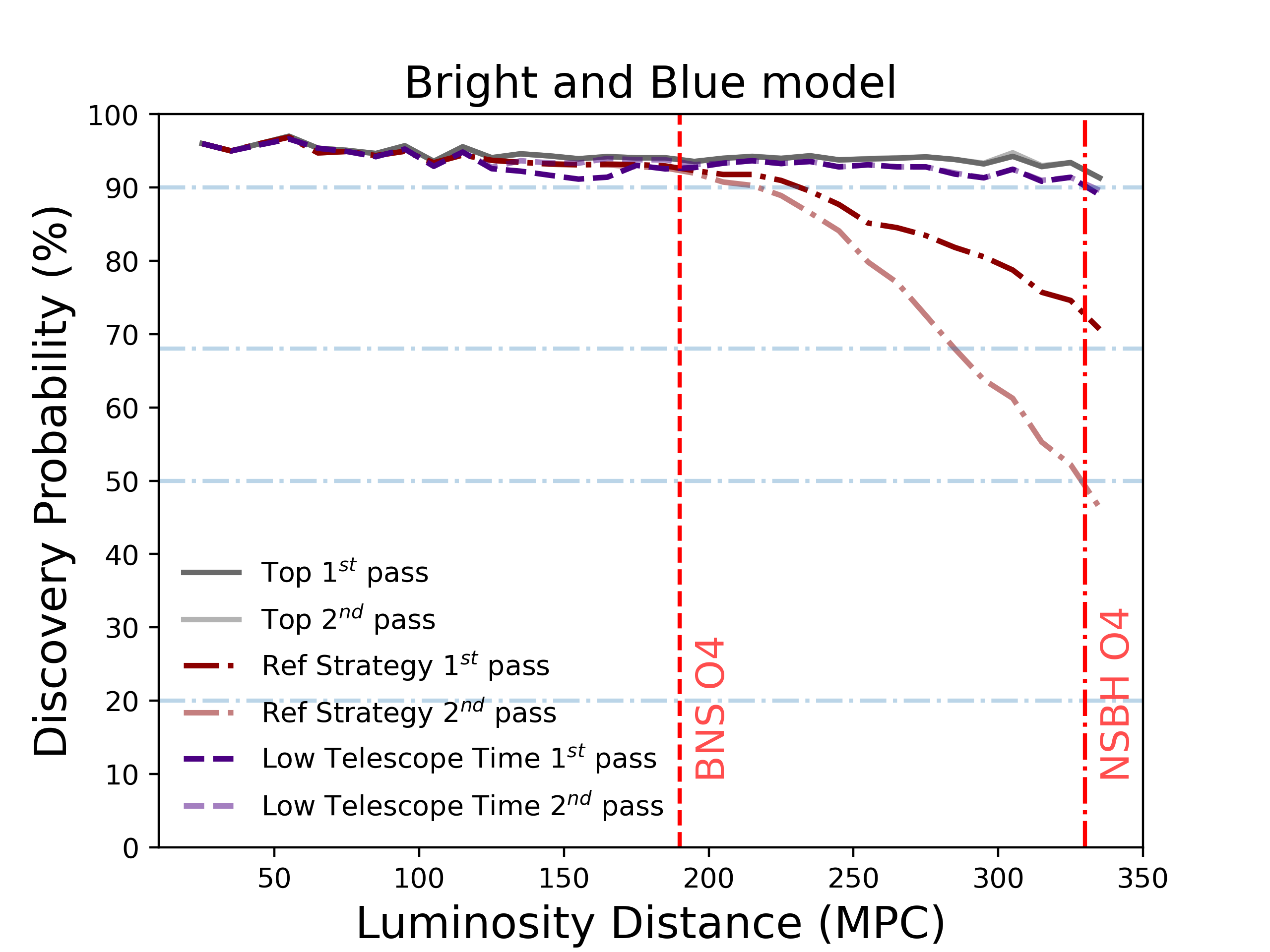}

	\includegraphics[width=0.95\linewidth]{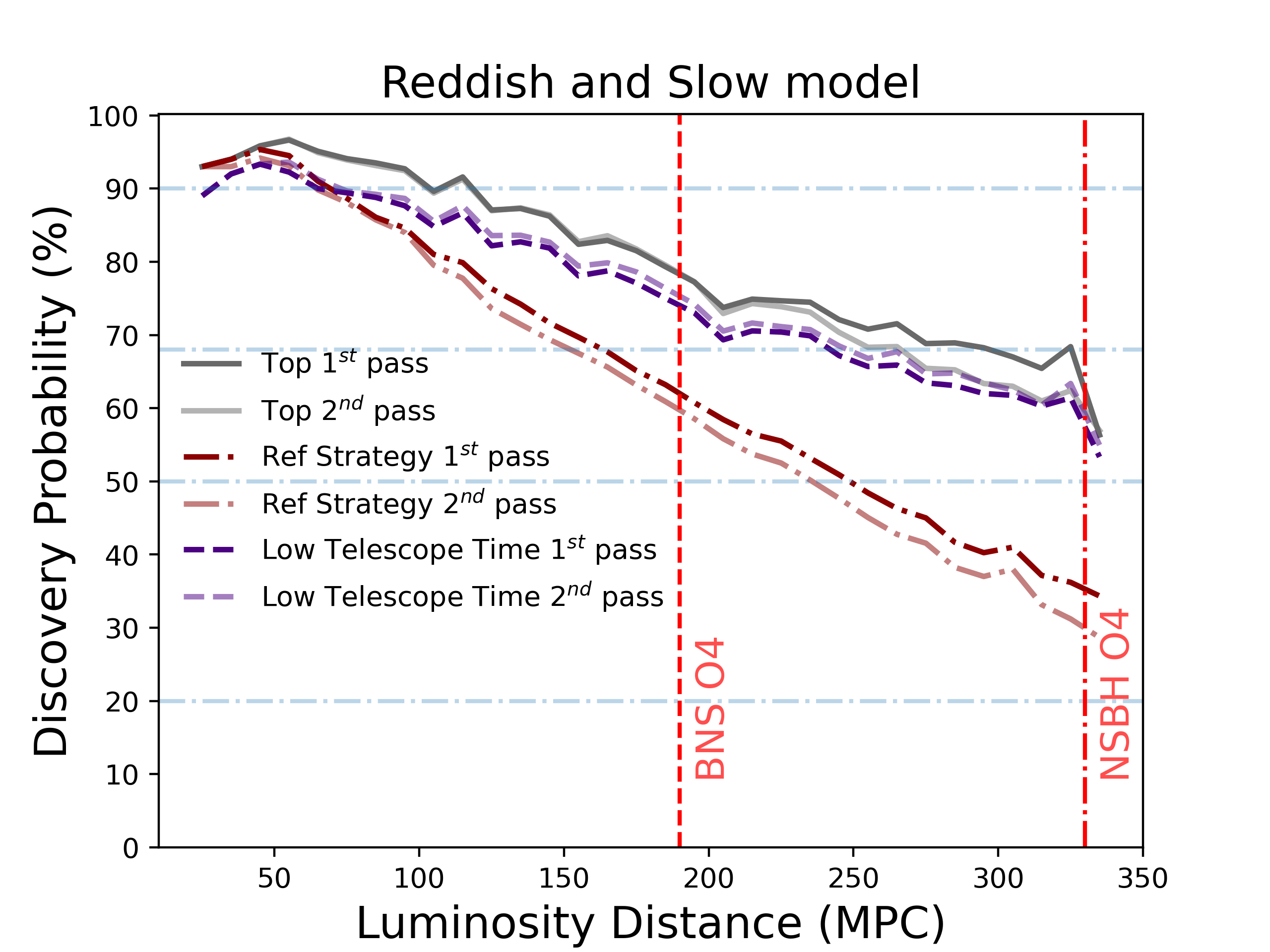}
 	\includegraphics[width=0.95\linewidth]{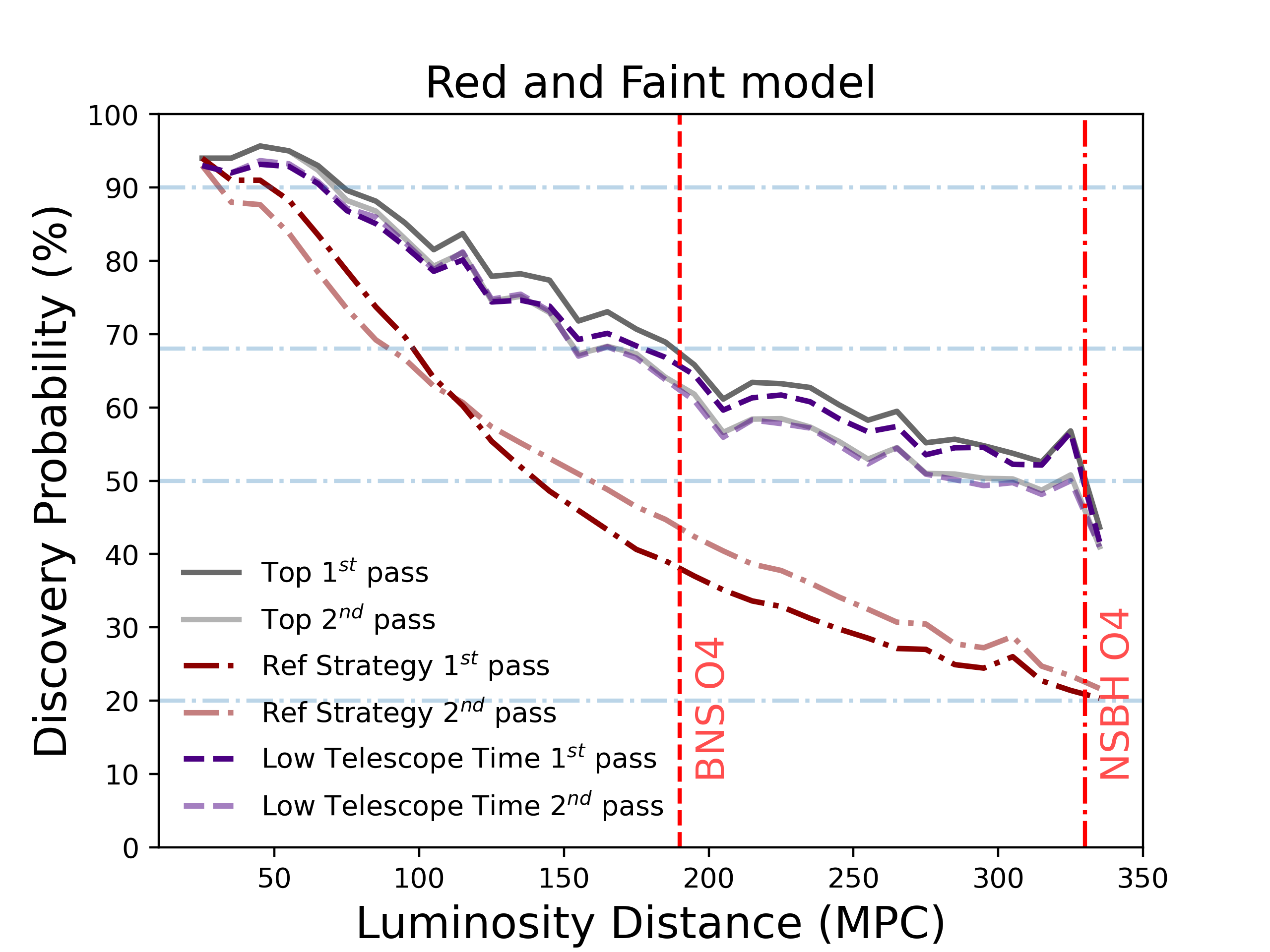}
    \caption {
	 \label{fig:first-second-pass}
        Discovery probability vs distance to event for the \lTT\ observing strategy in the first and second passes compared to the reference strategy for KNe described by model \bluebright\ (\textit{upper}), with wide priors \reddishslow\  (\textit{middle}), and \redfaint\ (\textit{bottom}). 
	 }
\end{figure}

\subsection{Impact of the two-detection requirement}


Detecting a counterpart twice in DECam images is an important step in our experience, as it allows one to distinguish extragalactic transients from asteroids or other objects within a short period of time. This requirement is common, for example \cite{zhu_2021}, \cite{sagues_2021}, and \cite{petrov2022} all demand two detections for a confirmation of a kilonova transient. 
Also important is covering the sky area with multiple filters to distinguish the KN from other transients.

The DECam search  does not happen in isolation, and we do emphasize the importance of the broader GW community during these times. While our programs' main line is to proceed to confirmation via a second image, smaller telescopes may chose to follow-up first image preliminary candidates in order to efficiently reduce the candidate list that will be sent to expensive spectroscopic efforts.  There are reasons to be interested in both $P_d$ and $P_c$.

In Figure~\ref{fig:first-second-pass}, we show  $P(\mathrm{discovery})$ in the first detection, $P_{d,1}$, compared to the second detection, $P_{d,2}$.  In the \lTT\ strategy the detection probability remains nearly constant between pass 1 and pass 2. This is not the case for the reference strategy. For the \reddishslow\ model there is a nearly constant offset between $P_{d,1}$ and $P_{d,2}$ favoring within a few percent the second pass. It is worth noticing that this is our model that peaks at later times, and therefore the second pass might happen closest to the peak. However, the difference is inside the $68\%$ confidence interval as presented in Figure  \ref{fig:prob-v-dist}. For the \redfaint\ model the $P_{d,1}$ is higher at $d <100$ Mpc, and $P_{d,2}$ is higher at $d > 100$ Mpc. For the \bluebright\ model, we see the reference strategy become less efficient than the \lTT\ strategy at $d > 220 $ Mpc, as both $P_{d,1}$ and $P_{d,2}$ drop with increasing distance. We infer that the \lTT\ strategy strongly prefers to balance $P_{d,1}$ and $P_{d,2}$.

\subsection{Other Strategy options}
Our choice of optimization has flexibility.
We might place a high priority on the earliest possible discovery, or we may have lost several nights due to weather conditions and need to find an optimal approach for the first clear night several days after trigger.   
Let us, therefore, explore three other optimizations:
\begin{enumerate}
\item 
\textit{Early Discovery (ED)} is the $\Theta_i$ that produces the earliest confirmation limited by the $P_c$ from \lTT($5\%$) for each $S_i$.
\item 
\textit{Late Discovery (LD)} is the $\Theta_i$ that produces the latest confirmation limited by the $P_c$ from \lTT($5\%$) for each $S_i$.
This family is intended to find a competitive strategy when one cannot observe during some early/intermediate nights or in case the event was not confirmed in the first days.  
\item 
\textit{Half Nights (HN)}.
This family is intended to find a competitive strategy when one cannot observe half the night- if the object rises or sets for example, or the telescope is only allocated for half nights. Thus, from the subset limited by the $P_c$ from \lTT($5\%$) for each $S_i$ configurations  we constrained the strategy to have both passes in less than $4$ hours if they are in the same night or each of the passes takes $4$ hours individually if they are in different nights. In this exercise we consider the observation starts in the first half of the night.

\end{enumerate}

\begin{figure}[htp!]
    \centering
    \includegraphics[width=0.95\linewidth]{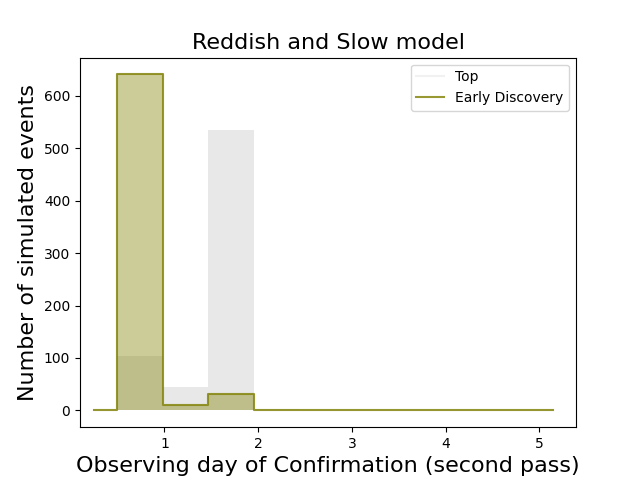}
    \includegraphics[width=0.95\linewidth]{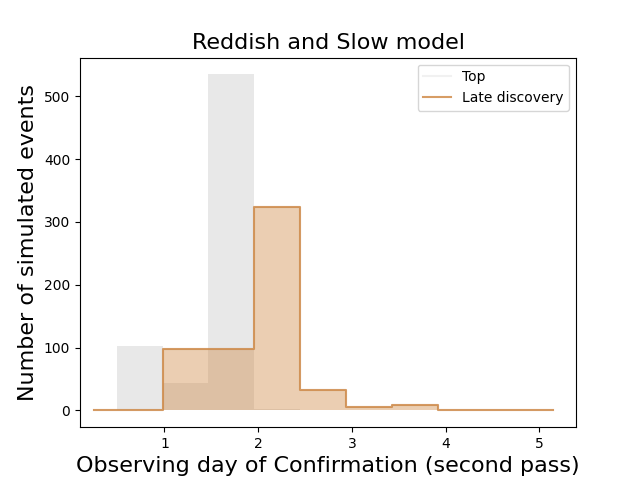}

    \caption{
        \label{fig:observing_day}
        Distribution of confirmation day using the \reddishslow\ model. The upper panel depicts the time of confirmation probability (i.e. how many days it took to observe the area twice) for our \textit{ED} strategy and the bottom panel depicts the \textit{LD} strategy. Most simulated events are confirmed by the first day. For reference we also show the distribution using the \textit{Top} strategy, which has no restrictions on when to perform the follow up.
	}
\end{figure}

All those configurations are restricted in telescope time. In Figure~\ref{fig:observing_day} we show the distributions of the time of second pass completion relative to the merger. It is in the second pass that we achieve a confirmation. The \textit{ED} scenario not surprisingly has earlier times than \textit{Top}, and, by design, earlier times than \textit{LD}.  The late discovery strategy has a different optimization and thus a different use. If the first night or two are not useful for observations, then the \textit{LD} strategy is useful for pursing the discovery at late times. It peaks around $2.5$ days after the merger.  Note that this strategy is optimized on $P_c$, so does not describe the case where the event is unobservable due to clouds for a night or two, but rather is working the scenario where the object is detected but cannot be confirmed for several days. Figure~\ref{fig:observing_day} suggests that since $P_d$ is on the first night, the most likely night to capture the confirmation pass is the third night. \textit{LD} is representative of the kinds of strategies that would be necessary to deal with weather. The \textit{Half Night} strategy enforces a limitation to the amount of time spent in pass 1, and the resulting performance is similar to \lTT\ in terms of both $P_c$ and telescope time expenditures, although without the guarantee of being within $10\%$ of the best strategy, \Top.

\begin{figure}[htp!]
    \centering
    \includegraphics[width=0.95\linewidth]{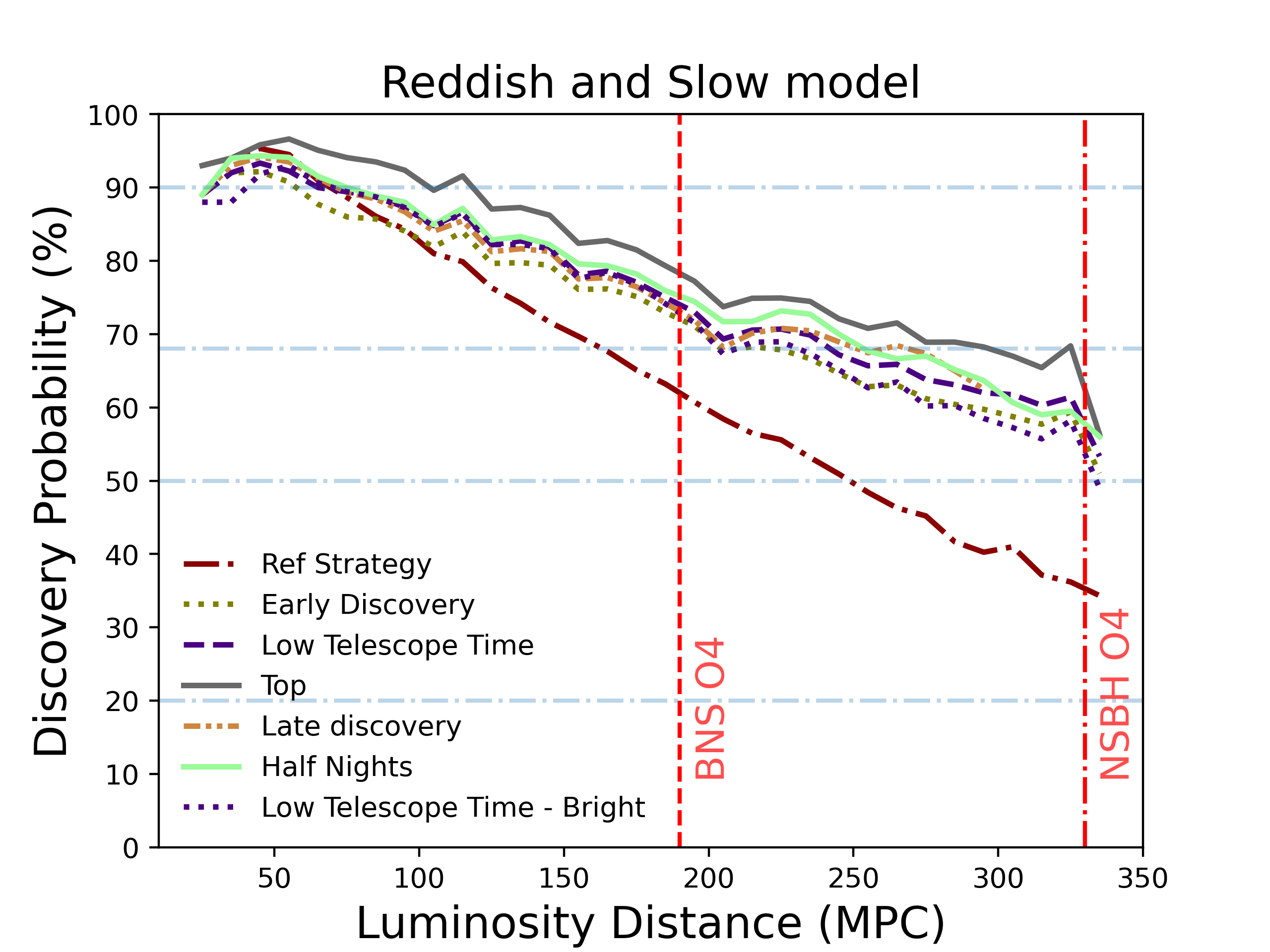}
    \includegraphics[width=0.95\linewidth]{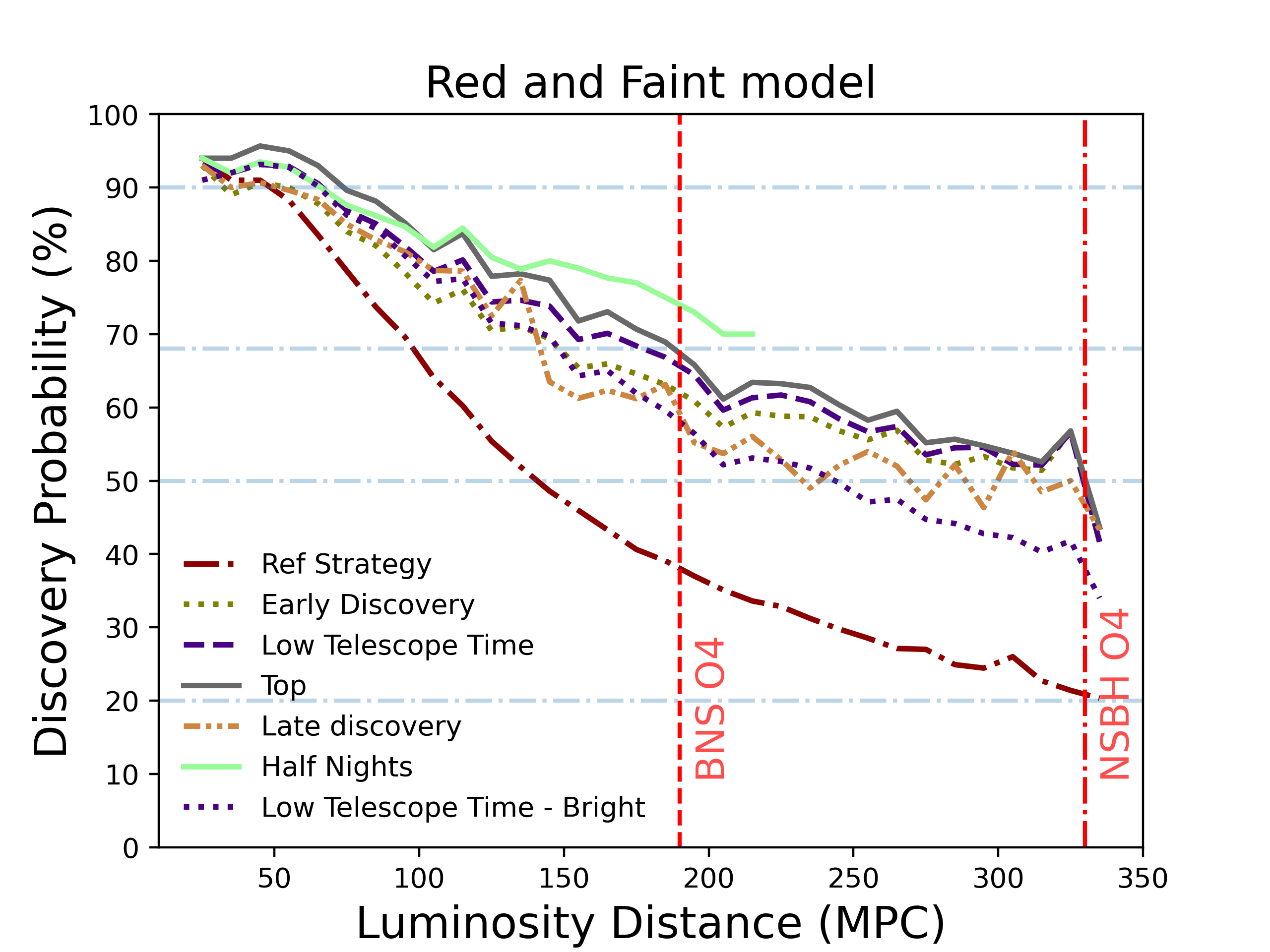}
    \caption{
        \label{fig:flexibility}
	 We illustrate the flexibility of the strategy families. We present the curves constrained with $5\%$ of \Top\ strategy, e.g. \lTT\ ($5\%$) and derived strategies. Upper: \reddishslow. Lower: \redfaint. Half Nights are not always available. For \reddishslow\ 395 half nights strategies produced detections, while for \redfaint\ 137  half nights strategies produced detections. As the exposure times skew deep, the success rate is likely anti-correlated with sky area. 
	 }
\end{figure}
 
 Figure~\ref{fig:flexibility} shows the discovery probability versus distance for the strategies discussed here. For the \reddishslow\ model, all of our strategies have roughly equally performance (but see the \Top\ outperforming \lTT\ by the amount it is allowed to) although with different telescope time cost. This is true to $d=180$ Mpc, but after that, the \textit{LD} strategy becomes less efficient. For the \redfaint\ model, we see, interestingly, that the \lTT\ and \Top\ strategies have equal performance. Finally, the \textit{Half Night} strategy performs very well until $d=180$ Mpc, then becomes infeasible.

For the \reddishslow\ model,
the \textit{ED} strategy for exposure time distribution is more or less equally split among the possible outer/inner exposure time pairs in Table~\ref{tab:obs_params} up to the outer region exposure time of 600 secs, with the most likely outer exposure time of 300 seconds. There are no outer exposure time greater than 600 seconds nor inner exposure times greater than 2400 seconds. 
The \textit{LD} strategy's most likely exposure time is 1200 seconds, and the strategy tries to cover the largest inner region possible. The outer region has the most likely exposure time of 300 seconds and is never longer than 600 seconds. The strategy is to cover the largest inner region possible with relatively shallow outer region exposures. The second pass exposure time is weighted deeper than the first pass but the most likely exposure time remains 1200 seconds. 
The \textit{Half Night} strategy for exposure time distribution  is 31\% of simulations using 1200 seconds, and 17\% using 2400, most often in an 0.7 inner core region. The outer region uses 300 and 600 second exposures over half the time.

For the \redfaint\ model,
the \textit{ED} strategy for exposure time distribution is  deeper than for the same in \reddishslow, preferring 600, 1200, 2400 seconds instead of 600 and 1200, never using 90 second exposure times for the inner core like 15\% of the \reddishslow\ simulations do.
The \textit{LD} strategy simulations 57\% of the time use 600, 1200, or 2400 second exposures but uses a wide range of inner areas.  The outer region uses all exposure times from 60 to 2400 seconds but is most heavily weighted towards 300, 600, and especially 1200 seconds.
The \textit{Half Night} strategy exposure time distribution  is complicated, using a wide variety of inner areas and exposure times skewing deep at 1200 and 2400 second exposures.  The outer region is similar, although here, as most often throughout the strategies, the outer region coverage of 0.9 is preferred.

\begin{table}[htp!]
    \centering
    \caption{
        \label{tab:cumulative2} 
    Average telescope time per event in hours required for two detections, discovery and confirmation, as in Table~\ref{tab:cumulative}. 
    }
    \begin{tabular}{lrrr}

\hline \hline 
    Strategy & \multicolumn{3}{c}{Telescope time (hours)}  \\
    \hline \hline
        \hline
    \bluebright &  $50\%$ &  $90\%$ & $100\%$ \\
    
    \hline
    Early Discovery & 0.4 & 1.0 & 1.4  \\
    Late Discovery$^{\star}$ & 0.4 & 1.0 & 1.4  \\
    Half Nights$^{\star}$ & 0.4 & 1.0 & 1.3 \\
    Low TT - Bright Night & 0.5 & 1.2 & 1.6 \\
    
    \reddishslow \\
    \hline
    Early Discovery & 3.3 & 4.3 & 4.4  \\
    Late Discovery$^{\star}$ & 3.1 & 3.9 & 4.0  \\
    Half Nights$^{\star}$ & 2.9 & 4.0 & 4.0 \\
    Low TT - Bright Night & 3.2 & 4.7 & 5.3  \\
    
    \hline
    \redfaint \\
    \hline
    Early Discovery & 6.7 & 7.7 & 7.8  \\
    Late Discovery$^{\star}$ & 6.6 & 9.1 & 9.1  \\
    Half Nights$^{\star}$ & 2.6 & 3.9 & 4.0 \\
    Low TT - Bright Night & 7.6 & 9.6 & 9.9  \\
    \hline \hline
    \end{tabular}\tablecomments{$^{\star}$  Due to the strong constraints in Half Night and Late Discovery, they are not always available for a given GW event simulation. 
    For \bluebright, \reddishslow, and \redfaint models, these strategies are only defined for 602, 496, and 131 simulations respectively in the half night scenario. While Late discovery is defined for 532, 511, and 251}
\end{table}


\begin{table*}[htp!]
    \begin{adjustwidth}{-2cm}{}
    \centering
    \begin{tabular}{@{}ccccccc@{}} 
    \toprule
    Strategy & Detection Prob. & Filter & Exposure Time & Integrated Prob. Area & Total Telescope time  & Confirmation Day \\
     & (\%)   & & outer, inner (sec) & outer, inner & ($2$ detections, hours) & (days after alert)\\
    \hline
    Top & 75 & r & 3600, 5400 & 0.9, 0.5 & 14.7 & 1.5 \\
    LTT & 68 & z & 300, 1200 & 0.9, 0.7 & 3.2 & 1.0 \\

    HN & 69 & z & 600, 1200 & 0.9, 0.7 & 3.5 & 1.0 \\

    ED & 65 & i & 600, 1200 & 0.9, 0.8 & 2.9 & 0.5 \\ 
    LD & 65 & i & 600, 1200 & 0.9, 0.7 & 3.5 & 2.0\\
    Ref & 48 & i & 90 & 0.9 & 0.5 & 1.5 \\ \hline
    \end{tabular}%
    \end{adjustwidth}
    \caption{
        \label{tab:example-event}
        Excerpt of output csv for example simulated event. Table displays the highest probability configuration for each strategy using scenario 2 (using deeper exposures in higher probability sky area) for the \bluebright model. Exposure time outer, inner represent the time used for the shallow and deep area used out to the area listed in column Integrated Prob. Area outer, inner. This summary information will be used in the decision tree of which strategy to use for an actualy LVC event trigger.  Dual exposure times and probability areas describe the exposure time and area covered for the 'deep' inner and 'shallow' outer region exposures of Table~\ref{tab:obs_params}.
        }
\end{table*}

\section{Real observations}\label{section:real_observations}


We have presented a variety of strategies optimized for a variety of purposes.
Here we describe how to use them when an observing team receives a LVK alert.

We can compute $P(\mathrm{confirmed})$ for the grid of $\Theta$ in Table~\ref{tab:obs_params} (both scenarios) using estimations of \teff\  for the upcoming night. Given the $\Theta_i$, we can choose a strategy to follow, placing the \Top, \lTT, \textit{ED}, or \textit{Half Night} optimizations on as appropriate for the event. The optimization gives us the single $\Theta_i$ for each of the strategy families.

The strategy computation takes about 30 minutes on a single core. This includes the four strategy families (\Top, \lTT, \textit{ED}, \textit{Half Night}), as well as considerations about bright/dark time. The time to complete the computation can be brought down to $<1$ second by using the simulations as an approximation for the real event. Here our approach would be to choose a $\Theta_i$ by nearest neighbor search or to build a simple neural net on the simulation parameters and $P(\mathrm{confirmed})$ to chose $\Theta_i$.

The choice of the KN model is important.
For example, if we are very early on-sky, with great observational conditions for just a couple of hours, or some limitation in telescope access is imposed upon us, we might also consider a fast detection of a blue flash (\bluebright\ model). This strategy could be interesting in particular for \lTT\ due to the cost of deep exposures over a wide area.

There is also a flow of decision making external to what we have described. Whether or not the there is a short GRB, we choose distinct approaches. In the case that there is, we are likely looking for a GW170817-like event (\bluebright). 
If not, then a conservative model is indicated (\reddishslow), as we might not be looking for an event with a high inclination angle. If the alert indicates it is a NSBH merger, then the KN is likely more consistent with the \redfaint\ model.   After this if there is a possible half night strategy, they are low budget and high performing by definition. It also gives time for us to use spectroscopy to confirm quickly. 
In an observing run where we expect 1 BNS merger every month, we could plan on allocating telescope time through the run considering the amount of telescope time remaining versus where the particular event lies in a SNR distribution \cite{chen2014} or among a population of simulated events. This would lead to spending more toward the end of the run if considerable time was left, or spending more if it was a particularly good event compared to the simulations.

\subsection{Plan for Usage}

To demonstrate how using this code might work during a real GW follow up campaign, we randomly select one of the simulations used in this analysis. For this exercise, we are assuming the LVK trigger announcement was received towards the beginning of the night, meaning we will aim to be on sky within a few hours of merger. The general logic flow for determining the observing plan goes as follows:
\begin{enumerate}
    \item Assess our external factors: 
    \begin{enumerate}
        \item How many hours are we allotted for the night and telescope availability, 
        \item Sky conditions,
        \item Check if there was a GRB reported in the area at the same time (as was the case with GW170817). This last criterion is important for choosing a KN model to use. 
    \end{enumerate}
    \item Using the external factors as input, run the code.
    \item Assess output csv, see Table~\ref{tab:example-event}. 
    Here we would decided if this is an event that we would like to follow-up, considering the discovery probability.
    \item Choose filters, exposure times, and area coverage from strategy and compile observing plan.
\end{enumerate}

While most external factors are easy to identify, choosing which KN model to follow may not be obvious and will depend on the particular science goal of the project. For instance, if one's science goal is to have the most complete set of KN, then using the faintest KN model is useful and thus \redfaint\ is the favored model. For simplicity, in this example we will assume a GRB was reported within the LVK sky area, thus favoring \bluebright.  Once the code has generated an output csv file, we can compare strategies. 

The configuration with the highest detection probability for each strategy is displayed in Table~\ref{tab:example-event}. 
Here we can see for this event the \Top\ yields the highest probability of detection, as expected. In this example, however, the \Top\ strategy more has a factor of almost $\sim 5$ in the amount of telescope time needed for any other strategy for only $\sim 7\%$ more probability of detection.

%% file: discussion.tex
In this paper our science goal is to maximize the number of KN with GW measured $d_L$ and securely identified redshifts. We put high weight on completeness of detection of the KN, given the considerable uncertainty in the KN population. Our strategies go deep. 
The strategy chosen by our optimization in \lTT~mode and \reddishslow\ kilonova reach $m_{\mathrm{lim}}(10\sigma)$ of $r \ge 24.4$ and $z \ge 22.9$ for the $\ge 1200$ second exposures (see Table~\ref{tab:mlim}) for the inner regions of $47\%$  (as defined in Table\ref{tab:obs_params})  from our set of $611$ simulations, regardless of distance (see  Figure~\ref{fig:exposures}).

 We can gain insight into our results by comparing how the literature handles a set of 3 questions:
 \begin{enumerate}
     \item the modeling of GW merger event distance and sky area distribution,
     \item the range of KN model physical parameters, including inclination angle
     \item the telescope, search and cadence, detectability versus distance and $m_{\mathrm{lim}}$.
 \end{enumerate}

{\it The GW event properties:}
 We simulate a merging NS-NS population expected in O4 given the expected LIGO sensitivities, drawing from the NS population, simulating the GW waveform and projecting onto a GW observatory network, and converting the GW observations to skymaps. Our draw, simulate, project, and form skymaps approach has been performed before (\citet{petrov2022, chen2021}), though it sometimes is done without skymaps \citep{zhu_2021}. There are simplified approaches. One can just assume a KN population to observe \citep{sagues_2021,setzer2022, chase2022}. Or one can use toy GW models, such as a low-significance event skymap at 200 Mpc \citep{coughlin_2020}. All of these approaches assume the KN properties are independent of the GW event properties, other than distance. \cite{colombo2022} performs the more sophisticated analysis of connecting the KN properties to the GW event properties by going through the chirp mass and mass ratio, which informs $M_\mathrm{ej}$ and $v_\mathrm{ej}$. Of the analyses that go through the draw, simulate, and project methodology, all assume the SN $> 8$ or network SN$>12$ in selection except for \citet{petrov2022} which derive the effective SN threshold from the published GWTC events. They find that there will likely be more events with $90\%$ credible sky areas $>300$ sq-degrees than we simulate, but do not include in our followup identification analysis. For the analyses that extend to jet production (e.g., \cite{zhu_2021, colombo2022}) about $10\%$ of events are sufficiently pole on to for jets to be observed.
  It is worth noticing, however, that the simulations used are based on the assumptions about the sensitivity of the LVK detectors. The predicted sensitivities are roughly equivalent to those anticipated around mid-2022 for the upcoming observing cycle O4. The actual operational conditions diverge from these expected design sensitivity. The Virgo detector, for instance, has not joined in the O4 from its beginning, and it remains uncertain whether it will achieve the proposed sensitivity for this cycle.

Moreover, the KAGRA detector has participated in the run for a limited period, but at a reduced binary neutron star (BNS) inspiral range than that estimated here (80 Mpc). The adjustments in observational capabilities were publicly disclosed only when this work was nearing completion. As a result, the predictions offered in this study may be optimistic when evaluated in the context of the present O4 run. Nonetheless, these forecasts retain their significance, in terms of methodology for guiding observations, even if they do not reflect the current O4 schedule.

 {\it Kilonova physical properties and inclination angles:} We have only one well studied KN in the literature, so there is considerable uncertainty in the KN population. The literature has three ways of modeling KN- using model atmospheres as building blocks \citep[e.g.][]{kasen2017origin}), full physical models of KN atmospheres \citep[e.g.][]{bulla2019, wollaeger_2021}), and using scaling relations and fitting functions from relativisitc numerical simulations (e.g., \citet{dietrich2017, coughlin2019}). The first two approaches are parameterized by at least three variables (for us, $M_\mathrm{ej}, v_\mathrm{ej}, \log{X_\mathrm{lan}}$). The Bulla models, which include inclination angle explicitly, are widely used in the relevant literature (e.g., \citet{coughlin_2020, sagues_2021, zhu_2021, petrov2022}). Often, analyses in the literature will set up a grid of KN parameters and inclination angles and proceed to simulate detection  of each entry in the grid, either for a fixed distance of for NS merger population. Typically the fraction of models detected is termed the detection efficiency. This makes the detection efficiency depend on the model space in unfortunate ways. Consider the case of inclination angle dependent KN properties. One can set up a grid of inclination angles. Better would be to use the probability distribution function of inclination angles for an  random isotropic inclination sample, $PDF(i) \propto \sin(i)$. This weights edge on, i.e. red and faint, KN more than a grid is likely to. The inclination angles sample selected by a GW detector network search isn't isotropic. The amplitude of the strain detected is generically described by $PDF(i) = 0.076076(1+6\, cos^2(i) + cos^4(i))^{3/2} \sin(i)$, that it, GW observatories prefer inclination angles near $30^\circ$ \citep{schutz2011}. For the Bulla models, this could be accounted for by using Bayesian priors on the models.  It would be interesting to have a version of the \cite{setzer2022} KN population absolute magnitude distribution weighted by the expected GW network inclination angle distribution. In our work, we use Bayesian model averaging on our calculated detection probabilities.  We do not model polar vs equatorial directly, but one could map the appropriate blue and red Bayesian model average into those. Our \redfaint\ average absolute magnitude is about 0.4 mags fainter than the brighter peak of \cite{setzer2022}.

 {\it Detection efficiency: } Our study is for DECam on the Blanco 4m with its 3 sq-degree field of view, covering the $90\%$ credible sky area using real cadences in 2 passes to ensure a confirmation. A complete search and discovery simulation over full LVK simulated skymaps has been done by \citet{coughlin_2020} and \cite{petrov2022}. As the 1m class telescopes have very large fields of view, up to the ZTF 47 sq-degrees, much more common in the literature is to assume a KN model, usually a GW170817 analog, and ask what exposure time or limiting magnitude is necessary to detect it. If instead of Bayesian model averaging of the detection probability we had used Bayesian model average absolute magnitudes, our study would be very different. The average absolute magnitudes for our \bluebright , \reddishslow,  and  \redfaint\ models for $r$-band are $-16.3, -15.2, -14.2$ respectively.  In 100 seconds, DECam reaches $r=23.0$, sufficient to detect $M_r=-15.2$ to 500 Mpc. In 1200 seconds, DECam reaches $r=24.4$, sufficient to detect $M_r = -14.2$ to 500 Mpc, and  GW170817's $M_r=-15.5$ to 1 Gpc. Why then do events at $d_L=200$ Mpc, have in our study detection probabilities $P_d$ of 90\%, 73\%, \& 60\%  for the \bluebright, \reddishslow, and \redfaint\ models, respectively, using the \lTT\ strategy, and routinely require 60-90 second exposures for \bluebright and 300-1200 second exposures for the last two models.  The effect of using eq~\ref{eq:prob_detect} is to extend the search to lower $M\_\mathrm{ej}$ and $\log{X_\mathrm{lan}}$ and thus lower luminosities. Our method of accounting for uncertainty in the KN population is driving our results. If we adopt the \citet{coughlin2018} model parameters and uncertainties, our detection efficiencies would increase and exposure times decrease. Using the GW170817 absolute magnitude for detection is likely over-optimistic, as suggested by \citet{colombo2022} placing GW170817 at the $75^{th}$ percentile bright, and \citet{setzer2022} placing it at $95^{th}$ percentile bright. The studies using 1m class telescopes are most likely to assume GW170817 analogs, though \citet{petrov2022} uses both that and a lower luminosity, red model, and \citet{sagues_2021} does a careful analysis of viewing angle dependent models.
The studies assuming the Rubin Observatory \citep{cowperthwaite2019, chen2021}, or a variety/network of telescopes \citep{coughlin_2020, chase2022} tend to analyze detection probabilities for lower luminosity events. In summary, one will have to be careful comparing our detection probabilities with others in the literature, which often use a single luminous model or evaluate and average detection efficiencies over a grid of models using a uniform prior. In our language, ``chance of finding it'' in Fig~\ref{fig:prob-v-dist} is to be interpreted as ``fraction of models detected given our priors on the space of models''.

\begin{deluxetable}{@{}lrrrrr@{}}[htp!]


\label{table:gwtc}
\tablecaption{LVC GWTC events (O1-O3) containing neutron stars and with  $M_2<5 M\odot$ and SNR$ > 8$.}
\tablehead{\colhead{id} & \colhead{$M_1$} & \colhead{$M_2$} & \colhead{distance} & \colhead{SNR} & \colhead{class}\\ 
\colhead{} & \colhead{($M_\odot$)} & \colhead{($M_\odot$)} & \colhead{(Mpc)} & \colhead{}& \colhead{}   } 

\startdata
GW170817 & 1.5 & 1.3 & 40 & 33 & BNS  \\ 
GW190425 & 2.0 & 1.4 & 160 & 13 & BNS \\ 
--\\
GW190814 & 23.2 & 2.6 & 240 & 22 & BH-NS\\ 
GW200105\_162426$^*$ & 9.0 & 1.9 & 270 & 14 & BH-NS\\ 
GW200115\_042309 & 5.9 & 1.4 & 290 & 11 & BH-NS\\ 
GW190426\_152155 & 5.7 & 1.5 & 370 & 10  & BH-NS\\ 
GW191219\_163120  & 31.1 & 1.2 & 550 & 9 & BH-NS\\ 
GW190917\_114630 & 9.3 & 2.1 & 720 & 10 & BH-NS\\ 
GW200210\_092254 & 24.1 & 2.8 & 940 & 8 & BH-NS\\ 
\enddata
\tablecomments{SNR is the matched filter SNR. 
GW200105\_162426$^*$ has $p_{\mathrm{astro}}=0.3$ is thus considered a marginal candidate. Only 2 events have a 90\% confidence sky area of $< 30$ sq-deg, 2 at $< 300$ sq-deg, and 3 at $<400 $ sq-deg. The median sky area is 1700 sq-deg. 
Data from \url{https://www.gw-openscience.org/eventapi/html/allevents/}
}
\end{deluxetable}

\subsection{Applicability to NSBH and mass gap events}

For standard siren studies, NSBH mergers are just as valuable as BNS mergers, as long as they produce electromagnetic counterparts.
NSBH merger events have higher distances for similar SNR than BNS mergers, as can be seen in Table~\ref{table:gwtc},

The dominant factor for their use in standard siren cosmology is the probability of a KN given a NS-BH merger. No counterpart to a NSBH merger has been detected (e.g. \cite{rob2020, Anand_2020, kawaguchi_2020_bh}). \cite{zhu_2021} argue that no detectable KN counterpart was expected for NSBH mergers in O3. \cite{Kawaguchi_2020} and \cite{darbha_2021} study the brightness of KN from NSBH mergers, and \cite{drozda_2020} does the same for mass-gap objects. The summary is that only a fraction of NSBH events will produce KN, primarily those mergers with low mass ratios and high spin. 
  
  The \cite{Kawaguchi_2020} models have absolute magnitudes that peak for $r, i$ at $-14.5$, $-15.0$ respectively. There is a spread of about 1 magnitude fainter in the i-band absolute magnitude, going fainter as the binary mass ratio increases and the effective spin gets smaller. Our \redfaint\ model has an $i$-band absolute magnitude of $-14.5$, midway through the range of \cite{Kawaguchi_2020}.   
  \cite{petrov2022} adopt the \cite{bulla2019} models, broken into BNS and NSBH models both optimistic ($M_\mathrm{ej} = 0.05, 0.08$) and conservative ($M_\mathrm{ej} = 0.01, 0.01$). The optimistic BNS model has an absolute magnitude in the r-band of -16.0, the conservative NSBH model of -14.8; these correspond well to our \bluebright\ and \redfaint\ models, respectively. 
 Taking our \redfaint\ model as appropriate for dynamical ejecta dominated NSBH mergers, our \lTT\ strategy has $50\%$ detection probability out to 330 Mpc (see Figure~\ref{fig:prob-v-dist}). Our strategies are sufficient to obtain the majority of NSBH events that have  EM counterparts if they are at distances $\leq 330$ Mpc.  

\subsection{Blanco/DECam and Rubin LSST}

It is of interest to compare the strategies defined here with the program outlined in
\cite{chen2021}. They assume inclination independence and a GW170817-like KN, and argue for two filter observations. 
The program  conservatively assumes $30\%$ of A+ events by dedicating 7 hours of Rubin Observatory time in 30s exposures, capturing 12/year. Our expectation is that the Rubin ToO program will use 3\% of the available LSST time, so on order of 100 hours which can pursue all BNS events in LVK O4 assuming 1/month and 8 hours per event, so pursuing light curves. Alternatively and more likely, Rubin will choose to observe the 50\% best events by sky area in both the BNS and the NSBH categories. In this scenario a good use of DECam/Blanco would be to follow-up the others that have sky area $< 300$ sq-degrees. Table~\ref{tab:cumulative} suggests this would be viable. 

In fact, if the results of \cite{petrov2022} hold, then there will be many merger events containing NS that have sky areas greater than 300 sq-degrees; our simulations would have to be extended by another $\sim360$ events all with sky area $>300 $ sq-degrees to match their statistics. For the bright siren cosmology every NS event is important. We demonstrate here that the Blanco/DECam especially in combination with the Zwicky observatory and its counterparts PS1, OAJ, LS4, are capable of following up the sources with sky area $<300$ sq-degrees. The optimal use of the Vera C. Rubin Observatory, with its immense etendue, is to followup the LVK sources with $> 300$ sq-degrees. The combination of sky coverage and depth is unmatched. \cite{petrov2022} predicts the median sky coverage for BNS events in O4 is $1820^{+190}_{-170}$ sq-degrees and the median luminosity distance is $352\pm10$ sq-degrees, and the NSBH median distance further away. For the Rubin FoV of 9.6 sq-degrees, the number of exposures to cover the sky area once is $\approx 200$, which at 100 second exposures can be done in less than 6 hours, assuming a \reddishslow\ model and 1.2 mags deeper $m_0$ for Eq.~\ref{eq:mlim}. Likely one could build a two-visit strategy that would take 10 hours per event, allowing Rubin to followup 10 additional events per year without light curves. The Rubin time-domain ecosystem of data, brokers, and routine spectroscopic follow-up is likely to minimize positives, though perhaps not until after O4.

As discussed in  \citet{rob2020,alyssa2020,douglas2021}, the need for coordination with spectroscopic telescopes is vital in identifying the true counterpart. 
Given there has only been one confirmed optical counterpart, there is uncertainty in the expected light curve from photometric data.

%% file: main.bbl
\begin{thebibliography}{}
\providecommand\natexlab[1]{#1}
\providecommand\JournalTitle[1]{#1}
\providecommand{\eprint}[1][]{\url{#1}}

\bibitem[{{Abbott} {et~al.}(2017){Abbott} \& {Abbott} {et~al.}}]{GRB170817A}
{Abbott}, B.~P., {Abbott}, R., {Abbott}, T.~D., {et~al.} 2017, \href{http://dx.doi.org/10.3847/2041-8213/aa920c}{\JournalTitle{\apjl}, 848, L13}, \eprint arXiv:{1710.05834}

\bibitem[{Abbott {et~al.}(2017{\natexlab{a}})Abbott \& Abbott {et~al.}}]{Abbott_2017}
Abbott, B.~P., Abbott, R., Abbott, T.~D., {et~al.} 2017{\natexlab{a}}, \href{http://dx.doi.org/10.3847/2041-8213/aa91c9}{\JournalTitle{The Astrophysical Journal}, 848, L12}

\bibitem[{Abbott {et~al.}(2017{\natexlab{b}})Abbott \& Abbott {et~al.}}]{abbott2017multi}
Abbott, B.~P., Abbott, R., Abbott, T., {et~al.} 2017{\natexlab{b}}, \JournalTitle{Astrophys. J. Lett}, 848, L12

\bibitem[{Abbott {et~al.}(2018)Abbott \& Abbott {et~al.}}]{abbott2018prospects}
Abbott, B.~P., Abbott, R., Abbott, T., {et~al.} 2018, \JournalTitle{Living Reviews in Relativity}, 21, 3

\bibitem[{Abbott {et~al.}(2020)Abbott \& Abbott {et~al.}}]{Abbott2020}
Abbott, B.~P., Abbott, R., Abbott, T.~D., {et~al.} 2020, \href{http://dx.doi.org/10.1007/s41114-020-00026-9}{\JournalTitle{Living Reviews in Relativity}, 23}

\bibitem[{{Abbott} {et~al.}(2021){Abbott} \& {Abbott} {et~al.}}]{2NSBH}
{Abbott}, R., {Abbott}, T.~D., {Abraham}, S., {et~al.} 2021, \href{http://dx.doi.org/10.3847/2041-8213/ac082e}{\JournalTitle{\apjl}, 915, L5}, \eprint arXiv:{2106.15163}

\bibitem[{Abbott {et~al.}(2021)}]{LIGOScientific:2021psn}
Abbott, R. {et~al.} 2021, \JournalTitle{Phys. Rev. X}, \eprint arXiv:{2111.03634}

\bibitem[{Abdalla {et~al.}(2022)Abdalla \& Abellán {et~al.}}]{cosmointer}
Abdalla, E., Abellán, G.~F., Aboubrahim, A., {et~al.} 2022, \href{http://dx.doi.org/https://doi.org/10.1016/j.jheap.2022.04.002}{\JournalTitle{Journal of High Energy Astrophysics}, 34, 49}

\bibitem[{Anand {et~al.}(2020)Anand \& Coughlin {et~al.}}]{Anand_2020}
Anand, S., Coughlin, M.~W., Kasliwal, M.~M., {et~al.} 2020, \href{http://dx.doi.org/10.1038/s41550-020-1183-3}{\JournalTitle{Nature Astronomy}, 5, 46}

\bibitem[{Anand {et~al.}(2021)Anand \& Coughlin {et~al.}}]{anand2021optical}
Anand, S., Coughlin, M.~W., Kasliwal, M.~M., {et~al.} 2021, \JournalTitle{Nature Astronomy}, 5, 46

\bibitem[{{Andreoni} {et~al.}(2017){Andreoni} \& {Ackley} {et~al.}}]{2017PASA...34...69A}
{Andreoni}, I., {Ackley}, K., {Cooke}, J., {et~al.} 2017, \href{http://dx.doi.org/10.1017/pasa.2017.65}{\JournalTitle{\pasa}, 34, e069}, \eprint arXiv:{1710.05846}

\bibitem[{{Andreoni} {et~al.}(2019){Andreoni} \& {Goldstein} {et~al.}}]{igor2019}
{Andreoni}, I., {Goldstein}, D.~A., {Anand}, S., {et~al.} 2019, \href{http://dx.doi.org/10.3847/2041-8213/ab3399}{\JournalTitle{\apjl}, 881}, \eprint arXiv:{1906.00806}

\bibitem[{Andreoni {et~al.}(2019)Andreoni \& Goldstein {et~al.}}]{andreoni2019growth190814}
Andreoni, I., Goldstein, D.~A., Kasliwal, M.~M., {et~al.} 2019, \JournalTitle{arXiv preprint arXiv:1910.13409}

\bibitem[{{Arcavi} {et~al.}(2017){Arcavi} \& {Hosseinzadeh} {et~al.}}]{arcavi}
{Arcavi}, I., {Hosseinzadeh}, G., {Howell}, D.~A., {et~al.} 2017, \href{http://dx.doi.org/10.1038/nature24291}{\JournalTitle{\nat}, 551, 64}, \eprint arXiv:{1710.05843}

\bibitem[{{Ashton} {et~al.}(2020){Ashton} \& {Ackley} \& {Maga{\~n}a Hernandez} \& {Piotrzkowski}}]{ashton}
{Ashton}, G., {Ackley}, K., {Maga{\~n}a Hernandez}, I., \& {Piotrzkowski}, B. 2020, \JournalTitle{arXiv e-prints}, arXiv:2009.12346, \eprint arXiv:{2009.12346}

\bibitem[{{Balasubramanian} {et~al.}(2021){Balasubramanian} \& {Corsi} {et~al.}}]{Balasubramanian21}
{Balasubramanian}, A., {Corsi}, A., {Mooley}, K.~P., {et~al.} 2021, \href{http://dx.doi.org/10.3847/2041-8213/abfd38}{\JournalTitle{\apjl}, 914, L20}, \eprint arXiv:{2103.04821}

\bibitem[{{Bom} \& {Palmese}(2023)}]{bom23}
{Bom}, C.~R. \& {Palmese}, A. 2023, \href{http://dx.doi.org/10.48550/arXiv.2307.01330}{\JournalTitle{arXiv e-prints}, arXiv:2307.01330}, \eprint arXiv:{2307.01330}

\bibitem[{{Bulla}(2019)}]{bulla2019}
{Bulla}, M. 2019, \href{http://dx.doi.org/10.1093/mnras/stz2495}{\JournalTitle{\mnras}, 489, 5037}, \eprint arXiv:{1906.04205}

\bibitem[{Buonanno {et~al.}(2009)Buonanno \& Iyer \& Ochsner \& Pan \& Sathyaprakash}]{Buonanno_2009}
Buonanno, A., Iyer, B.~R., Ochsner, E., Pan, Y., \& Sathyaprakash, B.~S. 2009, \href{http://dx.doi.org/10.1103/physrevd.80.084043}{\JournalTitle{Physical Review D}, 80}

\bibitem[{{Chase} {et~al.}(2022){Chase} \& {O'Connor} {et~al.}}]{chase2022}
{Chase}, E.~A., {O'Connor}, B., {Fryer}, C.~L., {et~al.} 2022, \href{http://dx.doi.org/10.3847/1538-4357/ac3d25}{\JournalTitle{\apj}, 927, 163}, \eprint arXiv:{2105.12268}

\bibitem[{{Chen} {et~al.}(2021){Chen} \& {Cowperthwaite} \& {Metzger} \& {Berger}}]{chen2021}
{Chen}, H.-Y., {Cowperthwaite}, P.~S., {Metzger}, B.~D., \& {Berger}, E. 2021, \href{http://dx.doi.org/10.3847/2041-8213/abdab0}{\JournalTitle{\apjl}, 908, L4}, \eprint arXiv:{2011.01211}

\bibitem[{Chen {et~al.}(2018)Chen \& Fishbach \& Holz}]{Chen2018}
Chen, H.-Y., Fishbach, M., \& Holz, D.~E. 2018, \href{http://dx.doi.org/10.1038/s41586-018-0606-0}{\JournalTitle{Nature}, 562, 545}

\bibitem[{{Chen} \& {Holz}(2014)}]{chen2014}
{Chen}, H.-Y. \& {Holz}, D.~E. 2014, \JournalTitle{arXiv e-prints}, \eprint arXiv:{1409.0522}

\bibitem[{{Colombo} {et~al.}(2022){Colombo} \& {Salafia} {et~al.}}]{colombo2022}
{Colombo}, A., {Salafia}, O.~S., {Gabrielli}, F., {et~al.} 2022, \href{http://dx.doi.org/10.3847/1538-4357/ac8d00}{\JournalTitle{\apj}, 937, 79}, \eprint arXiv:{2204.07592}

\bibitem[{{Coughlin} {et~al.}(2019){Coughlin} \& {Dietrich} \& {Margalit} \& {Metzger}}]{coughlin2019}
{Coughlin}, M.~W., {Dietrich}, T., {Margalit}, B., \& {Metzger}, B.~D. 2019, \href{http://dx.doi.org/10.1093/mnrasl/slz133}{\JournalTitle{\mnras}, 489, L91}, \eprint arXiv:{1812.04803}

\bibitem[{{Coughlin} {et~al.}(2018){Coughlin} \& {Dietrich} {et~al.}}]{coughlin2018}
{Coughlin}, M.~W., {Dietrich}, T., {Doctor}, Z., {et~al.} 2018, \href{http://dx.doi.org/10.1093/mnras/sty2174}{\JournalTitle{\mnras}, 480, 3871}, \eprint arXiv:{1805.09371}

\bibitem[{{Coughlin} {et~al.}(2020){Coughlin} \& {Dietrich} {et~al.}}]{coughlin_2020}
{Coughlin}, M.~W., {Dietrich}, T., {Antier}, S., {et~al.} 2020, \href{http://dx.doi.org/10.1093/mnras/staa1925}{\JournalTitle{\mnras}, 497, 1181}, \eprint arXiv:{2006.14756}

\bibitem[{Coulter {et~al.}(2017{\natexlab{a}})Coulter \& Foley {et~al.}}]{coulter2017swope}
Coulter, D., Foley, R., Kilpatrick, C., {et~al.} 2017{\natexlab{a}}, \JournalTitle{Science}, 358, 1556

\bibitem[{Coulter {et~al.}(2017{\natexlab{b}})Coulter \& Foley {et~al.}}]{Coulter1556}
Coulter, D.~A., Foley, R.~J., Kilpatrick, C.~D., {et~al.} 2017{\natexlab{b}}, \href{http://dx.doi.org/10.1126/science.aap9811}{\JournalTitle{Science}, 358, 1556}

\bibitem[{Cowperthwaite {et~al.}(2017)Cowperthwaite \& Berger {et~al.}}]{cowperthwaite2017electromagnetic}
Cowperthwaite, P., Berger, E., Villar, V., {et~al.} 2017, \JournalTitle{The Astrophysical Journal Letters}, 848, L17

\bibitem[{{Cowperthwaite} {et~al.}(2019){Cowperthwaite} \& {Villar} \& {Scolnic} \& {Berger}}]{cowperthwaite2019}
{Cowperthwaite}, P.~S., {Villar}, V.~A., {Scolnic}, D.~M., \& {Berger}, E. 2019, \href{http://dx.doi.org/10.3847/1538-4357/ab07b6}{\JournalTitle{\apj}, 874, 88}, \eprint arXiv:{1811.03098}

\bibitem[{{Darbha} \& {Kasen}(2020)}]{darbha_2020}
{Darbha}, S. \& {Kasen}, D. 2020, \href{http://dx.doi.org/10.3847/1538-4357/ab9a34}{\JournalTitle{\apj}, 897, 150}, \eprint arXiv:{2002.00299}

\bibitem[{{Darbha} {et~al.}(2021){Darbha} \& {Kasen} \& {Foucart} \& {Price}}]{darbha_2021}
{Darbha}, S., {Kasen}, D., {Foucart}, F., \& {Price}, D.~J. 2021, \href{http://dx.doi.org/10.3847/1538-4357/abff5d}{\JournalTitle{\apj}, 915, 69}, \eprint arXiv:{2103.03378}

\bibitem[{{Del Pozzo}(2012)}]{delpozzo}
{Del Pozzo}, W. 2012, \href{http://dx.doi.org/10.1103/PhysRevD.86.043011}{\JournalTitle{\prd}, 86, 043011}, \eprint arXiv:{1108.1317}

\bibitem[{{D{\'\i}az} {et~al.}(2017){D{\'\i}az} \& {Macri} {et~al.}}]{toros}
{D{\'\i}az}, M.~C., {Macri}, L.~M., {Garcia Lambas}, D., {et~al.} 2017, \href{http://dx.doi.org/10.3847/2041-8213/aa9060}{\JournalTitle{\apjl}, 848, L29}, \eprint arXiv:{1710.05844}

\bibitem[{{Dietrich} \& {Ujevic}(2017)}]{dietrich2017}
{Dietrich}, T. \& {Ujevic}, M. 2017, \href{http://dx.doi.org/10.1088/1361-6382/aa6bb0}{\JournalTitle{Classical and Quantum Gravity}, 34, 105014}, \eprint arXiv:{1612.03665}

\bibitem[{{Drout} {et~al.}(2017){Drout} \& {Piro} {et~al.}}]{drout2017}
{Drout}, M.~R., {Piro}, A.~L., {Shappee}, B.~J., {et~al.} 2017, \href{http://dx.doi.org/10.1126/science.aaq0049}{\JournalTitle{Science}, 358, 1570}, \eprint arXiv:{1710.05443}

\bibitem[{{Drozda} {et~al.}(2020){Drozda} \& {Belczynski} \& {O'Shaughnessy} \& {Bulik} \& {Fryer}}]{drozda_2020}
{Drozda}, P., {Belczynski}, K., {O'Shaughnessy}, R., {Bulik}, T., \& {Fryer}, C.~L. 2020, \JournalTitle{arXiv e-prints}, \eprint arXiv:{2009.06655}

\bibitem[{Finke {et~al.}(2021)Finke \& Foffa \& Iacovelli \& Maggiore \& Mancarella}]{finke2021cosmology}
Finke, A., Foffa, S., Iacovelli, F., Maggiore, M., \& Mancarella, M. 2021, Cosmology with LIGO/Virgo dark sirens: Hubble parameter and modified gravitational wave propagation, \eprint arXiv:{2101.12660}

\bibitem[{{Finn} \& {Chernoff}(1993)}]{finn1993}
{Finn}, L.~S. \& {Chernoff}, D.~F. 1993, \href{http://dx.doi.org/10.1103/PhysRevD.47.2198}{\JournalTitle{\prd}, 47, 2198}, \eprint arXiv:{gr-qc/9301003}

\bibitem[{{Flaugher} {et~al.}(2015){Flaugher} \& {Diehl} {et~al.}}]{decam}
{Flaugher}, B., {Diehl}, H.~T., {Honscheid}, K., {et~al.} 2015, \href{http://dx.doi.org/10.1088/0004-6256/150/5/150}{\JournalTitle{\aj}, 150, 150}, \eprint arXiv:{1504.02900}

\bibitem[{{Garcia} {et~al.}(2020){Garcia} \& {Morgan} {et~al.}}]{alyssa2020}
{Garcia}, A., {Morgan}, R., {Herner}, K., {et~al.} 2020, \href{http://dx.doi.org/10.3847/1538-4357/abb823}{\JournalTitle{\apj}, 903, 75}, \eprint arXiv:{2007.00050}

\bibitem[{{Gillanders} {et~al.}(2022){Gillanders} \& {Smartt} \& {Sim} \& {Bauswein} \& {Goriely}}]{gillanders_2022}
{Gillanders}, J.~H., {Smartt}, S.~J., {Sim}, S.~A., {Bauswein}, A., \& {Goriely}, S. 2022, \JournalTitle{arXiv e-prints}, \eprint arXiv:{2202.01786}

\bibitem[{{Goldstein} {et~al.}(2017){Goldstein} \& {Veres} {et~al.}}]{goldstein2017}
{Goldstein}, A., {Veres}, P., {Burns}, E., {et~al.} 2017, \href{http://dx.doi.org/10.3847/2041-8213/aa8f41}{\JournalTitle{\apjl}, 848, L14}, \eprint arXiv:{1710.05446}

\bibitem[{Goldstein {et~al.}(2019)Goldstein \& Andreoni {et~al.}}]{goldstein2019growth_190426}
Goldstein, D.~A., Andreoni, I., Nugent, P.~E., {et~al.} 2019, \href{http://dx.doi.org/10.3847/2041-8213/ab3046}{\JournalTitle{The Astrophysical Journal}, 881, L7}

\bibitem[{{Gottlieb} {et~al.}(2018){Gottlieb} \& {Nakar} \& {Piran}}]{gottlieb2018}
{Gottlieb}, O., {Nakar}, E., \& {Piran}, T. 2018, \href{http://dx.doi.org/10.1093/mnras/stx2357}{\JournalTitle{\mnras}, 473, 576}, \eprint arXiv:{1705.10797}

\bibitem[{{Graham} {et~al.}(2020){Graham} \& {Ford} {et~al.}}]{Graham}
{Graham}, M.~J., {Ford}, K.~E.~S., {McKernan}, B., {et~al.} 2020, \href{http://dx.doi.org/10.1103/PhysRevLett.124.251102}{\JournalTitle{\prl}, 124}, \eprint arXiv:{2006.14122}

\bibitem[{{Gray} {et~al.}(2023){Gray} \& {Beirnaert} {et~al.}}]{gray2023}
{Gray}, R., {Beirnaert}, F., {Karathanasis}, C., {et~al.} 2023, \href{http://dx.doi.org/10.48550/arXiv.2308.02281}{\JournalTitle{arXiv e-prints}, arXiv:2308.02281}, \eprint arXiv:{2308.02281}

\bibitem[{Herner {et~al.}(2020)Herner \& Annis {et~al.}}]{HERNER2020100425}
Herner, K., Annis, J., Brout, D., {et~al.} 2020, \href{http://dx.doi.org/https://doi.org/10.1016/j.ascom.2020.100425}{\JournalTitle{Astronomy and Computing}, 33, 100425}

\bibitem[{{Hotokezaka} \& {Nakar}(2020)}]{hotokezaka_2020}
{Hotokezaka}, K. \& {Nakar}, E. 2020, \href{http://dx.doi.org/10.3847/1538-4357/ab6a98}{\JournalTitle{\apj}, 891, 152}, \eprint arXiv:{1909.02581}

\bibitem[{{Kasen} {et~al.}(2017){Kasen} \& {Metzger} \& {Barnes} \& {Quataert} \& {Ramirez-Ruiz}}]{kasen2017elements}
{Kasen}, D., {Metzger}, B., {Barnes}, J., {Quataert}, E., \& {Ramirez-Ruiz}, E. 2017, \href{http://dx.doi.org/10.1038/nature24453}{\JournalTitle{\nat}, 551, 80}, \eprint arXiv:{1710.05463}

\bibitem[{Kasen {et~al.}(2017)Kasen \& Metzger \& Barnes \& Quataert \& Ramirez-Ruiz}]{kasen2017origin}
Kasen, D., Metzger, B., Barnes, J., Quataert, E., \& Ramirez-Ruiz, E. 2017, \JournalTitle{Nature}, 551, 80

\bibitem[{{Kawaguchi} {et~al.}(2020{\natexlab{a}}){Kawaguchi} \& {Shibata} \& {Tanaka}}]{kawaguchi_2020_bh}
{Kawaguchi}, K., {Shibata}, M., \& {Tanaka}, M. 2020{\natexlab{a}}, \href{http://dx.doi.org/10.3847/1538-4357/ab8309}{\JournalTitle{\apj}, 893, 153}, \eprint arXiv:{2002.01662}

\bibitem[{{Kawaguchi} {et~al.}(2020{\natexlab{b}}){Kawaguchi} \& {Shibata} \& {Tanaka}}]{Kawaguchi_2020}
{Kawaguchi}, K., {Shibata}, M., \& {Tanaka}, M. 2020{\natexlab{b}}, \href{http://dx.doi.org/10.3847/1538-4357/ab61f6}{\JournalTitle{\apj}, 889, 171}, \eprint arXiv:{1908.05815}

\bibitem[{{Kessler} {et~al.}(2009){Kessler} \& {Bernstein} {et~al.}}]{kessler_2009}
{Kessler}, R., {Bernstein}, J.~P., {Cinabro}, D., {et~al.} 2009, \href{http://dx.doi.org/10.1086/605984}{\JournalTitle{\pasp}, 121, 1028}, \eprint arXiv:{0908.4280}

\bibitem[{{Kessler} {et~al.}(2019){Kessler} \& {Narayan} {et~al.}}]{kessler_2019}
{Kessler}, R., {Narayan}, G., {Avelino}, A., {et~al.} 2019, \href{http://dx.doi.org/10.1088/1538-3873/ab26f1}{\JournalTitle{\pasp}, 131, 094501}, \eprint arXiv:{1903.11756}

\bibitem[{{Kilpatrick} {et~al.}(2017){Kilpatrick} \& {Foley} {et~al.}}]{Kilpatrick_2017}
{Kilpatrick}, C.~D., {Foley}, R.~J., {Kasen}, D., {et~al.} 2017, \href{http://dx.doi.org/10.1126/science.aaq0073}{\JournalTitle{Science}, 358, 1583}, \eprint arXiv:{1710.05434}

\bibitem[{Kilpatrick {et~al.}(2021)Kilpatrick \& Coulter {et~al.}}]{Kilpatrick_2021}
Kilpatrick, C.~D., Coulter, D.~A., Arcavi, I., {et~al.} 2021, \href{http://dx.doi.org/10.3847/1538-4357/ac23c6}{\JournalTitle{The Astrophysical Journal}, 923, 258}

\bibitem[{{Kilpatrick} {et~al.}(2022){Kilpatrick} \& {Fong} {et~al.}}]{kilpatrick22}
{Kilpatrick}, C.~D., {Fong}, W.-f., {Blanchard}, P.~K., {et~al.} 2022, \href{http://dx.doi.org/10.3847/1538-4357/ac3e59}{\JournalTitle{\apj}, 926, 49}, \eprint arXiv:{2109.06211}

\bibitem[{{Knox} \& {Millea}(2020)}]{knox2020}
{Knox}, L. \& {Millea}, M. 2020, \href{http://dx.doi.org/10.1103/PhysRevD.101.043533}{\JournalTitle{\prd}, 101, 043533}, \eprint arXiv:{1908.03663}

\bibitem[{{LIGO Scientific Collaboration}(2018)}]{lalsuite}
{LIGO Scientific Collaboration}. 2018, {LIGO} {A}lgorithm {L}ibrary - {LALS}uite, free software (GPL)

\bibitem[{{LIGO Scientific Collaboration} \& {Virgo Collaboration}(2019)}]{LVC_O2_StS}
{LIGO Scientific Collaboration} \& {Virgo Collaboration}. 2019, A gravitational-wave measurement of the Hubble constant following the second observing run of Advanced LIGO and Virgo, \eprint arXiv:{1908.06060}

\bibitem[{{Lipunov} {et~al.}(2017){Lipunov} \& {Gorbovskoy} {et~al.}}]{lipunov}
{Lipunov}, V.~M., {Gorbovskoy}, E., {Kornilov}, V.~G., {et~al.} 2017, \href{http://dx.doi.org/10.3847/2041-8213/aa92c0}{\JournalTitle{\apjl}, 850, L1}, \eprint arXiv:{1710.05461}

\bibitem[{{Morgan} {et~al.}(2020){Morgan} \& {Soares-Santos} {et~al.}}]{rob2020}
{Morgan}, R., {Soares-Santos}, M., {Annis}, J., {et~al.} 2020, \href{http://dx.doi.org/10.3847/1538-4357/abafaa}{\JournalTitle{\apj}, 901, 83}, \eprint arXiv:{2006.07385}

\bibitem[{Neilsen {et~al.}(2016)Neilsen \& Bernstein \& Gruendl \& Kent}]{osti_1250877}
Neilsen, E., Bernstein, G., Gruendl, R., \& Kent, S. 2016, \href{http://dx.doi.org/10.2172/1250877}{\JournalTitle{Fermilab Technical Notes}}

\bibitem[{Nicholl {et~al.}(2017)Nicholl \& Berger {et~al.}}]{Nicholl_2017}
Nicholl, M., Berger, E., Kasen, D., {et~al.} 2017, \href{http://dx.doi.org/10.3847/2041-8213/aa9029}{\JournalTitle{The Astrophysical Journal}, 848, L18}

\bibitem[{{Nissanke} {et~al.}(2010){Nissanke} \& {Holz} \& {Hughes} \& {Dalal} \& {Sievers}}]{nissanke2010}
{Nissanke}, S., {Holz}, D.~E., {Hughes}, S.~A., {Dalal}, N., \& {Sievers}, J.~L. 2010, \href{http://dx.doi.org/10.1088/0004-637X/725/1/496}{\JournalTitle{\apj}, 725, 496}, \eprint arXiv:{0904.1017}

\bibitem[{Oates {et~al.}(2021)Oates \& Marshall {et~al.}}]{SwiftO3_Oates2021}
Oates, S.~R., Marshall, F.~E., Breeveld, A.~A., {et~al.} 2021, \href{http://dx.doi.org/10.1093/mnras/stab2189}{\JournalTitle{Monthly Notices of the Royal Astronomical Society}, 507, 1296}

\bibitem[{{O'Donnell}(1994)}]{1994ApJ...422..158O}
{O'Donnell}, J.~E. 1994, \href{http://dx.doi.org/10.1086/173713}{\JournalTitle{\apj}, 422, 158}

\bibitem[{{Palmese} {et~al.}(2021){Palmese} \& {Bom} \& {Mucesh} \& {Hartley}}]{palmese21_sts}
{Palmese}, A., {Bom}, C.~R., {Mucesh}, S., \& {Hartley}, W. 2021, \JournalTitle{arXiv e-prints}, \eprint arXiv:{2111.06445}

\bibitem[{Palmese {et~al.}(2021)Palmese \& Fishbach \& Burke \& Annis \& Liu}]{Palmese_2021}
Palmese, A., Fishbach, M., Burke, C.~J., Annis, J., \& Liu, X. 2021, \href{http://dx.doi.org/10.3847/2041-8213/ac0883}{\JournalTitle{The Astrophysical Journal Letters}, 914, L34}

\bibitem[{{Palmese} {et~al.}(2020){Palmese} \& {deVicente} {et~al.}}]{palmese20_sts}
{Palmese}, A., {deVicente}, J., {Pereira}, M.~E.~S., {et~al.} 2020, \href{http://dx.doi.org/10.3847/2041-8213/abaeff}{\JournalTitle{\apjl}, 900, L33}, \eprint arXiv:{2006.14961}

\bibitem[{{Petrov} {et~al.}(2022){Petrov} \& {Singer} {et~al.}}]{petrov2022}
{Petrov}, P., {Singer}, L.~P., {Coughlin}, M.~W., {et~al.} 2022, \href{http://dx.doi.org/10.3847/1538-4357/ac366d}{\JournalTitle{\apj}, 924, 54}, \eprint arXiv:{2108.07277}

\bibitem[{{Planck Collaboration} {et~al.}(2020){Planck Collaboration} \& {Aghanim} {et~al.}}]{planckcolab}
{Planck Collaboration}, {Aghanim}, N., {Akrami}, Y., {et~al.} 2020, \href{http://dx.doi.org/10.1051/0004-6361/201833910}{\JournalTitle{\aap}, 641, A6}, \eprint arXiv:{1807.06209}

\bibitem[{Riess {et~al.}(2021)Riess \& Casertano {et~al.}}]{riess_2021}
Riess, A.~G., Casertano, S., Yuan, W., {et~al.} 2021, \href{http://dx.doi.org/10.3847/2041-8213/abdbaf}{\JournalTitle{The Astrophysical Journal}, 908, L6}

\bibitem[{{Sagu{\'e}s Carracedo} {et~al.}(2021){Sagu{\'e}s Carracedo} \& {Bulla} \& {Feindt} \& {Goobar}}]{sagues_2021}
{Sagu{\'e}s Carracedo}, A., {Bulla}, M., {Feindt}, U., \& {Goobar}, A. 2021, \href{http://dx.doi.org/10.1093/mnras/stab872}{\JournalTitle{\mnras}, 504, 1294}, \eprint arXiv:{2004.06137}

\bibitem[{{Savchenko} {et~al.}(2017){Savchenko} \& {Ferrigno} {et~al.}}]{savchenko2017}
{Savchenko}, V., {Ferrigno}, C., {Kuulkers}, E., {et~al.} 2017, \href{http://dx.doi.org/10.3847/2041-8213/aa8f94}{\JournalTitle{\apjl}, 848}, \eprint arXiv:{1710.05449}

\bibitem[{{Sch{\"o}neberg} {et~al.}(2022){Sch{\"o}neberg} \& {Abell{\'a}n} \& {S{\'a}nchez} \& {Witte} \& {Poulin} \& {Lesgourgues}}]{olympics_2021}
{Sch{\"o}neberg}, N., {Abell{\'a}n}, G.~F., {S{\'a}nchez}, A.~P., {et~al.} 2022, \href{http://dx.doi.org/10.1016/j.physrep.2022.07.001}{\JournalTitle{\physrep}, 984, 1}, \eprint arXiv:{2107.10291}

\bibitem[{{Schutz}(1986)}]{schutz}
{Schutz}, B.~F. 1986, \href{http://dx.doi.org/10.1038/323310a0}{\JournalTitle{\nat}, 323, 310}

\bibitem[{{Schutz}(2011)}]{schutz2011}
{Schutz}, B.~F. 2011, \href{http://dx.doi.org/10.1088/0264-9381/28/12/125023}{\JournalTitle{Classical and Quantum Gravity}, 28, 125023}, \eprint arXiv:{1102.5421}

\bibitem[{{Setzer} {et~al.}(2022){Setzer} \& {Peiris} \& {Korobkin} \& {Rosswog}}]{setzer2022}
{Setzer}, C.~N., {Peiris}, H.~V., {Korobkin}, O., \& {Rosswog}, S. 2022, \JournalTitle{arXiv e-prints}, \eprint arXiv:{2205.12286}

\bibitem[{Shandonay {et~al.}(2022)Shandonay \& Morgan {et~al.}}]{Shandonay_2022}
Shandonay, A., Morgan, R., Bechtol, K., {et~al.} 2022, \href{http://dx.doi.org/10.3847/1538-4357/ac3760}{\JournalTitle{The Astrophysical Journal}, 925, 44}

\bibitem[{Singer \& Price(2016)}]{bayestar}
Singer, L.~P. \& Price, L.~R. 2016, \href{http://dx.doi.org/10.1103/PhysRevD.93.024013}{\JournalTitle{Phys. Rev. D}, 93, 024013}

\bibitem[{Singer {et~al.}(2016)Singer \& Chen {et~al.}}]{Singer_2016}
Singer, L.~P., Chen, H.-Y., Holz, D.~E., {et~al.} 2016, \href{http://dx.doi.org/10.3847/2041-8205/829/1/l15}{\JournalTitle{The Astrophysical Journal}, 829, L15}

\bibitem[{{Singer} {et~al.}(2016{\natexlab{a}}){Singer} \& {Chen} {et~al.}}]{singer2016}
{Singer}, L.~P., {Chen}, H.-Y., {Holz}, D.~E., {et~al.} 2016{\natexlab{a}}, \href{http://dx.doi.org/10.3847/2041-8205/829/1/L15}{\JournalTitle{\apjl}, 829, L15}, \eprint arXiv:{1603.07333}

\bibitem[{{Singer} {et~al.}(2016{\natexlab{b}}){Singer} \& {Chen} {et~al.}}]{Singer_supp}
{Singer}, L.~P., {Chen}, H.-Y., {Holz}, D.~E., {et~al.} 2016{\natexlab{b}}, \href{http://dx.doi.org/10.3847/0067-0049/226/1/10}{\JournalTitle{\apjs}, 226, 10}, \eprint arXiv:{1605.04242}

\bibitem[{Soares-Santos {et~al.}(2017)Soares-Santos \& Holz {et~al.}}]{soares2017electromagnetic}
Soares-Santos, M., Holz, D., Annis, J., {et~al.} 2017, \JournalTitle{The Astrophysical Journal Letters}, 848, L16

\bibitem[{{Soares-Santos} {et~al.}(2017){Soares-Santos} \& {Holz} {et~al.}}]{marcelle2017}
{Soares-Santos}, M., {Holz}, D.~E., {Annis}, J., {et~al.} 2017, \href{http://dx.doi.org/10.3847/2041-8213/aa9059}{\JournalTitle{\apjl}, 848, L16}, \eprint arXiv:{1710.05459}

\bibitem[{{Soares-Santos} {et~al.}(2019){Soares-Santos} \& {Palmese} {et~al.}}]{darksiren1}
{Soares-Santos}, M., {Palmese}, A., {Hartley}, W., {et~al.} 2019, \href{http://dx.doi.org/10.3847/2041-8213/ab14f1}{\JournalTitle{\apjl}, 876, L7}, \eprint arXiv:{1901.01540}

\bibitem[{{Stewart} {et~al.}(2022){Stewart} \& {Lo} {et~al.}}]{stewart2022}
{Stewart}, A.~R., {Lo}, L.-T., {Korobkin}, O., {et~al.} 2022, \JournalTitle{arXiv e-prints}, \eprint arXiv:{2201.01865}

\bibitem[{{The LIGO Scientific Collaboration} {et~al.}(2021){The LIGO Scientific Collaboration} \& {the Virgo Collaboration} \& {the KAGRA Collaboration} \& {Abbott} \& {Abbott} \& {Zweizig}}]{gwtc3}
{The LIGO Scientific Collaboration}, {the Virgo Collaboration}, {the KAGRA Collaboration}, {et~al.} 2021, \JournalTitle{arXiv e-prints}, \eprint arXiv:{2111.03606}

\bibitem[{{Thielemann} {et~al.}(2017){Thielemann} \& {Eichler} \& {Panov} \& {Wehmeyer}}]{thielemann2017}
{Thielemann}, F.~K., {Eichler}, M., {Panov}, I.~V., \& {Wehmeyer}, B. 2017, \href{http://dx.doi.org/10.1146/annurev-nucl-101916-123246}{\JournalTitle{Annual Review of Nuclear and Particle Science}, 67, 253}, \eprint arXiv:{1710.02142}

\bibitem[{{Tucker} {et~al.}(2022){Tucker} \& {Wiesner} {et~al.}}]{douglas2021}
{Tucker}, D.~L., {Wiesner}, M.~P., {Allam}, S.~S., {et~al.} 2022, \href{http://dx.doi.org/10.3847/1538-4357/ac5b60}{\JournalTitle{\apj}, 929, 115}, \eprint arXiv:{2109.13351}

\bibitem[{Utsumi {et~al.}(2017)Utsumi \& Tanaka {et~al.}}]{10.1093/pasj/psx118}
Utsumi, Y., Tanaka, M., Tominaga, N., {et~al.} 2017, \href{http://dx.doi.org/10.1093/pasj/psx118}{\JournalTitle{Publications of the Astronomical Society of Japan}, 69}, 101

\bibitem[{Valentino {et~al.}(2021)Valentino \& Mena {et~al.}}]{Di_Valentino_2021}
Valentino, E.~D., Mena, O., Pan, S., {et~al.} 2021, \href{http://dx.doi.org/10.1088/1361-6382/ac086d}{\JournalTitle{Classical and Quantum Gravity}, 38, 153001}

\bibitem[{{Villar} {et~al.}(2017){Villar} \& {Guillochon} {et~al.}}]{villar2017}
{Villar}, V.~A., {Guillochon}, J., {Berger}, E., {et~al.} 2017, \href{http://dx.doi.org/10.3847/2041-8213/aa9c84}{\JournalTitle{\apjl}, 851, L21}, \eprint arXiv:{1710.11576}

\bibitem[{{Wollaeger} {et~al.}(2021){Wollaeger} \& {Fryer} {et~al.}}]{wollaeger_2021}
{Wollaeger}, R.~T., {Fryer}, C.~L., {Chase}, E.~A., {et~al.} 2021, \href{http://dx.doi.org/10.3847/1538-4357/ac0d03}{\JournalTitle{\apj}, 918, 10}, \eprint arXiv:{2105.11543}

\bibitem[{{Zhu} {et~al.}(2021){Zhu} \& {Wu} {et~al.}}]{zhu_2021}
{Zhu}, J.-P., {Wu}, S., {Yang}, Y.-P., {et~al.} 2021, \href{http://dx.doi.org/10.3847/1538-4357/ac19a7}{\JournalTitle{\apj}, 921, 156}, \eprint arXiv:{2106.15781}

\bibitem[{{Zhu} {et~al.}(2023){Zhu} \& {Wu} {et~al.}}]{zhu-wu2021}
{Zhu}, J.-P., {Wu}, S., {Yang}, Y.-P., {et~al.} 2023, \href{http://dx.doi.org/10.3847/1538-4357/aca527}{\JournalTitle{\apj}, 942, 88}, \eprint arXiv:{2110.10469}

\end{thebibliography}
